\documentclass[a4paper,twocolumn,11pt,accepted=2023-02-10]{quantumarticle}

\pdfoutput=1

\newcommand{\myinv}[1]{#1^{\scalebox{0.9}[1.0]{-}1}}
\newcommand{\tr}{{\rm Tr}}
\newtheorem{definition}{Definition}
\usepackage{tikz}
\usepackage[numbers, sort&compress]{natbib}
\usepackage{amsmath,amsfonts,amssymb}
\newcommand{\id}{\mathbb{I}}

\usetikzlibrary{decorations.pathreplacing,decorations.markings,calc,3d,positioning,quotes}

\tikzset{
  tensor/.style={
    inner sep = 0.055cm,
    shape = circle,
    draw,
    fill
  },
  t/.style={
    inner sep = 0.03cm,
    shape = circle,
    draw,
    fill
  },
}

\tikzset{
  hopf2/.style={
    line cap=round,
    line width=1.5mm,
  },
  epsilon2/.style={
   draw = red,
   thin,
   densely dotted,
   rounded corners,
   fill opacity=0.2,
   fill=red,
  },
  every picture/.style = {
    baseline={([yshift=-.5ex]current bounding box.center)},
    font=\scriptsize
  },
  irrep/.style={
    anchor=south,
    font = \tiny,
    inner sep=2pt
  }
}

\tikzset{
  every label/.style = {
    text depth=0pt,
    text height=1ex,
  },
}

\tikzset{
  pics/V/.style args={#1/#2/#3}{
    code = {
	    \coordinate (-mid) at (-0.5,0);
	    \coordinate (-up) at (0.5,0.3);
	    \coordinate (-down) at (0.5,-0.3);
	    \coordinate (-top) at (0,0.45);
	    \coordinate (-bottom) at (0,-0.45);
	    \draw[hopf2] (0,-0.35)--(0,0.35);
	    \draw (0,0.3)--(-up);
	    \draw (0,-0.3)--(-down);
	    \draw (0,0)--(-mid);
	    \node[irrep] at (-0.25,0) {$#1$};
	    \node[irrep] at (0.25,0.3) {$#2$};
	    \node[irrep] at (0.25,-0.3) {$#3$};
    }
  },
  pics/W/.style args={#1/#2/#3}{
    code = {
	    \coordinate (-mid) at (0.5,0);
	    \coordinate (-up) at (-0.5,0.3);
	    \coordinate (-down) at (-0.5,-0.3);
	    \coordinate (-top) at (0,0.45);
	    \coordinate (-bottom) at (0,-0.45);
	    \draw[hopf2] (0,-0.35)--(0,0.35);
	    \draw (0,0.3)--(-up);
	    \draw (0,-0.3)--(-down);
	    \draw (0,0)--(-mid);
	    \node[irrep] at (0.25,0) {$#1$};
	    \node[irrep] at (-0.25,0.3) {$#2$};
	    \node[irrep] at (-0.25,-0.3) {$#3$};
    }
  },
      pics/V1/.style args={#1/#2/#3}{
    code = {
	    \coordinate (-mid) at (-0.4,0);
	    \coordinate (-up) at (0.5,0.3);
	    \coordinate (-down) at (0.5,-0.3);
	    \coordinate (-top) at (0,0.45);
	    \coordinate (-bottom) at (0,-0.45);
	  	    \draw[red] (0,0.3)--(-up);
	    \draw[red] (0,-0.3)--(-down);
	    \draw[red] (0,0)--(-mid);
	      \draw[hopf2] (0,-0.35)--(0,0.35);
	    \node[irrep] at (-0.25,0) {$#1$};
	    \node[irrep] at (0.25,0.3) {$#2$};
	    \node[irrep] at (0.25,-0.3) {$#3$};
    }
  },
  pics/W1/.style args={#1/#2/#3}{
    code = {
	    \coordinate (-mid) at (0.4,0);
	    \coordinate (-up) at (-0.5,0.3);
	    \coordinate (-down) at (-0.5,-0.3);
	    \coordinate (-top) at (0,0.45);
	    \coordinate (-bottom) at (0,-0.45);
	   \draw[red] (0,0.3)--(-up);
	    \draw[red] (0,-0.3)--(-down);
	    \draw[red] (0,0)--(-mid);
	    	    \draw[hopf2] (0,-0.35)--(0,0.35);
	    \node[irrep] at (0.25,0) {$#1$};
	    \node[irrep] at (-0.25,0.3) {$#2$};
	    \node[irrep] at (-0.25,-0.3) {$#3$};
    }
  },
    pics/V2/.style args={#1/#2/#3}{
    code = {
	    \coordinate (-mid) at (-0.4,0);
	    \coordinate (-up) at (0.5,0.3);
	    \coordinate (-down) at (0.5,-0.3);
	    \coordinate (-top) at (0,0.45);
	    \coordinate (-bottom) at (0,-0.45);
	    \draw[red] (0,0.3)--(-up);
	    \draw (0,-0.3)--(-down);
	    \draw (0,0)--(-mid);
	    \draw[hopf2,gray] (0,-0.35)--(0,0.35);
	    \node[irrep] at (-0.25,0) {$#1$};
	    \node[irrep] at (0.25,0.3) {$#2$};
	    \node[irrep] at (0.25,-0.3) {$#3$};
    }
  },
  pics/W2/.style args={#1/#2/#3}{
    code = {
	    \coordinate (-mid) at (0.4,0);
	    \coordinate (-up) at (-0.5,0.3);
	    \coordinate (-down) at (-0.5,-0.3);
	    \coordinate (-top) at (0,0.45);
	    \coordinate (-bottom) at (0,-0.45);
	   \draw[red] (0,0.3)--(-up);
	    \draw (0,-0.3)--(-down);
	    \draw (0,0)--(-mid);
	    \draw[hopf2,gray] (0,-0.35)--(0,0.35);
	    \node[irrep] at (0.25,0) {$#1$};
	    \node[irrep] at (-0.25,0.3) {$#2$};
	    \node[irrep] at (-0.25,-0.3) {$#3$};
    }
  },
     pics/V3/.style args={#1/#2/#3}{
    code = {
	    \coordinate (-mid) at (-0.4,0);
	    \coordinate (-up) at (0.5,0.3);
	    \coordinate (-down) at (0.5,-0.3);
	    \coordinate (-top) at (0,0.45);
	    \coordinate (-bottom) at (0,-0.45);
	    \draw[red] (0,0.3)--(-up);
	    \draw (0,-0.3)--(-down);
	    \draw (0,0)--(-mid);
	    \filldraw[fill=gray!30, draw=black] (-0.07,-0.4) rectangle (0.07,0.4);
	    \node[irrep] at (-0.25,0) {$#1$};
	    \node[irrep] at (0.25,0.3) {$#2$};
	    \node[irrep] at (0.25,-0.3) {$#3$};
    }
  },
  pics/W3/.style args={#1/#2/#3}{
    code = {
	    \coordinate (-mid) at (0.4,0);
	    \coordinate (-up) at (-0.5,0.3);
	    \coordinate (-down) at (-0.5,-0.3);
	    \coordinate (-top) at (0,0.45);
	    \coordinate (-bottom) at (0,-0.45);
	   \draw[red] (0,0.3)--(-up);
	    \draw (0,-0.3)--(-down);
	    \draw (0,0)--(-mid);
	    	    \filldraw[fill=gray!30, draw=black] (-0.07,-0.4) rectangle (0.07,0.4);
	    \node[irrep] at (0.25,0) {$#1$};
	    \node[irrep] at (-0.25,0.3) {$#2$};
	    \node[irrep] at (-0.25,-0.3) {$#3$};
    }
  },
    pics/W2big/.style args={#1/#2/#3}{
    code = {
	    \coordinate (-mid) at (0.4,0);
	    \coordinate (-up) at (-0.5,0.45);
	    \coordinate (-down) at (-0.5,-0.45);
	    \coordinate (-top) at (0,0.45);
	    \coordinate (-bottom) at (0,-0.45);
	   \draw[red] (0,0.45)--(-up);
	    \draw (0,-0.45)--(-down);
	    \draw (0,0)--(-mid);
	    \draw[hopf2,gray] (0,-0.5)--(0,0.5);
	    \node[irrep] at (0.25,0) {$#1$};
	    \node[irrep] at (-0.25,0.45) {$#2$};
	    \node[irrep] at (-0.25,-0.45) {$#3$};
    }
  },
  pics/V2p/.style args={#1/#2/#3}{
    code = {
	    \coordinate (-mid) at (-0.1,0);
	    \coordinate (-up) at (0.4,0.3);
	    \coordinate (-down) at (0.1,-0.3);
	    \coordinate (-top) at (0,0.45);
	    \coordinate (-bottom) at (0,-0.45);
	    \draw[red] (0,0.3)--(-up);
	    \draw (0,-0.3)--(-down);
	    \draw (0,0)--(-mid);
	    \draw[hopf2,gray] (0,-0.35)--(0,0.35);
	    \node[irrep] at (-0.2,-0.2) {$#1$};
	    \node[irrep] at (0.25,0.3) {$#2$};
	    \node[irrep] at (0.25,-0.4) {$#3$};
    }
  },
  pics/W2p/.style args={#1/#2/#3}{
    code = {
	    \coordinate (-mid) at (0.1,0);
	    \coordinate (-up) at (-0.4,0.3);
	    \coordinate (-down) at (-0.1,-0.3);
	    \coordinate (-top) at (0,0.45);
	    \coordinate (-bottom) at (0,-0.45);
	   \draw[red] (0,0.3)--(-up);
	    \draw (0,-0.3)--(-down);
	    \draw (0,0)--(-mid);
	    \draw[hopf2,gray] (0,-0.35)--(0,0.35);
	    \node[irrep] at (0.2,-0.2) {$#1$};
	    \node[irrep] at (-0.25,0.3) {$#2$};
	    \node[irrep] at (-0.25,-0.4) {$#3$};
    }
  }
 }

\begin{document}

\title{Classifying phases protected by matrix product operator symmetries using matrix product states}

\author{ Jos\'e Garre-Rubio}
\affiliation{\mbox{University of Vienna, Faculty of Mathematics, Oskar-Morgenstern-Platz 1, 1090 Wien, Austria}}

\author{Laurens Lootens}
\affiliation{Department of Physics and Astronomy, Ghent University, Krijgslaan 281, S9, 9000 Ghent, Belgium}

\author{ Andr\'as Moln\'ar}
\affiliation{\mbox{University of Vienna, Faculty of Mathematics, Oskar-Morgenstern-Platz 1, 1090 Wien, Austria}}

\begin{abstract}

We classify the different ways in which matrix product states (MPSs) can stay invariant under the action of matrix product operator (MPO) symmetries. This is achieved through a local characterization of how the MPSs, that generate a ground space, remain invariant under a global MPO symmetry. This characterization yields a set of quantities satisfying the coupled pentagon equations, associated with a module category over the fusion category that describes the MPO symmetry. Equivalence classes of these quantities provide complete invariants for an MPO symmetry protected phase: they are robust under continuous deformations of the MPS tensor, and two phases with the same equivalence class can be connected by a symmetric gapped path. Our techniques match and extend the known renormalization fixed point classifications and facilitate the numerical study of these systems. For MPO symmetries described by a group, we recover the symmetry protected topological order classification for unique and degenerate ground states. Moreover, we study the interplay between time reversal symmetry and an MPO symmetry and we also provide examples of our classification, together with explicit constructions based on groups. Finally, we elaborate on the connection between our setup and gapped boundaries of two-dimensional topological systems, where MPO symmetries also play a key role.
\end{abstract}

\maketitle

\section{Introduction}

The classification of symmetric phases plays a fundamental role in condensed matter physics. Two systems are in the same phase if their Hamiltonians can be connected through a smooth gapped path of local and symmetric Hamiltonians. Along such a smooth path, the physical properties do not change abruptly \cite{Hastings05}. The phases which are invariant under on-site symmetries arising in this setup are called symmetry protected topological (SPT) phases \cite{Pollmann12prot}. Different SPT phases cannot be connected via a smooth, gapped and symmetric path, and as such are protected by the global on-site symmetry. In stark contrast to Landau's paradigm of local symmetry breaking, these SPT phases cannot be distinguished via a local order parameter. Remarkably, SPT phases not only exhibit exotic behaviours, such as degeneracies in the entanglement spectrum, fractional edges modes and non-trivial string orders \cite{Pollmann10}, but they also are powerful resources for applications like measurement-based one-way computation \cite{Raussendorf01}.

In the case of one dimensional spin systems which remain invariant under global on-site symmetries, a complete classification has been achieved \cite{Chen11,Schuch11,ogata2021classification}. 
That success arises from a fruitful interplay between quantum information and quantum matter by using matrix product states (MPSs) \cite{Fannes92,PerezGarcia07}, the family of tensor networks that satisfy an area law for the entanglement entropy and efficiently approximates the ground states of local gapped Hamiltonians \cite{Hastings07A,Hastings07B}. It turns out that the different SPT phases which are symmetric under a group $G$, are in one-to-one correspondence with the distinct projective representations of $G$, and MPSs naturally allow to derive such characterization \cite{Chen11}. This is because the on-site global symmetry can be encoded in a local transformation acting on the entanglement space, the so-called virtual level, of the tensors that constitute the MPS \cite{PerezGarcia07}. Remarkably, this local transformation on the entanglement space captures the topological information about the phase, that is, the projective representation. The success of MPSs as an interplay between quantum information and quantum matter extends the SPT classification: it has been used to extend the measurement-based quantum computation beyond the one-way model \cite{Gross07} as well as to design numerical and experimental probes to detect SPT phases \cite{Haegeman12,Pollmann12,Elben19}.

The goal of this manuscript is to generalize the approach of the SPT phase classification using MPSs in the case where the symmetry is not an on-site representation of a group. The generalization is twofold: first, we do not require the global symmetry to be on-site and second, we allow the symmetry to represent an algebra, not only a group. The only restriction we impose on the global symmetry is for it to be realized as a matrix product operator (MPO).

Generalized symmetries of (1+1)d systems, also in the form of MPOs, have been studied before for concrete models, revealing novel features not present in standard SPT phases. MPO symmetries of abelian groups give rise to anomalous domain wall excitations \cite{Roose19, Wang15A} and {\it topological symmetries} based on truncated $su(2)$ deformations studied in anyonic spin chains \cite{Feiguin07, Trebst08, Gils13} protect gapless phases present in those systems. Interestingly, symmetries of non-truncated $su(2)$ deformations have recently been argued to host SPT phases\cite{Quella20,Quella21}. Furthermore, symmetries based on fusion categories, whose MPO representation can be found in Refs. \cite{Sahinoglu14,Bultinck17A, Lootens21A}, are referred to as topological defects or categorical symmetries, and they have become ubiquitous in the description of both topologically ordered systems in (2+1)d and their (1+1)d boundary theories \cite{Kitaev12,hauru2016topological,aasen2016topological,vanhove2018mapping,Thorngren19,Aasen20,Ji21,vanhove2021topological,lootens2021category, Inamura22}. 

Our classification of MPO-symmetric MPSs, when the MPO represents a fusion category, coincides with the classification of (1+1)d gapped boundaries of (2+1)d topologically ordered systems. In particular, Refs.\ \cite{Kitaev12,Thorngren19} use a categorical approach that matches our classification in terms of MPO symmetries, resulting in the classification of  module categories of fusion categories. However, our approach allows to study these phases directly in their lattice realizations, which, together with their tensor network representations, facilitates the numerical investigation of these systems. Importantly, due to the nature of the techniques used, our classification extends to systems outside of renormalization group fixed points, {\it i.e.}\ to non-zero correlation length states. Additionally, in our study, we also study MPSs which are invariant under general MPOs, not necessarily representing boundary symmetries of (2+1)d topologically ordered states. In fact, we look for a more general definition of MPO invariant MPSs that in particular cover the cases of MPO representations of two-dimensional boundaries and could potentially be used in other situations.

\subsection*{Approach and summary of the results}

In this manuscript we use tensor networks to describe both the ground space and the symmetry operators. We start by giving a definition of a subspace, spanned by a set of MPSs, which remains invariant under the action of an MPO algebra. This MPO algebra is characterized by the data of a fusion category, in the form of $F$-symbols that satisfy the well-known pentagon equations \cite{Sahinoglu14,Bultinck17A}. The invariance of the MPSs under the MPOs results in a local condition: the action of the MPO tensors on the MPS 
 tensors translates into the virtual level of the latter. This local characterization of the global symmetry allows us to derive a set of quantities, which we call ${L}$-symbols, satisfying consistency equations in the form of coupled pentagon equations involving both the $F$ and the $L$ symbols. Given the ${F}$-symbols of the MPO, these equations classify the ${L}$-symbols through their inequivalent solutions and in particular, the possible solutions determine the possible ground state degeneracies compatible with the MPO symmetry. 

When the MPO symmetry is a (unitary) representation of a group $G$, where the $F$-symbol is a $3$-cocycle, the solutions of the coupled pentagon equations have a very compact form. It turns out that the solutions are labeled by: first, a subgroup $H\subset G$ that trivializes the $3$-cocycle of the MPO and second, a $2$-cocycle class in $\mathcal{H}^2(H,\mathbb{C}^*)$. The subgroup $H$ determines the ground state degeneracy $|G/H|$ and the $2$-cocycle class characterizes the SPT phase of the unbroken symmetry. 

This structure reduces itself to the standard SPT classification in situations where the MPO representation of the group is on-site, so that the $3$-cocycle is trivial and the local characterization becomes the well-known transformation of the MPS tensor: the on-site operator is translated to virtual symmetry operators. Then, the coupled pentagon equation corresponds to the $2$-cocycle condition, which classifies the projective representations of the virtual operators. 

Once the algebraic characterization is done, we proceed to prove that (equivalence classes of) ${L}$-symbols are the proper invariants that completely classify gapped phases protected by MPO symmetries. We provide the formal definition of the symmetric phase we use: the existence of a smooth and gapped Hamiltonian path, that commutes with the MPO symmetry, connecting two Hamiltonians. First, we show that a continuous deformation of the MPS tensor (corresponding to a continuous gapped path of the parent Hamiltonian) preserves the ${L}$-symbols. Second, we construct a well-behaved path connecting any two such systems with the same $F$ and $L$-symbols.

The structure of the paper is as follows. We start by providing the necessary background of MPSs and MPO in section \ref{sec:data}. In Section \ref{sec:inter} we present the main result of the paper: we define what it means for an MPS to be invariant under MPO symmetries, we show how the invariance is locally characterized, we construct the $L$-symbols, that characterize these phases, and the pentagon equations that they satisfy. In Section \ref{groupcase} we particularize to MPO algebras based on finite groups and we connect with the standard SPT classification of Ref.\ \cite{Schuch11} for on-site symmetries. We also discuss the case of MPSs which are symmetric under MPO (unitary) representations of a finite group and the interplay between time reversal symmetry and an MPO symmetry.

In Section \ref{sec:classif}, we formulate the extra conditions for an MPO to be a physical symmetry, we define what it means for two systems to be in the same phase under MPO symmetries, we show how Hamiltonians can be symmetrized, we prove that the $L$-symbols proposed in Section \ref{sec:inter} are invariant under continuous deformations of the MPS tensors and we construct a smooth and symmetric path connecting two MPSs with the same $L$-symbols. We also construct for every phase in our classification an MPS boundary which is invariant under the virtual MPO symmetries of the (2+1)d bulk tensor network state, that is, we explicitly connect to the classification of gapped boundary theories of (2+1)d topological phases. 

In Section \ref{sec:examples} we provide solutions of the ${L}$-symbols for different MPO symmetries and we construct explicit MPOs and MPSs for the group case. We finish the manuscript with Section \ref{sec:outlook}, where we comment on possible applications of our work.

\section{ Matrix product states and matrix product operators}\label{sec:data}

An MPS is defined via a $d\times D\times D$  tensor $A$, where $d$ is the dimension of the local Hilbert space $\mathbb{C}^d$ and $D$ is the bond dimension of the virtual space. The relevance of these states is that the ground states of local gapped Hamiltonians can be approximated with an MPS of small bond dimension (in term of the system size) \cite{Hastings07A}. Without loss of generality (w.l.o.g.) we consider MPSs that are translationally invariant in the bulk, where each tensor in the MPS is the same.

In this manuscript we define an MPS on $n$ sites with boundary $X\in \mathcal{M}_D$ and tensor $A$ as
$$|\psi^n_{A,X}\rangle = \sum_{\{l\}} \tr[XA^{l_1}{\cdots} A^{l_n}]|{l_1}, \dots, {l_n}\rangle,$$
where $A=\sum_l A^l \otimes |l\rangle$ and $A^l$ is a $D\times D$ matrix for each $l=1,\dots,d$. 
We say that the MPS is defined with arbitrary boundary conditions (arbitrary BC) if any matrix $X$ is allowed in the boundary.
We say that the MPS has periodic boundary conditions (PBC) if we only allow for boundary matrices proportional to the identity.

MPSs with arbitrary boundary conditions give rise to subspaces that have the same dimension independent of system size and that correspond to representations of coalgebras, see \cite{molnar22} for details. We define the following subspace by considering all matrices at the boundary of a $n$-site MPS:
\begin{equation}\label{MPSssubs}
\mathcal{S}^n_A= \left\{ |\psi^n_{A,X}\rangle, X\in \mathcal{M}_{D} \right\}.
\end{equation}

We consider block-injective tensors, $A^l = \bigoplus_{\alpha } A^l_\alpha$ where each  $A^l_\alpha$ is a $D_\alpha \times D_\alpha$ matrix and, $\alpha \in \{x,y,z,\dots\}$ denotes the block labels. Note that any MPS can be brought into a block-diagonal form \cite{PerezGarcia07}  and that  after blocking a {\it small} number of sites \cite{Sanz10}, the blocks become injective. Because any two injective MPSs become either orthogonal or proportional in the thermodynamic limit \cite{Cirac17A}, we restrict w.l.o.g. to MPSs whose blocks are injective and orthogonal between each other.

Importantly, to any block-injective MPS a gapped local and frustration free Hamiltonian, called {\it parent} Hamiltonian, can be associated. The ground space of the PBC parent Hamiltonian is spanned by the individual blocks of the MPS, $\{ |\psi_{A_\alpha, \id} \rangle, \alpha \in x, y, z, \dots \}$, so that its degeneracy coincides with the number of blocks. For example, the MPS tensor $A^0=|0\rangle \langle 0|, A^1=|1\rangle \langle 1|$ corresponds to the GHZ state and its parent Hamiltonian has a two-fold GS degeneracy spanned by $|00\cdots 0\rangle$ and $|11\cdots 1\rangle$. This degeneracy can be seen as symmetry breaking: there is always a symmetry action that permutes between the ground states i.e. between the different blocks of the MPS. In the previous example the symmetry is just $\sigma_x^{\otimes N}$. If we do not impose PBC, {\it i.e.}\ we are in the arbitrary BC where the Hamiltonian term between site $1$ and $n$ is not present, the ground space coincides with $\mathcal{S}^n_A$ and has degeneracy $\sum_\alpha D_\alpha^2$. 

Matrix product operators (MPOs) are operators that can be constructed using local tensors for any system size. We define an MPO by placing a $d\times d\times \chi \times \chi$ tensor $T$ at every site and placing a boundary condition given by a $\chi \times \chi$ matrix $B$:
$$ O^n_{T, B} = \sum_{ \{ l_k, p_k \} }  \tr[B T^{l_1 p_1} {\cdots} T^{l_n p_n} ] | l_1 {\cdots} l_n \rangle \langle  p_1 {\cdots} p_n |, $$
where $l,p=1,\dots, d$ and $T^{l,p}$ is defined through $T=\sum_{l,p}T^{l,p}\otimes |l\rangle \langle p|  $. As for MPSs, we assume w.l.o.g. that the MPO tensor $T$ is block-injective and that there are no repeated blocks. We denote the blocks as $\{a,b,c,\dots \}$, the block tensors as $T_a$ and its bond dimension as $\chi_a$. 

In this paper we consider MPOs that form an algebra with arbitrary boundary conditions, more precisely, we define the algebra $\mathcal{A}^n_T= \left\{ O^n_{T,B},B\in \mathcal{M}_\chi \right\}$ that satisfies

 {\it \bf(Closedness condition)} For every $X,Y\in \mathcal{M}_\chi$ there is a $Z\in \mathcal{M}_\chi$ such that for all $n$:
\begin{equation}\label{algcond}
 O^n_{T,X}   \cdot  O^n_{T,Y} = O^n_{T,Z},
 \end{equation}
or diagramatically:
\begin{equation*}
		\begin{tikzpicture}
		  \draw[red] (0,0) rectangle (2.55,-0.5);
		  \draw[red] (-0.2,0.3) rectangle (2.75,-0.7);
		  \foreach \x in {0.75,2.25}{
		    \node[tensor] (t\x) at (\x,0) {};
		    \node[tensor] (t\x) at (\x,0.3) {};
		    \draw (\x,-0.3) --++ (0,0.9);
		    \node[] at (\x+0.25,0.5) {$T$};
      }
		  \node[tensor,label=below:$Y$]  at (0.25,0) {};
		  \node[tensor,label=above:$X$]  at (0.25,0.3) {};
		  \node[fill=white] at (1.5,0) {$\dots$};
		  \node[fill=white] at (1.5,0.3) {$\dots$};
		\end{tikzpicture}  \   = \ 
		\begin{tikzpicture}
		  \draw[red] (0,0) rectangle (2.55,-0.5);
		  \foreach \x in {0.75,2.25}{
		    \node[tensor] (t\x) at (\x,0) {};
		    \node[] at (\x+0.25,0.25) {$T$};
		    \draw (\x,-.3) --++ (0,0.6);
		  }
		  \node[tensor,label=below:$Z$]  at (0.25,0) {};
		  \node[fill=white] at (1.5,0) {$\dots$};
		\end{tikzpicture} \ .
	\end{equation*}

The closedness condition implies the existence of two sets of unique (up to a gauge) rank-three tensors $W_{ab}^{c,\mu}$ and ${\hat{W}}_{ab}^{c,\mu}$, the so-called fusion tensors, that satisfy $\sum_m T^{l,m}_a  T^{m,p}_b = \sum_{c,\mu}{\hat{W}}_{ab}^{c,\mu}{\cdot} T^{l,p}_c {\cdot} {{W}}_{ab}^{c,\mu}$ (see appendix \ref{ap:proofs} for a proof). The fusion tensors also satisfy the orthogonality relations ${{W}}_{ab}^{c,\mu} {\hat{W}}_{ab}^{d,\nu} = \delta_{c,d} \delta_{\mu,\nu}  \id_{\chi_c}$. The index $\mu$ runs from $1$ to $N_{ab}^c$, the multiplicity of the product of $a$ and $b$ resulting in $c$. We notice that the MPOs $O_a \equiv O_{T_a,\id}$ satisfy $O_a {\cdot} O_b = \sum_c N_{ab}^c O_c$ for any system size. Diagramatically, the local decomposition can be written as
\begin{equation}\label{fusiontensors}
  \begin{tikzpicture}
    \draw[red]  (-0.5,0)--(0.5,0);
    \draw[red]  (-0.5,-0.5)--(0.5,-0.5);
    \node[irrep] at (-0.35,0) {$a$};
    \node[irrep] at (-0.35,-0.5) {$b$};
    \node[tensor] (t) at (0,0) {};
    \node[tensor] (t) at (0,-0.5) {};
    \draw (0,-0.8) -- (0,0.3);
  \end{tikzpicture} =
  \sum_{c,\mu}
  \begin{tikzpicture}[baseline=-1mm]
    \node[tensor] (t) at (0,0) {};
    \draw (0,-0.3) -- (0,0.3);
    \pic (v) at (0.5,0) {V1=c/a/b};
    \pic (w) at (-0.5,0) {W1=c/a/b};
     \node[anchor=south,inner sep=2pt] at (v-top) {$\mu$};
    \node[anchor=south,inner sep=2pt] at (w-top) {$\mu$};
  \end{tikzpicture} \ .
\end{equation}
and the orthogonality relations as
\begin{equation}
   \begin{tikzpicture}\label{eq:orthoW}
    \pic (v) at (0,0) {V1=c/a/b};
    \pic (w) at (0.6,0) {W1=d//};
     \node[anchor=south, inner sep=2pt] at (v-top) {$\mu$};
    \node[anchor=south, inner sep=2pt] at (w-top) {$\nu$};
   \end{tikzpicture} 
= \delta_{c,d} \delta_{\mu,\nu}  \id_c \ .
\end{equation}
Associativity of the product in the algebra implies the existence of ${F}$-symbols: 
\begin{equation}\label{Fsymbolsdef}
 \begin{tikzpicture}[baseline=0mm]
    \pic (w1) at (0,0) {V1=d/e/c};
    \pic (w2) at (w1-up) {V1=/a/b};
        \node[anchor=south,inner sep=5pt] at (w1-top) {$\nu$};
    \node[anchor=south,inner sep=5pt] at (w2-top) {$\mu$};
  \end{tikzpicture} =
  \sum_{f,\chi,\eta}
  \left({F}_{abc}^d\right)_{f\chi\eta}^{e \mu \nu}
  \begin{tikzpicture}[baseline=-3mm]
    \pic (w1) at (0,0) {V1=d/a/f};
    \pic (w2) at (w1-down) {V1=/b/c};
        \node[anchor=north,inner sep=5pt] at (w1-bottom) {$\eta$};
    \node[anchor=north,inner sep=5pt] at (w2-bottom) {$\chi$};
   \end{tikzpicture}
   \ \ ,
\end{equation}
that satisfy the so-called pentagon equations obtained from decomposing the product of four elements in equivalent ways:
\begin{align}\label{pentagon0F}
 \sum_{h\sigma\psi\rho} & \left( {F}_{abc}^g\right)_{h\psi\sigma}^{f\alpha\beta} \left( {F}_{ahd}^e\right)_{k\rho\lambda}^{g\sigma\gamma} \left( {F}_{bcd}^k\right)_{l\nu\mu}^{h\psi\rho} \notag \\
   & =  \sum_\delta \left( {F}_{fcd}^e\right)_{l\nu\delta}^{g\beta \gamma} \left( {F}_{abl}^e\right)_{k\mu\lambda}^{f\alpha\delta} \ .
\end{align}

Eq.\ \eqref{fusiontensors} is left invariant under transformations on the fusion tensors of the form $ {W}_{ab}^{c,\mu}  \to \sum_{\mu'} \left(Y_{ab}^c \right)_{\mu'}^\mu{W}_{a b}^{c,\mu'} $ and  $\hat{W}_{a b}^{c, \mu} \to \sum_{\mu'} {\left( {\hat{Y}_{a b}^c} \right)_{\mu'}^\mu} \hat{W}_{a b}^{c,\mu' }$, where $\hat{Y}$ denotes the inverse of $Y$. So $ \left({F}_{a b c}^d\right)_{f\chi\eta}^{e \mu \nu}$ and the ${F}$-symbol arising from the previous transformations of the fusion tensors,
 $$ \sum_{\chi',\eta',\mu',\nu'}
  \left(Y_{ab}^e \right)_{\mu'}^\mu
  \left(Y_{e c}^d \right)_{\nu'}^\nu  
     \left({\hat{Y}_{b c}^f}\right)_{\chi'}^\chi
        \left({\hat{Y}_{a f}^d}\right)_{\eta'}^\eta
  \left({F}_{a b c}^d\right)_{f\chi\eta}^{e \mu \nu} 
$$
are considered equivalent. Then ${F}$-symbols are classified by the pentagon equations \eqref{pentagon0F} when quotienting by the above transformations.

It is worth to mention that to obtain the pentagon equations, only a finite number of injective blocks and the closedness condition are required, no extra assumptions are imposed on the MPO algebra. Generalized symmetries are usually associated to {\it fusion categories}, see for example Ref.\ \cite{Thorngren19} and references therein, which have an identity element, the notion of dual element, a pivotal structure, etc. MPO representations of fusion categories have been characterized previously in Refs.\ \cite{Bultinck17A, Lootens21A}, where the fusion tensors satisfy extra properties. However, we will not follow this approach and we will derive our results as general as possible to allow to use them outside the framework of condensed matter physics. Only when we focus on symmetries of quantum systems (Section \ref{sec:classif}) we will state further assumptions.

\section{MPS symmetric under MPOs}\label{sec:inter}
An MPO symmetric MPS is defined as the invariance of the subspace $\mathcal{S}_A^n$ under the action of the algebra $\mathcal{A}_T^n$ for all system sizes $n$:
$$\mathcal{A}_T \cdot \mathcal{S}_A \subset \mathcal{S}_A. $$ 
Concretely, for any two elements of $\mathcal{S}_A^n$ and $\mathcal{A}_T^n$, generated by matrices $X\in \mathcal{M}_D$ and $B\in \mathcal{M}_\chi$ respectively, there is always an element of $\mathcal{S}_A^n$ independent of $n$, {\it i.e.}\ a matrix $Y \in \mathcal{M}_D$ such that
\begin{equation}\label{eq:compatible}
			\begin{tikzpicture}
		  \draw (0,0) rectangle (2.55,-0.5);
		  \draw[red] (-0.2,0.3) rectangle (2.75,-0.7);
		  \foreach \x in {0.75,2.25}{
		    \node[tensor,label=below:$A$] (t\x) at (\x,0) {};
		    \node[tensor] (t\x) at (\x,0.3) {};
		    \draw (\x,0) --++ (0,0.6);
		    \node[] at (\x+0.25,0.5) {$T$};
      }
		  \node[tensor,label=below:$X$]  at (0.25,0) {};
		  \node[tensor,label=above:$B$]  at (0.25,0.3) {};
		  \node[fill=white] at (1.5,0) {$\dots$};
		  \node[fill=white] at (1.5,0.3) {$\dots$};
		\end{tikzpicture}  \   = \ 
		\begin{tikzpicture}
		  \draw (0,0) rectangle (2.55,-0.5);
		  \foreach \x in {0.75,2.25}{
		    \node[tensor,label=below:$A$] (t\x) at (\x,0) {};
		    \draw (\x,0) --++ (0,0.3);
		  }
		  \node[tensor,label=below:$Y$]  at (0.25,0) {};
		  \node[fill=white] at (1.5,0) {$\dots$};
		\end{tikzpicture} \ .
	\end{equation}
This global condition can be translated into the existence of a set of rank-three tensors that decompose the action of the MPO tensor on the MPS tensor (the proof of appendix \ref{ap:proofs} can also be applied here). We name them {\it action tensors} and they are defined by their action onto each block,
$V_{ax}^{y,i}: \mathbb{C}^{\chi_a} \otimes \mathbb{C}^{D_x} \to\mathbb{C}^{D_y} $ so that $\sum_lT^{m,l}_a  A^l_x = \sum_{y,i} \hat{V}_{ax}^{y,i} {\cdot} A^m_y {\cdot} {{V}}_{ax}^{y,i}$, where $i=1,\cdots, M_{a,x}^y$ and $ M_{a,x}^y$ is the multiplicity of the block $y$ after the action of $a$ on the block $x$. This is represented graphically as 
\begin{equation}\label{fusiontensors2}
  \begin{tikzpicture}
    \draw[red] (-0.5,0)--(0.5,0);
     \node[irrep] at (0.55,0) {$a$};
        \node[irrep] at (0.55,-0.5) {$x$};
    \draw (-0.5,-0.5)--(0.5,-0.5);
    \node[tensor] (t) at (0,0) {};
    \node[tensor] (t) at (0,-0.5) {};
    \draw (0,-0.5) -- (0,0.3);
  \end{tikzpicture} =
  \sum_{i,y}
  \begin{tikzpicture}
    \pic[] (v) at (0.5,0) {V2=y/a/x};
    \pic[] (w) at (-0.5,0) {W2=y/a/x};
    \node[anchor=north,inner sep=2pt] at (v-bottom) {$i$};
    \node[anchor=north,inner sep=2pt] at (w-bottom) {$i$};
    \node[tensor] (t) at (0,0) {};
    \draw (0,0) -- (0,0.3);
  \end{tikzpicture} \ .
\end{equation}
The action tensors also satisfy the following orthogonality relations:  ${{V}}_{ax}^{y,i} \hat{V}_{ax}^{z,j} = \delta_{y,z} \delta_{i,j}  \id_{D_y}$, represented as
\begin{equation}
   \begin{tikzpicture}\label{eq:orthoV}
    \pic (v) at (0,0) {V2=y/a/x};
    \pic (w) at (0.6,0) {W2=z//};
     \node[anchor=north, inner sep=2pt] at (v-bottom) {$i$};
    \node[anchor=north, inner sep=2pt] at (w-bottom) {$j$};
   \end{tikzpicture} 
   = \ \delta_{y,z} \delta_{i,j}  \id_y \ .
\end{equation}
Associativity implies that there are two equivalent ways of decomposing the action of two MPO blocks $a,b$ over an MPS block $x$: $(a\times b) {\cdot} x = a{\cdot}(b {\cdot} x) $ which gives rise to the following relation
\begin{equation}\label{rawrels}
  \sum_{\mu,i,c} \
  \begin{tikzpicture}
    \pic[] (w1) at (0,0) {W2=y/c/x};
    \pic (w2) at (w1-up) {W1=/a/b};
    \pic[] (v1) at (1,0) {V2=y/c/x};
    \pic (v2) at (v1-up) {V1=/a/b};
    \node[anchor=north, inner sep=2pt] at (w1-bottom) {$i$};
    \node[anchor=south, inner sep=2pt] at (w2-top) {$\mu$};
    \node[anchor=north, inner sep=2pt] at (v1-bottom) {$i$};
    \node[anchor=south, inner sep=2pt] at (v2-top) {$\mu$};
  \end{tikzpicture} =
  \sum_{i,j,z} \
  \begin{tikzpicture}
    \pic (w1) at (0,0) {W2=y/a/z};
    \pic[] (w2) at (w1-down) {W2=/b/x};
    \pic (v1) at (1,0) {V2=y/a/z};
    \pic[] (v2) at (v1-down) {V2=/b/x};
    \node[anchor=north,inner sep=2pt] at (w1-bottom) {$j$};
    \node[anchor=north,inner sep=2pt] at (w2-bottom) {$i$};
    \node[anchor=north,inner sep=2pt] at (v1-bottom) {$j$};
    \node[anchor=north,inner sep=2pt] at (v2-bottom) {$i$};
  \end{tikzpicture} \ .
\end{equation}
Using the orthogonality relations, this is equivalent to the existence of a set of non-zero constants, that we call ${L}$-symbols, that relate the two decompositions:
  \begin{equation}\label{eq:F_symbol2}
    \begin{tikzpicture}
    \pic (w1) at (0,0) {V2=y/a/z};
    \pic[] (w2) at (w1-down) {V2=/b/x};
    \node[anchor=north,inner sep=2pt] at (w1-bottom) {$i$};
    \node[anchor=north,inner sep=2pt] at (w2-bottom) {$j$};
    \end{tikzpicture} 
    =
    \sum_{c,\mu,k} \left({L}_{abx}^y\right)_{c,k \mu}^{z,ij}
    \begin{tikzpicture}[baseline=0.5mm]
        \pic[] (w1) at (0,0) {V2=y/c/x};
    \pic (w2) at (w1-up) {V1=/a/b};
    \node[anchor=north,inner sep=2pt] at (w1-bottom) {$k$};
    \node[anchor=south,inner sep=2pt] at (w2-top) {$\mu$};
    \end{tikzpicture}
    \ ,
\end{equation}
where the ${L}$-symbol is defined as follows 
    \begin{equation}\label{1Fsymbol}
    \left({L}_{abx}^y \right)_{c,k \mu}^{z,ij} \delta_{y,y'} \id_y =
    \begin{tikzpicture}
     \pic (v1) at (0,0) {V2=y'/a/z};
     \pic[] (v2) at (v1-down) {V2=/b/x};
    \pic[] (w1) at (2.2,-0.3) {W2=y/c/x};
    \pic (w2) at (w1-up) {W1=//};
            \node[anchor=north,inner sep=2pt] at (w1-bottom) {$k$};
    \node[anchor=south,inner sep=2pt] at (w2-top) {$\mu$};
    \node[anchor=north,inner sep=2pt] at (v1-bottom) {$i$};
    \node[anchor=north,inner sep=2pt] at (v2-bottom) {$j$};
    \draw[red] (v1-up)-- (w2-up);
    \draw[] (v2-down)-- (w1-down);
     \draw[red] (v2-up) to [out=0,in=180] (w2-down);
    \end{tikzpicture} \ .
    \end{equation}

We can apply also associativity of the product and use the two types of $F$-symbols, ${L}$ and ${F}$, to decomposed the action of three MPO blocks $a,b,c$ acting on an MPS block $x$. This results in the coupled pentagon equations:

\begin{align}\label{coupledpent}
\sum_{d,n,\eta,\chi}  & \left({L}_{abt}^{y} \right)_{d,n\eta}^{z,ij} 
 \left({L}_{dcx}^{y} \right)^{t,nk}_{e,m\chi}  
 \left({F}_{abc}^{e} \right)_{f,\nu \mu}^{d,\chi \eta}
 \notag \\
   & =  \sum_{l }
\left( {L}_{bcx}^{z} \right)_{f,l \mu}^{t,jk}   
\left( {L}_{afx}^{y} \right)_{e,m \nu }^{z,il} \ .
\end{align}

From the associativity relations, the multiplicities $M_{a,x}^y$ and $N_{ab}^c$ satisfy the following equation:
$$\sum_{c} N_{ab}^c M_{c,x}^y = \sum_{z} M_{a,z}^y  M_{b,x}^z \ .$$

It is instructive to evaluate Eq.\ \eqref{eq:compatible} for MPO and MPSs with PBC:
$$ O_a \cdot |\psi_{A_x,\id}\rangle =  \sum_y M_{a,x}^y |\psi_{A_y,\id}\rangle$$
or equivalently in a graphical representation,
\begin{equation*}
			\begin{tikzpicture}
		  \draw (0.4,0) rectangle (2.55,-0.5);
		  \draw[red] (0.2,0.3) rectangle (2.75,-0.7);
		  \foreach \x in {0.75,2.25}{
		    \node[tensor,label=below:$A_x$] (t\x) at (\x,0) {};
		    \node[tensor] (t\x) at (\x,0.3) {};
		    \draw (\x,0) --++ (0,0.6);
		    \node[] at (\x+0.25,0.5) {$T_a$};
      }
		  \node[fill=white] at (1.5,0) {$\dots$};
		  \node[fill=white] at (1.5,0.3) {$\dots$};
		\end{tikzpicture}  \   = \ \sum_y M_{a,x}^y \
		\begin{tikzpicture}
		  \draw (0.4,0) rectangle (2.55,-0.5);
		  \foreach \x in {0.75,2.25}{
		    \node[tensor,label=below:$A_y$] (t\x) at (\x,0) {};
		    \draw (\x,0) --++ (0,0.3);
		  }
		  \node[fill=white] at (1.5,0) {$\dots$};
		\end{tikzpicture} \ .
	\end{equation*}

\subsection{Gauge transformations of the ${L}$-symbols}

The action tensors are defined up to the gauge transformations $V_{ax}^{y,i} \to \sum_{i'} \left(X_{ax}^y \right)_{i'}^i V_{ax}^{y,i'}$ and  $\hat{V}_{ax}^{y,i} \to \sum_{i'} \left({\hat{X}_{ax}^y}\right)_{i'}^i \hat{V}_{ax}^{y,i'}$ that leave invariant  Eq.\ \eqref{fusiontensors2}. These transformations give rise to the equivalence relation of ${L}$-symbols. Explicitely,
 $\left({L}_{abx}^y \right)_{c,k \mu}^{z,ij}$ is equivalent to any ${L}$-symbol of the form
 $$ \sum_{i',j',k',\mu'} \left(X_{az}^y \right)_{i'}^i 
 \left(X_{bx}^z \right)_{j'}^j 
 \left({\hat{X}_{cx}^y}\right)_{k'}^k
   \left(\hat{Y}_{ab}^c \right)_{\mu'}^\mu
 \left({L}_{a b x}^y \right)_{c, k' \mu'}^{z, i' j'},$$
where the matrix $Y$ comes from the gauge transformation of the fusion tensor in the ${L}$-symbol, see Eq\eqref{1Fsymbol}. Therefore, the ${L}$-symbols are classified by the coupled pentagon equations \eqref{coupledpent} quotiented by the previous gauge transformations.

It is interesting to note that there is always a possible solution of the coupled pentagon equations for the ${L}$-symbols by choosing as many MPS blocks as MPO blocks. This corresponds to associate any MPO block $a$ with an MPS block $x_a$ and the action $a{\cdot} x_b = \sum_{c} N_{a,b}^c  x_c$. Then, the ${L}$-symbols are equal to the ${F}$-symbols using the previous correspondence. This solution corresponds to the maximally degenerate  ground space.

\subsection{Injective MPSs cannot be invariant under non-trivial MPOs}\label{nonuniqueGS}

Let us consider an MPS with a single injective block $x$ and let us assume that there is no multiplicity in the MPO action, {\it i.e.} that $M_{a,x}^x=1$. Using Eq.\ \eqref{rawrels} we can conclude that there is a set of constants $ \lambda_{a,b}^{c,\mu} = \left(\hat{{L}}_{abx}^x \right)_{c \mu}^{x}$ that satisfy
  \begin{equation*}
    \begin{tikzpicture}[baseline=0.5mm]
        \pic[] (w1) at (0,0) {V2=x/c/x};
    \pic (w2) at (w1-up) {V1=/a/b};
    \node[anchor=south,inner sep=2pt] at (w2-top) {$\mu$};
    \end{tikzpicture}
    =
 \lambda_{a,b}^{c,\mu} \ 
 \begin{tikzpicture}
    \pic (w1) at (0,0) {V2=x/a/x};
    \pic[] (w2) at (w1-down) {V2=/b/x};
    \end{tikzpicture} 
    \ .
\end{equation*}
The pentagon equation \eqref{coupledpent} then, can be expressed using these constants $\lambda$ as
\begin{equation*}
 \left({F}_{abc}^{e} \right)^{d,\mu \nu}_{f,\chi \eta}
=
\frac{\lambda_{ab}^{d \mu} 
\lambda_{dc}^{e\nu}  
}{\lambda_{bc}^{f\chi}   
\lambda_ {af}^{e \eta } } ,
\end{equation*}
which means that the ${F}$-symbols of the MPO belongs to the trivial class, {\it i.e.} they are gauge equivalent to $1$. This implies that there is no injective MPS without multiplicity which remains invariant under a non-trivial MPO. This was proven for the first time in Ref.\ \cite{Chen11A} for MPO algebras based on groups, and for fusion categories with no multiplicities in Ref.\ \cite{reviewPEPS}.

That implies that if a Hamiltonian is invariant under a non-trivial MPO algebra, it cannot host unique gapped ground state without multiplicity in the form of a MPS. In section \ref{sec:examples} we will show a non-trivial MPO with unique ground state with multiplicity bigger than one.

\section{MPO algebras based on groups}\label{groupcase}
In this section we particularize to MPO algebras  whose blocks are labeled by elements of a finite group $G$ and they satisfy (see Eq.\ \eqref {fusiontensors})
\begin{equation}\label{fusiontensorG}
  \begin{tikzpicture}
    \draw[red]  (-0.5,0)--(0.5,0);
    \draw[red]  (-0.5,-0.5)--(0.5,-0.5);
    \node[irrep] at (-0.35,0) {$g$};
    \node[irrep] at (-0.35,-0.5) {$h$};
    \node[tensor] (t) at (0,0) {};
    \node[tensor] (t) at (0,-0.5) {};
    \draw (0,-0.8) -- (0,0.3);
  \end{tikzpicture} =
  \begin{tikzpicture}[baseline=-1mm]
    \node[tensor] (t) at (0,0) {};
    \draw (0,-0.3) -- (0,0.3);
    \pic (v) at (0.5,0) {V1=gh/g/h};
    \pic (w) at (-0.5,0) {W1=gh/g/h};
  \end{tikzpicture}\ .
\end{equation}
The importance of this case is that it covers the SPT classification, where the symmetry is on-site, $O_g = (u_g)^{\otimes n} \ \forall g\in G$, so that the blocks are one dimensional and the fusion tensors are trivial. 

Eq.\ \eqref{fusiontensorG} implies $O_g O_h = O_{gh}$. In particular, there is a trivial MPO block $e\in G$ such that $O_e O_g =O_g O_e= O_g$ and an inverse $g^{-1}$ for every element $g$ satisfying $O_{g^{-1}} O_g =O_g O_{g^{-1}}= O_e$. We emphasize that outside the on-site case $O_e$, in general, is not the identity, instead, it is a projector onto the relevant PBC subspace so that the operators $\{O_g \}$ form a representation of $G$ on $O_e {\cdot} \mathcal{H}_{loc}^{\otimes n}$. In Section \ref{sec:examples}, using Ref.\ \cite{Bultinck17A}, we construct a family of such MPOs representing any group $G$ as well as MPSs which remain invariant under the action of these MPOs.

The $F$-symbols defined for this case have no indices coming from the multiplicities of the product, so that Eq.\ \eqref{Fsymbolsdef} restricts to

\begin{equation}\label{3cocygroup}
  \begin{tikzpicture}
    \pic (w1) at (0,0) {V1=/gh/k};
    \node[irrep, anchor = south] at (w1-mid) {$ghk$};
    \pic (w2) at (w1-up) {V1=/g/h};
  \end{tikzpicture} =
 \omega(g,h,k)
  \begin{tikzpicture}
    \pic (w1) at (0,0) {V1=/g/hk};
    \node[irrep, anchor = south] at (w1-mid) {$ghk$};
    \pic (w2) at (w1-down) {V1=/h/k};
   \end{tikzpicture} \ ,
\end{equation}
where we have denoted ${F}_{g,h,k}= \omega(g,h,k)$ since the pentagon equation \eqref{pentagon0F} is the 3-cocycle condition:
$$ \omega(g, h, k) \omega(g, h k,l) \omega(h, k, l) = \omega(g h, k, l)\omega(g, h, k l).$$
Two $3$-cocycles, $\omega$ and $\omega'$, belong to the same class if they can be related by a $3$-coboundary $\beta$:
 $$\omega'(g,h,k) \sim \omega(g,h,k) \frac{\beta_{h,k}\beta_{g,hk}}{\beta_{g,h}\beta_{gh,k}} \ .$$
Quotienting by this gauge freedom, $3$-cocycles are classified by the third cohomology group of $G$: $\mathcal{H}^3(G,\mathbb{C}^*)$. 

Let us consider a subspace generated by a MPS which is  invariant under the action of the  MPOs $\{O_g\}_{g\in G}$ in the sense of Eq.\ \eqref{eq:compatible}. We assume that $O_e$ acts trivially on the subspace generated by the MPS in particular, $M_{e,x}^x=1$ for every MPS block $x$. The multiplicities satisfy $M_{gh,x}^y = \sum_z M_{g,z}^y M_{h,x}^z$, so the matrices $M_g = \sum_{x,y}M_{g,x}^y |y\rangle \langle x|$ (that have non-negative integer entries) form a representation of $G$. From $\id = M_e = M_g {\cdot} M_{g^{-1}}$, it follows that $M_g$, for all $g\in G$, is a permutation matrix. Therefore, for every $g\in G$ and MPS block $x$ there is a unique block $y$ satisfying
\begin{equation}\label{MPOsymG}
  \begin{tikzpicture}
    \draw[red] (-0.5,0)--(0.5,0);
     \node[irrep] at (0.55,0) {$g$};
        \node[irrep] at (0.55,-0.5) {$x$};
    \draw (-0.5,-0.5)--(0.5,-0.5);
    \node[tensor] (t) at (0,0) {};
    \node[tensor] (t) at (0,-0.5) {};
    \draw (0,-0.5) -- (0,0.3);
  \end{tikzpicture} 
\ =  \ 
  \begin{tikzpicture}
    \pic[] (v) at (0.5,0) {V2=y/g/x};
    \pic[] (w) at (-0.5,0) {W2=y/g/x};
    \node[tensor] (t) at (0,0) {};
    \draw (0,0) -- (0,0.3);
  \end{tikzpicture} \ .
\end{equation}
Due to the uniqueness of $y$, we can define the product  $g{\cdot}x$ as $g{\cdot}x= y$ and also we denote the action tensor by $V_{g,x}$ instead of $V_{g,x}^y$. Therefore, Eq.\ \eqref{eq:F_symbol2} has the following form:
\begin{equation}\label{F1group}
    \begin{tikzpicture}
    \pic (w1) at (0,0) {V2=y/g/z};
    \pic[] (w2) at (w1-down) {V2=/h/x};
    \end{tikzpicture} 
    =
    {L}_{g,h}^{x} \ 
    \begin{tikzpicture}[baseline=0.5mm]
        \pic[] (w1) at (0,0) {V2=y/gh/x};
    \pic (w2) at (w1-up) {V1=/g/h};
    \end{tikzpicture} \ ,
\end{equation}
where we have written ${L}_{g,h,x}^{z, y}$ instead of ${L}_{g,h}^{x}$ since $z=h {\cdot} x$, and $y= g {\cdot} z$. With this notation the coupled pentagon equations, Eq.\ \eqref{coupledpent}, reduce to
\begin{equation}\label{pentagongroups}
{L}_{h,k}^{x} \cdot {L}_{g,hk}^{x}
 = \omega(g,h,k) \cdot  {L}_{g,h}^{k {\cdot} x} \cdot  {L}_{gh,k}^{x} .
 \end{equation}
 Eq.\ \eqref{MPOsymG} is invariant under 
 $ V_{g,x} \to  \gamma_{g,x}  V_{g,x}$ and  $\hat{V}_{g,x} \to  \myinv{\gamma}_{g,x} \hat{V}_{g,x}$, where $ \gamma_{g,x}\in \mathbb{C}^*$. Therefore, the different ${L}$-symbols are classified by Eq.\ \eqref{pentagongroups} when quotienting with the equivalence relation:
 \begin{equation}\label{gdgroup}
{L}_{g,h}^{x} \sim  \frac{ \gamma_{g,h{\cdot} x} \cdot \gamma_{h,x} }{\gamma_{gh,x} \cdot \beta_{g,h} } \cdot {L}_{g,h}^{x}, 
 \end{equation}
where ${\beta}_{g,h}$ is the gauge freedom of the fusion tensor $\hat{W}_{g,h}$ that appears in the definition of the ${L}$-symbol Eq.\ \eqref{F1group}:
$\hat{W}_{g,h}\to \myinv{\beta}_{g,h} \hat{W}_{g,h}$. 

It is always possible to find gauges, $\gamma$ and $\beta$, such that ${L}_{g,e}^{x}={L}_{e,g}^{x}=1$ for all MPS blocks $x$ and $g \in G$. This can be seen by the fact that a gauge transformation on the action tensors modifies ${L}_{e,e}^{x}$ to $\gamma_{e,x} {L}_{e,e}^{x}$, so one can set ${L}_{e,e}^{x}=1$ for all block $x$. For normalized 3-cocycles we have that $\omega(g,g',e)= \omega(g,e,g')=\omega(e,g,g')=1$ so that we obtain the following relations: ${L}_{g,e}^{x} = {L}_{g',e}^{x}$, ${L}_{e,g'}^{x} = {L}_{g,e}^{g'{\cdot} x}$ and ${L}_{e,gg'}^{x} = {L}_{e,g}^{g'{\cdot} x}$ for all $g,g'\in G$. In particular, ${L}_{e,e}^{x}={L}_{g,e}^{x}={L}_{e,g}^{x}$ and then, any ${L}$-symbol with an $e$ component can be set to one.

\subsection{Possible solutions and symmetry action decomposition} \label{physym}

In this section we show how to characterize the different solutions of Eq.\ \eqref{pentagongroups} given a group $G$ and $3$-cocycle class $\omega \in  \mathcal{H}^3 (G,\mathbb{C}^*)$. It turns out that the different solutions are encoded by pairs $(H,\psi)$, where $H$ is a subgroup of $G$ that trivializes $\omega$ and $\psi \in \mathcal{H}^2(H,\mathbb{C}^*)$. 

We consider ground state degeneracies that are robust under bounded-strength perturbations of the local terms of the parent Hamiltonian. This can be achieved by assuming that the group action of the MPO on the MPS blocks is transitive: for every $x$ and $y$ there is a $g\in G$ such that $y=g {\cdot} x$. This is because a transitive action prevents one from modifying the energy of an isolated GS by a local perturbation (and then being able to close the gap), see details in Refs.\cite{Schuch11,Bravyi11}. Let $H$ denote the biggest subgroup of $G$ that does not permute some block $x_0$, $h\in H$ if $M_{h x_0}^y=\delta_{y,x_0}$ Then, the number of MPS blocks is $|G/H|$.  We notice that the choice of $x_0$ is arbitrary but $H$, up to isomorphism, does not depend on that choice. 

Let us postpone the proof of the fact that $H$ has to trivialize $\omega$ and let us comment now on one of its consequences. Since the subgroup $H$ depends on the 3-cocycle, this imposes restrictions on $H$ and therefore, on the number of blocks of the MPS, {\it i.e.}\ in the degeneracy of the ground space, that an MPO with such a 3-cocycle can host. In particular, any non-trivial 3-cocycle cannot be trivialized for the whole group $G$ so there is no injective (single block) MPS which remains  invariant under this MPO, see Section \ref{nonuniqueGS}. This is in sharp contrast with on-site symmetries, trivial 3-cocycle, where any $H$ can be taken, in particular $H=G$ obtaining a unique ground state. In this sense, there is a minimum degeneracy $1 < {\rm deg }\le |G| $ protected by a non-trivial MPO symmetry.

Let us proceed to prove that $H$ trivializes $\omega$.  For $h_1,h_2,h_3 \in H$, we can check that Eq.\ \eqref{pentagongroups} simplifies to 
\begin{equation*}
{L}^{x_0}_{h_2,h_3} {\cdot} {L}^{x_0}_{h_1,h_2h_3}  
=  {L}^{x_0}_{h_1,h_2} {\cdot} {L}^{x_0}_{h_1h_2,h_3}{\cdot} \omega(h_1,h_2,h_3).
\end{equation*}
That is, $\omega(h_1,h_2,h_3)$, the ${F}$-symbol from the MPO representation, is a trivial 3-cocycle (since it can be written as a 3-coboundary) when restricted to $H$. So if a subgroup $H$ leaves some block invariant, then it also trivializes the 3-cocycle. In particular, this means that there is a gauge from the fusion tensors such that
$$ \omega(h_1,h_2,h_3){\cdot} \beta_{h_1,h_2}{\cdot}  \beta_{h_1h_2,h_3}{\cdot} \beta^{-1}_{h_2,h_3} {\cdot} \beta^{-1}_{h_1,h_2h_3} =1.$$ 
The previous gauge also modifies the ${L}$-symbols so that Eq.\ \eqref{pentagongroups}, restricted to $H$, is
$$\frac{{L}^{x_0}_{h_1,h_2}}{\beta_{h_1,h_2}}  {\cdot} \frac{{L}^{x_0}_{h_1h_2,h_3}}{\beta_{h_1h_2,h_3}}   =   \frac{{L}^{x_0}_{h_2,h_3}}{\beta_{h_2,h_3}}  {\cdot} \frac{{L}^{x_0}_{h_1,h_2h_3}}{\beta_{h_1,h_2h_3}} .$$
If we define $\psi_{x_0}(h_1,h_2) = {{L}^{x_0}_{h_1,h_2}}/{\beta_{h_1,h_2}} \ \forall \ h_1,h_2\in H$, the previous equation is a 2-cocycle condition for $\psi_{x_0}$. Once the gauge of the fusion tensors is fixed, the gauge freedom from the action tensors (restricted to $H$ and $x_0$) modify ${L}|_H$ as ${L}^{x_0}_{h_1,h_2} \sim {L}^{x_0}_{h_1,h_2}{\cdot}  \gamma_{h_1,{x_0}} {\cdot} \gamma_{h_2,{x_0}}/\gamma_{h_1h_2,{x_0}}$. Incorporating this equivalence relation into $\psi_{x_0}(h_1,h_2)$, we conclude that $\psi_{x_0}$ belongs to $\mathcal{H}^2(H,\mathbb{C}^*)$. Therefore, given a $H$ that trivializes $\omega$, the different solutions of Eq.\ \eqref{pentagongroups} are characterized by $\mathcal{H}^2(H,\mathbb{C}^*)$.

It is important to note that the permutation pattern of $G$ on the MPS blocks is encoded in the subgroup $H$. The same argument of Ref.\ \cite{Schuch11} for on-site symmetries is valid here. Let us show this by considering $G=\bigcup_{k_y\in G} k_y H$ to be the coset decomposition of $G$ by $H$, where we label the representative of the coset $k_y$ by the block $y$ that arises from the action of $k_y$ on $x_0$: $y=k_y{\cdot}x_0$. Then, the block $z$ obtained by acting with the element $g\in G$ on any block $x$ is given by $k_z{\cdot}x_0 =k_zhk_x^{-1}{\cdot}x  $, where $k_z$ is the unique element satisfying $g k_x = k_z h$ for a unique $h \in H$.

Once $H$ (that determines the number of blocks and how they are permuted by the MPO action) and $\psi$ (that characterizes $L|_H$ and then the internal action on each block) are specified given $G$ and $\omega$, all the components of the $L$-symbols can be obtained. This can be seen by the fact that we can provide a solution of Eq.\ \eqref{pentagongroups} for $L_{g_2,g_1}^{\alpha}$ in terms of the previous data:
\begin{equation}\label{Lexpre}
L_{g_1,g_2}^{\alpha}
 = \frac{\omega(g_1 g_2 k_{\alpha},\myinv{h}_2,\myinv{h}_1) \cdot \psi({h}_1,{h}_2)}{\omega(g_1,g_2 k_{\alpha},\myinv{h}_2 ) \cdot \omega(g_1,g_2,k_{\alpha})},
 \end{equation}
where $h_2, h_1 \in H$ are uniquely defined by $g_2 k_{\alpha} = k_{\beta} h_2$ and $g_1 k_{\beta} = k_{\gamma} h_1$. This result coincides with the classification of module categories of the fusion category ${\rm Vec}_G^\omega$, see \cite{ostrik2006module}. 
 
It is interesting to note that the solutions of the $L$-symbols, given $(G,\omega, H)$, do not form a group structure in general. Mathematically, this can be seen from the fact that the solutions of ${L}|_H$ are a $\mathcal{G}$-torsor where the group $\mathcal{G}$ is $\mathcal{H}^2(H,\mathbb{C}^*)$. A $\mathcal{G}$-torsor is a set with a group action such that any two elements of the set are connected by an action of the group $\mathcal{G}$. Given $\omega$ and $\beta$ by fixing some gauge, any solution satisfies ${L}=\psi {\cdot} \beta$. So two solutions correspond to two elements $\psi_1,\psi_2\in \mathcal{H}^2(H,\mathbb{C}^*)$; they satisfy ${L}_i=\psi_i {\cdot} \beta$  for $i=1,2$ with the same $\beta$. Then,  ${L}_1$ and  ${L}_2$ are connected by an element of $\mathcal{G}$ since ${L}_1=( \psi_1 /\psi_2) {L}_2$. The same holds for any two solutions of the ${L}$-symbols: they are related by an element of $\mathcal{G}$. The importance of being a $\mathcal{G}$-torsor is that besides the number of solutions is $|\mathcal{H}^2(H,\mathbb{C}^*)|$, the solutions are not isomorphic to $\mathcal{H}^2(H,\mathbb{C}^*)$ if $\beta$ is not strictly $1$. This implies in particular, that in general there is no canonical way to associate a phase to the identity element in $\mathcal{H}^2(H,\mathbb{C}^*)$.

In what follows we show how we can decompose the action of any element $g$ on any block $x$ in terms of the action of $H$ and $G/H$, the permutations, on $x_0$ and the fusion tensors of the MPO. We do so to compare with the classification of on-site symmetries with degenerate ground states in Ref.\ \cite{Schuch11}. We recall that for every $g$ and $x$, there are unique $h=h(g,x)$ and $k_y$ ($y=g{\cdot} x$) such that $g k_x = k_y h(g,x)\Rightarrow g=k_y h(g,x) k_x^{-1}$. Then, we can write Eq.\ \eqref{MPOsymG} as 
\begin{equation}\label{gxdecomp}
  \begin{tikzpicture}
    \draw[red] (-0.5,0)--(0.5,0);
     \node[irrep] at (0.55,0) {$g$};
        \node[irrep] at (0.55,-0.5) {$x$};
    \draw (-0.5,-0.5)--(0.5,-0.5);
    \node[tensor] (t) at (0,0) {};
    \node[tensor] (t) at (0,-0.5) {};
    \draw (0,-0.5) -- (0,0.3);
  \end{tikzpicture} 
=
  \begin{tikzpicture}
         \pic[] (v) at (0.5,0) {V2=y/k_y/x_0};
    \pic[] (v1) at (v-down) {V2=/h/x_0};
     \pic[]  at (v1-down) {V2=/ \ \myinv{k_x}/x};
      \pic[scale=0.7] (v4) at (+1.5,0.21) {W1=//};
      \pic[scale=0.85]  at (+2.1,-0.045) {W1=g/ k_yh/};
        \draw[red] (v4-up) to [out=180,in=0] (v-up);
     \pic[] (w) at (-0.5,0) {W2=y/k_y/x_0};
    \pic[] (w2) at (w-down) {W2=/h/x_0};
     \pic[] (w3) at (w2-down) {W2=/ \myinv{k_x}/x};
      \pic[scale=0.7] (v2) at (-1.5,0.21) {V1=//};
      \pic[scale=0.85] (v3) at (-2.1,-0.045) {V1=g/ \ k_yh/};
        \draw[red] (v2-up) to [out=0,in=180] (w-up);
    \node[tensor] (t) at (0,0) {};
    \draw (0,0) -- (0,0.3);
  \end{tikzpicture} 
\end{equation}
where it is important to note that the action tensor  $V_{\myinv{k_x},x}$ can be written in terms of the action tensors ${V}_{{k_x},x_0},{V}_{{e},x_0}$ and the fusion tensor $W_{\myinv{k_x},k_x}$ as follows:
\begin{equation*}
    \begin{tikzpicture}
    \pic (w1) at (0,0) {W2=x_0/\myinv{k_x}/x};
    \pic (v1) at (1,0) {V2=x_0/\myinv{k_x}/x};
  \end{tikzpicture} =    
  \begin{tikzpicture}
   \pic[] (w1) at (0,0) {W2=x_0/e/x_0};
    \pic[scale=0.75] (w2) at (w1-up) {W1=/\myinv{k_x}/};
    \pic[scale=0.75] (w3)  at (-1,-0.15) {V2=x/k_x/};
     \draw[black] (w3-down) to [out=0,in=180] (w1-down);
    \pic[] (v1) at (1,0) {V2=x_0/e/};
    \pic[scale=0.75]  (v2) at (v1-up) {V1=/\ \myinv{k_x}/};
    \pic[scale=0.75] (v3)  at (2,-0.15) {W2=x/k_x/x_0 \ };
    \draw[black] (v3-down) to [out=180,in=0] (v1-down);
  \end{tikzpicture} .
\end{equation*}

This means that the action of $g$ on $x$ can be decomposed into action tensors of the cosets $k_x,k_y$ acting on $x_0$ (which determine the permutation action) and the action tensor of $h(g,x)$ acting internally on $x_0$ (characterized by ${L}|_H$). 

Finally, we show how $L$ and $\omega$ appear as projective and associative factors on a suitable product define for the action tensors. To do so, we define a product of the action tensors as follows
\begin{equation}\label{defodot}
 V_{g,y}\odot V_{h, x} = \delta_{y,h{\cdot}x} \ 
    \begin{tikzpicture}
    \pic (w1) at (0,0) {V2=/g/y};
     \draw[] (w1-mid)--++(-0.15,0);
 \node[irrep] at (-0.3,0) {$g{\cdot}y$};
    \pic[] (w2) at (0.55,-0.3) {V2=/h/x};
 \pic[scale=0.75] (w3) at (1.1,0.225) {W1=//};
 \draw[red] (w1-up) to [out=0,in=180] (w3-up);
 \draw[red] (w3-mid)--++(0.15,0);
 \node[irrep] at (1.35,0.2) {$gh$};
    \end{tikzpicture} \ .
\end{equation}
Using Eq.\ \eqref{F1group}, the action tensors restricted to $H$ acting on any block $x$ satisfy
\begin{equation}\label{intactens} V_{h_1, {x}} \odot V_{h_2, {x}}= {L}_{{h_1},{h_2}}^{x}  V_{{h_1}{h_2}, {x}} \ , \end{equation}
for all $h_1 , h_2 \in H$. That is, $V|_H$ forms a projective representation whose projective factors, $L$, are associative up to $\omega|_H$, a $3$-coboundary. The action tensors satisfy 
\begin{align}\label{nonasso} V_{g_3,\alpha_2}\odot(V_{g_2,\alpha_1} \odot V_{g_1,\alpha_0})& = \\ \omega({g_3},{g_2},{g_1})&(V_{g_3,\alpha_2}\odot V_{g_2,\alpha_1})\odot V_{g_1,\alpha_0}, \notag \end{align}
where $\alpha_{i} = g_i {\cdot} \alpha_{i-1}$. Thus, the action tensors are associative, under the product defined in Eq.\ \eqref{defodot}, up to the phase factor $\omega$.

We notice that the whole action on the MPS is linear and associative but, the fact that the action on the virtual level can be split as $\hat{V} \otimes V$, allows for the non-trivial relations of Eq.\ \eqref{intactens} and Eq.\ \eqref{nonasso} which show up when considering one isolated boundary.

\subsection{Reduction to SPT: on-site representation of a group}

In this section we further particularize to on-site  representations of a group $G$: $U_g=u^{\otimes n}_g$. In particular, the $3$-cocycle associated to an on-site operator is trivial. In this case, the fusion tensors are matrices so they satisfy
\begin{equation}\label{Lgroupdef}
V_{g,h{\cdot} x} \cdot V_{h, x}  = {L}_{g,h}^{x} V_{gh, x}\ ,
 \end{equation}
so that their product is associative $V_{g,x}{\cdot}(V_{h,y} {\cdot} V_{k,z})= (V_{g,x}{\cdot} V_{h,y}){\cdot} V_{k,z}$
and Eq.\ \eqref{pentagongroups} is just
\begin{equation}\label{coupengroup}
{L}_{h,k}^{x} \cdot {L}_{g,hk}^{x}
 = {L}_{g,h}^{k{\cdot} x} \cdot  {L}_{gh,k}^{x} \ .
 \end{equation}

If the MPS which is invariant under the group contains a unique injective block, Eq.\ \eqref{Lgroupdef} is simplified to $V_{g}{\cdot} V_{h} = {L}_{g,h} V_{gh}$ 
and Eq.\ \eqref{coupengroup} has the following form: 
$${L}_{g,h}{L}_{gh,k}= {L}_{h,k}{L}_{g,hk}\ ,$$
{\it i.e.} the ${L}$-symbols form a 2-cocycle, classified by $\mathcal{H}^2(G,U(1))$.

For multiple MPS blocks, corresponding to a degenerate ground space, we also reproduce the result of Ref.\ \cite{Schuch11} where the classification is given by the induced representation using the decomposition of Eq.\ \eqref{gxdecomp}, that is,  first, by fixing the subgroup $H$ (which determines the degeneracy) and then, by specifying the projective representation of $H$ on every block and its permutation action. When we restrict Eq.\ \eqref{coupengroup} to the elements of $H$ we obtain
$${L}^x_{h_2,h_3} {\cdot} {L}^x_{h_1,h_2h_3}  
=  {L}^x_{h_1,h_2} {\cdot} {L}^x_{h_1h_2,h_3},$$
which accounts for the internal projective representation of each block $x$. Then, Eq.\ \eqref{coupengroup} reflects the compatibility between the projective action inside the internal blocks and the permutations.

The main differences between the on-site and the non-trivial MPO symmetries, is that for the latter there are restrictions on the degeneracies of the ground state, that the solutions could not have a group structure and that the action tensors do not have to be associative.

\subsection{Interplay between MPO and time reversal symmetry}\label{sec:TRS}

In this section we impose both MPO and time reversal symmetry (TRS) on the MPS and study the interplay between these two symmetries. TRS is represented as an anti-unitary operator $\mathcal{T}= v^{\otimes n} \mathcal{K}$ where $\mathcal{K}$ is the complex conjugation operator. We assume for simplicity that the operator $\mathcal{T}$ does not permute the MPS blocks. Then, the action of $\mathcal{T}$ on the $x$-block MPS tensor at the virtual level is given by $\sum_j v_{ij} {A_x^j}^*= S^{-1}_x {A_x^i} S_x$ where $S_x S_x^* = \beta_x(\mathcal{T}) \id$ and the phase factors are $ \beta_x(\mathcal{T})=\pm 1$ for each block $x$, see \cite{Chen11B,Pollmann10}. 

Through our analysis, we impose that TRS and the PBC MPOs commute, $\mathcal{T}  O_g = O_g \mathcal{T}$, what implies that there are matrices $q_g$ that satisfy 
$$\sum_{k,l} v_{ik} T^{k,l}_g v^{-1}_{lj} = q^{-1}_g {(T^{i,j}_g)}^* q_g \ ,$$
for all $g\in G$, where $T_g$ is the MPO tensor $T$ on the block labelled by $g$. Under this assumption and using Eq.\ \eqref{fusiontensorG}, we can conclude that the fusion tensors satisfy:
$$
  \begin{tikzpicture}
\pic[] (w) at (-1,0) {V1=gh/g/h};
  \end{tikzpicture}
  =  \beta_{g,h} \
  \begin{tikzpicture}[baseline=-1mm]
            \draw[red] (-0.5,-0.3) -- (0.3,-0.3);
         \draw[red] (-0.5,0.3) -- (0.3,0.3);
         \draw[red] (-1,0) -- (-1.3,0);
         \pic[] (w) at (-0.5,0) {V1=gh/g/h};
     \node[tensor,label=below:$q_h$] at (0,-0.3) {};
          \node[tensor,label=above:$q_g$] at (0,0.3) {};
   \node[tensor,label=below:$q_{gh}^{-1}$] at (-1,0) {};
   \node[] at (-0.5,0.5) {$^*$};
  \end{tikzpicture}
\ ,
$$
where $\beta_{g,h}\in \mathbb{C}^*$. Using the previous equation and Eq.\ \eqref{3cocygroup} we conclude that 
$$ \omega(g,h,k)=\omega^*(g,h,k) \frac{\beta_{g,h}\beta_{gh,k}}{\beta_{h,k}\beta_{g,hk}} \ ,$$
which means that $\omega$ can be gauge transformed to a real number.

We compare the local actions of $\mathcal{T}  O_g$ and $O_g \mathcal{T}$ and we obtain that
$$
  \begin{tikzpicture}
  \draw (-1,0)--(1,0);
   \node[tensor,label=below:$S_y$] at (0.5,0) {};
   \node[tensor,label=below:$S_y^{-1}$] at (-0.5,0) {};
   \node[] at (-1,0.5) {$^*$};
   \node[] at (1,0.5) {$^*$};
    \pic[] (v) at (1,0) {V2=y/g/x};
    \pic[] (w) at (-1,0) {W2=y/g/x};
    \node[tensor] (t) at (0,0) {};
    \draw (0,0) -- (0,0.3);
  \end{tikzpicture}
=
  \begin{tikzpicture}
    \pic[] (v) at (0.5,0) {V2=y/g/x};
    \pic[] (w) at (-0.5,0) {W2=y/g/x};
    \node[tensor] (t) at (0,0) {};
    \draw (0,0) -- (0,0.3);
     \node[tensor,label=below:$S_x$] at (1,-0.3) {};
   \node[tensor,label=below:$S_x^{-1}$] at (-1,-0.3) {};
   \draw (-1,-0.3) -- (-1.2,-0.3);
   \draw (1,-0.3) -- (1.2,-0.3);
  \end{tikzpicture} \ ,
$$
which implies
$$
  \begin{tikzpicture}
         \draw (-0.5,-0.3) -- (0.2,-0.3);
     \node[tensor,label=below:$S_x$] at (0,-0.3) {};
    \pic[] (w) at (-0.5,0) {V2=y/g/x};
   \node[tensor,label=below:$S_y^{-1}$] at (-1,0) {};
   \draw (-1,0) -- (-1.2,0);
  \end{tikzpicture}
  =  \gamma_{g,x} \
  \begin{tikzpicture}[baseline=-1mm]
   \node[] at (-1,0.5) {$^*$};
    \pic[] (w) at (-1,0) {V2=y/g/x};
  \end{tikzpicture}
\ ,
$$
where $\gamma_{g,x}$ is a non-zero phase factor. Using the previous equation for $g,h,gh\in G$ and Eq.\ \eqref{F1group} we obtain the following equation:
$$
\gamma_{h,x}  {\cdot}  \gamma_{g,h{\cdot} x} \ 
    \begin{tikzpicture}
    \node[] at (0,0.45) {$^*$};
    \node[] at (0.5,0.15) {$^*$};
    \pic (w1) at (0,0) {V2=y/g/z};
    \pic[] (w2) at (w1-down) {V2=/h/x};
    \end{tikzpicture} 
    \ = \
{L}_{g,h}^x \ {\cdot} \  \gamma_{gh, x} \ 
    \begin{tikzpicture}[baseline=0.5mm]
        \node[] at (0,0.45) {$^*$};
        \pic[] (w1) at (0,0) {V2=y/gh/x};
    \pic (w2) at (w1-up) {V1=/g/h};
    \end{tikzpicture} .
$$

This equation implies that 
\begin{equation}\label{TRSF1group}
 \left | {L}_{g,h}^x \right |^2 =   \gamma_{g h, x} / ( \gamma_{h,x} {\cdot}  \gamma_{g,h{\cdot} x})\ ,
 \end{equation}
so that the ${L}$-symbols squared are trivial since they can be decomposed as a gauge transformation. For on-site global symmetries, this relation reduces to the one obtained in Ref.\ \cite{Chen11B}. 

Besides the fact that this section focuses on the group case, the same reasoning can be applied to the interplay between TRS and MPO symmetries outside groups, so now let us briefly comment on the result for general MPO algebras. For the general case of an MPO algebra we obtain that the ${F}$-symbols are real and the following relation is satisfied
$$
  \begin{tikzpicture}
         \draw (-0.5,-0.3) -- (0.2,-0.3);
     \node[tensor,label=below:$S_x$] at (0,-0.3) {};
    \pic[] (w) at (-0.5,0) {V2=y/a/x};
 \node[anchor=north,inner sep=2pt] at (w-bottom) {$i$};
   \node[tensor,label=below:$S_y^{-1}$] at (-1,0) {};
   \draw (-1,0) -- (-1.2,0);
  \end{tikzpicture}
\  = \ \sum_{i'}  \left(X_{ax}^y \right)_{i'}^i \ 
  \begin{tikzpicture}[baseline=-1mm]
   \node[] at (-1,0.5) {$^*$};
    \pic[] (w) at (-1,0) {V2=y/a/x};
 \node[anchor=north,inner sep=2pt] at (w-bottom) {$i'$};
  \end{tikzpicture}
\ ,
$$
where
\begin{equation*}
   \begin{tikzpicture}
\node[tensor,label=below:$S_x$] at (0.5,-0.3) {};
\draw (1.4,0) -- (1.6,0);
\node[tensor,label=below:$S_y^{-1}$] at (1.4,0) {};
\node[] at (1,0.5) {$^*$};
    \pic (v) at (0,0) {V2=y/a/x};
    \pic (w) at (1,0) {W2=y//};
     \node[anchor=north, inner sep=2pt] at (v-bottom) {$i'$};
    \node[anchor=north, inner sep=2pt] at (w-bottom) {$i$};
   \end{tikzpicture} 
   = \left(X_{ax}^y \right)_{i'}^i {\cdot} D^{-1}_y {\cdot} \id_y  \ .
\end{equation*}
Using the previous relation on three elements of the algebra and Eq.\ \eqref{1Fsymbol} we obtain also that the ${L}$-symbols squared are trivial, {\it i.e.}\ they can be written as gauge transformations.

\subsection{Periodic boundary condition case }\label{sec:PBC}
In this section we consider PBC MPSs which are symmetric under PBC MPO representations of a finite group $G$, $U_g$. In particular, we only impose that $U_g U_h = U_{gh}$ for all $g,h\in G$ and that $U_e=\id$, so the representing MPOs can be unitaries. 
Algebraically, this is done by imposing a weaker condition than the closedness condition \eqref{algcond} for the MPO group algebra and a weaker condition than Eq.\ \eqref{eq:compatible} for the invariance of the MPS under the action of the MPO. To put it in other words, we only require \eqref{algcond} to hold for $X$ and $Y$ that are proportional to the identity in each block (and thus $Z$ has the same structure too). Similarly, we weaken the requirement of an MPS to be invariant under the MPO symmetry (Eq.\ \eqref{eq:compatible}) by requiring only that the equation holds for $B$ and $X$ that are both proportional to identity in each block (and thus $Z$ again has the same structure). 
We also assume that the MPS and MPO tensors are injective in every block. We anticipate that we will obtain that the same labels of the previous section, {\it i.e.}\ equivalence classes of the $F$-symbols and $L$-symbols, are valid to characterize the PBC MPO and their action on PBC MPSs respectively, but there are some subtleties in how to define these quantities.

In Ref.\ \cite{Molnar18B}, the authors proved that $U_g U_h = U_{gh}$ with $T_g$ injective implies the existence of fusion tensors that decompose the product of two tensors locally for any length $n$ as follows:
\begin{equation}\label{fusiontensorG2}
  \begin{tikzpicture}
  \node at (0.25,0.45) {$\overbrace{\hspace{0.8cm}}^{n}$};
      \draw[red] (0,-0.6) -- (1,-0.6);
    \draw[red] (0,0) -- (1,0);
     \pic (v) at (-0.5,-0.3) {V1=gh/g/h};
     \pic (w) at (1,-0.3) {W1=gh/g/h};
    \node[tensor] (t) at (0,0) {};
    \node[tensor] (t) at (0,-0.6) {};
        \node[tensor] (t) at (0.5,0) {};
    \node[tensor] (t) at (0.5,-0.6) {};
    \draw (0,-0.9) -- (0,0.3);
        \draw (0.5,-0.9) -- (0.5,0.3);
  \end{tikzpicture} =
  \begin{tikzpicture}[baseline=-1mm]
    \node at (0.25,0.45) {$\overbrace{\hspace{0.8cm}}^{n}$};
      \draw (0,-0.3) -- (0,0.3);
            \draw (0.5,-0.3) -- (0.5,0.3);
      \draw[red] (-0.3,0) -- (0.8,0);
    \node[tensor] (t) at (0,0) {};
     \node[tensor] (t) at (0.5,0) {};
   \node[irrep] at (0.65,0.05) {$gh$};
     \end{tikzpicture}\ .
\end{equation}
In particular, for $n=0$, this results in the orthogonality relations for the fusion tensors. 

Associativity of the MPOs implies that a $3$-cocycle can be associated to any of these group representations using the fusion tensors. However, we emphasize that since we do not impose the closedness condition \eqref{algcond}, Eq.\ \eqref{fusiontensorG} is not  satisfied. This means that, in general, there will be off-diagonal blocks in the product of two MPO tensors. These off-diagonal blocks will disappear after a finite number of concatenation: we denote by  $\ell$ the nilpotent length associated to that (which is of the order of the bond dimension), see Ref.\ \cite{Molnar18B} for details. It can be shown then that the following equation defines the $3$-cocycle $\omega$:
$$  
\begin{tikzpicture}
\node[irrep] at (-0.35,0.9) {$g$};
\node[irrep] at (-0.35,0.4) {$h$};
\node[irrep] at (-0.35,-0.1) {$k$};
  \node at (0.4,-0.45) {$\underbrace{\hspace{1cm}}_{\ge \ell}$};
      \pic[scale=1.25] at (1.75,0.625) {W1=ghk/g/hk};
      \pic[scale=0.83] at (1.2,0.25 ) {W1=/h/k};
      \draw[red] (1,1) -- (1.2,1);
\foreach \y in {0,0.5,1}{
		  \foreach \x in {0,0.4,0.8}{
		  \draw (\x,\y-0.25) -- (\x,\y+0.25);
		    \draw[red] (\x-0.25,\y) -- (\x+0.25,\y);
		    \node[tensor] at (\x,\y) {};   }}
\end{tikzpicture}
= \omega(g,h,k)
\begin{tikzpicture}
\node[irrep] at (-0.35,0.9) {$g$};
\node[irrep] at (-0.35,0.4) {$h$};
\node[irrep] at (-0.35,-0.1) {$k$};
  \node at (0.4,-0.45) {$\underbrace{\hspace{1cm}}_{\ge \ell}$};
 \pic[scale=1.25] at (1.75,0.375) {W1=ghk/gh/k};
      \pic[scale=0.83] at (1.2,0.75) {W1=/g/h};
      \draw[red] (1,0) -- (1.2,0);
\foreach \y in {0,0.5,1}{
		  \foreach \x in {0,0.4,0.8}{
		  \draw (\x,\y-0.25) -- (\x,\y+0.25);
		    \draw[red] (\x-0.25,\y) -- (\x+0.25,\y);
		    \node[tensor] at (\x,\y) {};   }}
\end{tikzpicture} \ .
$$

For this case we consider the subspace spanned by injective MPSs with PBC  $ \{ |\psi_{A_\alpha} \rangle \equiv |\psi_{A_\alpha,\id} \rangle \}_\alpha$, where $\alpha$ takes values from the block labels $\{x,y,z, \dots \}$.
We impose that for every element $g\in G $ and block $x$, there is a block $y\equiv g{\cdot} x$ such that
$$ U_g |\psi_{A_x} \rangle = |\psi_{A_y} \rangle ,$$
which ensures that the subspace generated by the MPSs is invariant under the action of the MPO. Using again the results of Ref.\ \cite{Molnar18B}, we conclude that there are action tensors that decompose locally the action of the MPO tensor on the MPS tensor for any length $n$:
\begin{equation}\label{mpoMPSsten}
  \begin{tikzpicture}[baseline=-3mm]
    \node at (0.25,0.45) {$\overbrace{\hspace{0.8cm}}^{n}$};
        \draw (0,-0.6) -- (1,-0.6);
    \draw[red] (0,0) -- (1,0);
     \pic (v) at (-0.5,-0.3) {V2=y/g/x};
      \pic (w) at (1,-0.3) {W2=y/g/x};
    \node[tensor] (t) at (0,0) {};
    \node[tensor] (t) at (0,-0.6) {};
        \node[tensor] (t) at (0.5,0) {};
    \node[tensor] (t) at (0.5,-0.6) {};
    \draw (0,-0.6) -- (0,0.3);
        \draw (0.5,-0.6) -- (0.5,0.3);

  \end{tikzpicture} =
  \begin{tikzpicture}[baseline=-1mm]
      \node at (0.25,0.45) {$\overbrace{\hspace{0.8cm}}^{n}$};
      \draw (0,0) -- (0,0.3);
            \draw (0.5,0) -- (0.5,0.3);
      \draw (-0.3,0) -- (0.8,0);
    \node[tensor] (t) at (0,0) {};
    \node[tensor] (t) at (0.5,0) {};
   \node[irrep] at (0.65,0.05) {$y$};
     \end{tikzpicture}\ .
\end{equation}
Just as above, there is a number $\ell'$ such that after blocking $\ell'$ tensors, there are no off-diagonal blocks present in the product of the MPO and MPS tensors. Then,  for a number of sites bigger than $\ell'$, the following phase factors that characterize the associativity of the product $(g\times h) {\cdot} x = g{\cdot}(h {\cdot} x) $ can be defined
$$  \begin{tikzpicture}
\node[irrep] at (-0.35,0.9) {$g$};
\node[irrep] at (-0.35,0.4) {$h$};
\node[irrep] at (-0.35,-0.1) {$x$};
  \node at (0.4,-0.45) {$\underbrace{\hspace{1cm}}_{\ge \ell'}$};

 \pic[scale=1.25] at (1.75,0.375) {W2=y/gh/x};
      \pic[scale=0.83] at (1.2,0.75) {W1=/g/h};
      \draw (-0.2,0) -- (1.2,0);
\foreach \y in {0.5,1}{
		  \foreach \x in {0,0.4,0.8}{
		  \draw (\x,\y-0.25) -- (\x,\y+0.25);
		    \draw[red] (\x-0.25,\y) -- (\x+0.25,\y);
		    \node[tensor] at (\x,\y) {};
		    \node[tensor] at (\x,0) {};  
		     \draw (\x,0) -- (\x,+0.25); }}
\end{tikzpicture} 
= {L}^x_{g,h}
\begin{tikzpicture}
\node[irrep] at (-0.35,0.9) {$g$};
\node[irrep] at (-0.35,0.4) {$h$};
\node[irrep] at (-0.35,-0.1) {$x$};
  \node at (0.4,-0.45) {$\underbrace{\hspace{1cm}}_{\ge \ell'}$};
      \pic[scale=1.25] at (1.75,0.625) {W2=y/g/z};
      \pic[scale=0.83] at (1.2,0.25 ) {W2=/h/x};
      \draw[red] (1,1) -- (1.2,1);
      \draw (-0.2,0)--(1,0);
\foreach \y in {0.5,1}{
		  \foreach \x in {0,0.4,0.8}{
		  \draw (\x,\y-0.25) -- (\x,\y+0.25);
		    \draw[red] (\x-0.25,\y) -- (\x+0.25,\y);
		    \node[tensor] at (\x,\y) {};
		    \node[tensor] at (\x,0) {};  
		     \draw (\x,0) -- (\x,+0.25); }}
\end{tikzpicture}
\ . $$

These ${L}$-symbols satisfy Eq.\ \eqref{pentagongroups} as in the previous section with the same equivalence relation Eq.\ \eqref{gdgroup}. 

It is worth to write down the analogous equations of \eqref{fusiontensorG} and \eqref{MPOsymG} for this PBC case. Here, there is no such a purely virtual transformation acting on $T_{gh}$ to obtain $T_g T_h$. Instead, there are some {\it tails} coming from the off-diagonals blocks of $T_g T_h$ that disappear after a big enough number of sites since they are nilpotent. Therefore, the following two equations are satisfied for a number of sites bigger than $\ell$ and $\ell'$ and, for any $n$:
\begin{equation}\label{oPBCMPO}
 \begin{tikzpicture}
\node[irrep] at (-0.25,0.5) {$g$};
\node[irrep] at (-0.25,0) {$h$};
\node[irrep] at (2.2,0.5) {$g$};
\node[irrep] at (2.2,0) {$h$};
    \node at (1,1) {$\overbrace{\hspace{0.7cm}}^{n}$};
    \node at (1.8,1) {$\overbrace{\hspace{0.7cm}}^{\ge \ell}$};
    \node at (0.2,1) {$\overbrace{\hspace{0.7cm}}^{\ge \ell}$};
\foreach \y in {0,0.5}{
		  \foreach \x in {0,0.4,0.8,1.2,1.6,2}{
		  \draw (\x,\y-0.3) -- (\x,\y+0.3);
		    \draw[red] (\x-0.3,\y) -- (\x+0.3,\y);
		    \node[tensor] at (\x,\y) {};
		    \node[tensor] at (\x,0) {};  
		      }}
\end{tikzpicture} 
=
\begin{tikzpicture}
    \node at (-1.2,0.5) {$\overbrace{\hspace{0.7cm}}^{\ge \ell}$};
    \node at (0,0.3) {$\overbrace{\hspace{0.7cm}}^{n}$};
    \node at (1.2,0.5) {$\overbrace{\hspace{0.7cm}}^{\ge \ell}$};
     \pic (w) at (-0.6,-0.3) {W1=gh/g/h};
\pic (v) at (0.6,-0.3) {V1=gh/g/h};
\foreach \y in {0,-0.6}{
		  \foreach \x in {-1.4,-1,1,1.4}{
		  \draw (\x,\y-0.3) -- (\x,\y+0.3);
		    \draw[red] (\x-0.3,\y) -- (\x+0.3,\y);
		    \node[tensor] at (\x,\y) {};
		      }}
		      		  \foreach \x in {-0.2,0.2}{
		  \draw (\x,-0.3-0.3) -- (\x,-0.3+0.3);
		    \draw[red] (\x-0.3,-0.3) -- (\x+0.3,-0.3);
		    \node[tensor] at (\x,-0.3) {};
		      }
\end{tikzpicture} 
\end{equation}
and
\begin{equation}\label{oPBCMPSs}
\begin{tikzpicture}
\node[irrep] at (-0.25,0.5) {$g$};
\node[irrep] at (-0.25,0) {$x$};
\node[irrep] at (2.2,0.5) {$g$};
\node[irrep] at (2.2,0) {$x$};
    \node at (1,1) {$\overbrace{\hspace{0.7cm}}^{n}$};
    \node at (1.8,1) {$\overbrace{\hspace{0.7cm}}^{\ge \ell'}$};
    \node at (0.2,1) {$\overbrace{\hspace{0.7cm}}^{\ge \ell'}$};
    \draw[red] (-0.3,0.5) -- (2+0.3,0.5);
    \draw (-0.3,0) -- (2+0.3,0);
\foreach \y in {0,0.5}{
		  \foreach \x in {0,0.4,0.8,1.2,1.6,2}{
		  \draw (\x,0) -- (\x,0.8);
		    \node[tensor] at (\x,\y) {};
		    \node[tensor] at (\x,0) {};  
		      }}
\end{tikzpicture} 
=
\begin{tikzpicture}
    \node at (-1.2,0.5) {$\overbrace{\hspace{0.7cm}}^{\ge \ell'}$};
    \node at (0,0.3) {$\overbrace{\hspace{0.7cm}}^{n}$};
    \node at (1.2,0.5) {$\overbrace{\hspace{0.7cm}}^{\ge \ell'}$};
     \pic (w) at (-0.6,-0.3) {W2=y/g/x};
\pic (v) at (0.6,-0.3) {V2=y/g/x};
		  \foreach \x in {-1.4,-1,1,1.4}{
		  \draw (\x,-0.6) -- (\x,+0.3);
		  \draw[red] (\x-0.3,0) -- (\x+0.3,0);
		\draw (\x-0.3,-0.6) -- (\x+0.3,-0.6);
	    \foreach \y in {0,-0.6}{ \node[tensor] at (\x,\y) {};} }
		\foreach \x in {-0.2,0.2}{
		   \draw (\x,-0.3) -- (\x,-0.3+0.3);
		   \draw[red] (\x-0.3,-0.3) -- (\x+0.3,-0.3);
		   \node[tensor] at (\x,-0.3) {};
		      }
\end{tikzpicture} \ .
\end{equation}
We remark that Eq.\ \eqref{oPBCMPO} and Eq.\ \eqref{fusiontensorG2} together are equivalent to $U_g U_h = U_{gh}$. Analogously, Eq.\ \eqref{oPBCMPSs} together with Eq.\ \eqref{mpoMPSsten} are equivalent to $ U_g |\psi_{A_x} \rangle = |\psi_{A_y} \rangle $.

The following lemma allows us to characterize the MPO tensors and the fusion tensors of group representations. 

{\bf Lemma}. An injective MPO representation of a group (not necessarily unitary), $U_g U_h = U_{gh}$ and $U_e=\id$, satisfy Eq.\ \eqref{fusiontensorG2}. If we further require that the fusion tensors fulfill Eq.\ \eqref{fusiontensorG}, which is equivalent to the closedness algebra condition for arbitrary BC \eqref{algcond}, then $U_g$ is on-site. In particular, the MPO representation is characterized by a trivial $3$-cocycle.

{\it Proof:} First we decompose the product of the three MPO tensors $T_g T_{g^{-1}} T_g$ in two equivalent ways using Eq.\ \eqref{fusiontensorG}. Then, we close the two upper (or lower) virtual indices. Finally, using that $T_e$ has to be the identity matrix (by using the Fundamental Theorem of injective MPSs \cite{Molnar18A}), we conclude that $T_g$ is on-site.

The previous lemma implies that for an injective MPO representation of a group, $U_g U_h = U_{gh}$ and $U_e=\id$, with non-trivial $3$-cocycle, the arbitrary BC MPO does not form a closed algebra in the sense of Eq.\ \eqref{algcond} (equivalent to Eq.\ \eqref{fusiontensorG}). This means that these MPOs are not covered by the formalism of Refs.\ \cite{Lootens21A, molnar22}, we left the connection for future work. Alternatively, one can drop the condition $U_e=\id$ and only ask for $U_e$ being a projector, then there is no obstruction to satisfy Eq.\ \eqref{fusiontensorG} and that the tensors being injective at the same time. In that case $U_g$ is a representation on the subspace projected by $U_e$. We will comment on these cases in the example section \ref{sec:expexam}.

\section{Classification of phases}\label{sec:classif}

Roughly speaking (see proper definition below), we will say that two symmetric Hamiltonians are in the same phase if they can be connected by a smooth, gapped and symmetric path. In this section we show that the previously defined equivalence classes of ${L}$-symbols are the invariants that classify the phases of symmetric Hamiltonians with exact MPSs ground spaces. To arrive to such result, some considerations are in order:

First, since we want to classify symmetric phases, we have to specify which kind of MPO symmetries we can compare. Note that in standard SPT phases, any system with the same symmetry group can be compared (systems with different symmetry groups are incomparable). We will define the freedom in the MPO algebra, when considered as the symmetry, in our definition of a symmetric phase.

Then, we will show that given a symmetric MPS ground space, a parent Hamiltonian which is symmetric under the same MPOs can always be chosen. We will see that a symmetric Hamiltonian with an MPS ground space belongs to the same phase as the symmetric parent Hamiltonian, so we can lift the classification of Hamiltonians to parent Hamiltonians of MPSs. 

Finally, we will prove that a continuous transformation on the MPS tensor (arising from a continuous gapped path of the parent Hamiltonian) does not change the class of the ${L}$-symbols. Then, two Hamiltonians that give rise to different solutions of the coupled pentagon equations cannot be connected by a smooth path and therefore, they do not belong to the same phase. Conversely, we will also show that two symmetric MPSs with the same class of $L$-symbols can be connected by a symmetric gapped path; we will explicitly construct such a path. This leads us to the main result of this paper:
\\

{\bf Main result.} Given two PBC Hamiltonians with exact MPS ground space left invariant under two MPO symmetries characterized by equivalent $F$-symbols; then, they are in the same phase if and only if they share the class of ${L}$-symbols.

\subsection{Fixing the symmetry of the Hamiltonian}\label{sec:fixsym}

In this section we focus on MPO algebras that correspond to physical symmetries of a Hamiltonian. This situation naturally imposes some desired extra conditions on the MPO blocks. For example, the existence of a unique identity element, the notion of a dual element, a pivotal structure, etc, which generalize the group case. The mathematical object capturing these properties is a fusion category \cite{etingof2017fusion}. Importantly, any fusion category corresponds to the representation theory of a (semisimple) weak Hopf algebra (WHA) \cite{ostrik2001module}. In Ref.\ \cite{molnar22}, the authors show what the properties of MPO algebras representing $C^*$-WHA are. Namely, the set of PBC MPO $\{ O_a \}$ is a representation of the fusion category $\mathcal{C}_\mathcal{A}$ characterized by the $F$-symbols of Eq.\ \eqref{Fsymbolsdef}.

In this work we are interested in classifying systems with periodic boundary conditions that are symmetric under MPO representations of fusion categories, that is, the PBC MPOs $\{ O_a, a\in \mathcal{C} \}$. However, we still require that the corresponding MPOs form an algebra with arbitrary boundary conditions (equivalent to Eq.\ \eqref{fusiontensors}), so we define {\it symmetry} as follows:

\begin{definition}\label{defsym}
We consider a Hamiltonian $H$ with PBC and a periodic boundary MPO representation of a fusion category $\{ O_a, a\in \mathcal{C} \}$ that satisfies Eq.\ \eqref{fusiontensors}, that is, a set of $F$-symbols can be defined. Then, we say that $H$ is symmetric under $\mathcal{C}$, characterized by the above considered class of the $F$-symbols, if $[H,O_a]=0 \ \forall a\in \mathcal{C}$. 
\end{definition}

That is, we can compare PBC Hamiltonians which are symmetric under the same fusion category but we allow these two MPOs to correspond to different $C^*$-WHA representations when arbitrary BC are considered (notice the equivalence between Eq.\ \eqref{algcond} and Eq.\ \eqref{fusiontensors}). We emphasize that the unitarity of the operators is not required and it would be satisfied only by on-site group representations. We remark that our definition of symmetry is broader than the standard SPT phase definition since we allow to compare on-site symmetries of groups with group-based MPOs having trivial $3$-cocycle (meaning they are not on-site as the ones constructed in section \ref{sec:expexam}).

MPO algebras and their connection to fusion categories have also been studied in \cite{Bultinck17A}. Recently in \cite{Lootens21A}, a more general categorical framework has been developed for MPO algebras, where they are described by a $(\mathcal{C},\mathcal{D})$-bimodule category $\mathcal{M}$. Importantly, given a bimodule category, one can construct a WHA from it and vice versa \cite{ostrik2001module,EGNObook,Bohm96}; the fusion categories $\mathcal{C}$ and $\mathcal{D}$ are recovered from the WHA $\mathcal{A}$ as $\text{Rep}(\mathcal{A}^*)$ and $\text{Rep}(\mathcal{A})$ respectively. Therefore, fixing the WHA $\mathcal{A}$ is a compact way of specifying the MPO symmetry. However, to construct explicit examples  the categorical formulation proves more useful, and we will use it to describe some of the examples in Section \ref{sec:examples}. Our main result establishes the fact that the possible phases protected by an MPO symmetry described by a $(\mathcal{C},\mathcal{D})$-bimodule category $\mathcal{M}$ correspond to the different left-module categories of the fusion category $\mathcal{C}$ \cite{etingof2003finite}. 

\subsection{Definition of symmetric phase}

Let us consider two local gapped Hamiltonians, $H_0$ and $H_1$, defined on a chain with periodic boundary conditions. Let $\mathcal{A}_0$ and $\mathcal{A}_1$ be two $C^*$-WHA with equivalent $F$-symbols, i.e. their fusion categories coincide $\mathcal{C}_{\mathcal{A}_0} =  \mathcal{C}_{\mathcal{A}_1} \equiv \mathcal{C}$. 
Then consider two Hamiltonians $H_p$ for $p=0,1$ that commute with the MPO representations of $\mathcal{C}_{\mathcal{A}_p}$, $\{O^p_a, a\in \mathcal{C} \}$, $[H_p,O^p_a]=0,\ \forall a\in \mathcal{C}$.

\begin{definition}\label{defphase}
We say that two Hamiltonians, $H_0$ and $H_1$, which are symmetric under the same fusion category $\mathcal{C}$ are in the same phase if 
\begin{itemize}
    \item there is a local gapped and continuous PBC Hamiltonian path $H(\gamma)$ that embeds both systems: $H(p)= H_p|_{\mathcal{H}^{\otimes N }_p}$, where $\mathcal{H}_p$ is the local Hilbert space,
    
    \item the Hamiltonian path commutes with an injective MPO representation of $\mathcal{C}$ that decomposes as $O_a= O^0_a \oplus O^1_a \oplus O^{\rm path}_a$ and satisfies Eq.\ \eqref{fusiontensors}.
\end{itemize} 
\end{definition}
Some comments are in order. It is important to note that with this definition we can compare systems with MPO symmetries with different bond dimensions, e.g. even on-site symmetries with non-on-site MPOs. We emphasize that our definition is weaker than the one of standard SPT phases since when comparing two on-site group symmetries we do not impose that the symmetry along the path be on-site. 

Our definition of symmetric phases with periodic boundary conditions is motivated by the standard SPT definition \cite{Chen11,Schuch11}. We believe that classifying phases with arbitrary BC can result in a way more restricted phase diagram. This is because when restricting to MPS ground spaces, the degeneracy would depend on the bond dimension of the MPS tensor (also the $C^*$-WHA depends on the bond dimension of the MPO). As a result, not too much freedom is allowed when trying to connect two systems in this setting. That is why the periodic BC case could give a richer and more meaningful scenario. 

\subsection{Restriction to MPSs as ground spaces}

We want to classify symmetric Hamiltonians that have exact MPS representations of their ground spaces. The symmetry of the Hamiltonian is inherited by the ground space, so that the MPS subspace remains invariant under $\{O^p_a, a\in \mathcal{C} \}$. We further assume that there are action tensors locally realizing this symmetry, i.e. Eq.\ \eqref{fusiontensors2} is satisfied (which is equivalent to Eq.\ \eqref{eq:compatible}). Then, a set of ${L}$-symbols can be defined satisfying the coupled pentagon equations \eqref{coupledpent}.

Also, we would like to connect such ground spaces with an interpolating MPS path. Then, we will restrict to embedding Hamiltonian paths (gapped and symmetric) that have an exact MPS ground space through the whole path, so that a set of $L$-symbols can also be defined all along. We are leaving the question of interpolating gapped paths beyond MPSs open. Besides that, we believe that the inequivalent classes of $L$-symbols classify symmetric phases under fusion categories outside tensor networks states. This is also justified by category theory results where the $L$-symbols correspond to the different module categories of a fusion category \cite{etingof2003finite, Thorngren19}.

\subsection{Symmetrization of the parent Hamiltonian}

Given the subspace spanned by a set of injective MPSs a parent Hamiltonian with PBC with such ground space can be constructed \cite{PerezGarcia07}. The parent Hamiltonian $H=\sum_i h_i$ is constructed by defining the translationally invariant local terms $h = \id - \Pi_S$, where $\Pi_\mathcal{S}$ is the orthogonal projector onto $\mathcal{S}^2_A$, on every adjacent sites including $n$ and $1$ (if we remove this term the Hamiltonian is defined with arbitrary BC and the subspace is $\mathcal{S}_A^n$ of Eq.\ \eqref{MPSssubs}).

When considering global symmetries of MPSs, one can always choose a parent Hamiltonian that commutes with the symmetries of the MPS \cite{Sanz09}. This is done by symmetrizing the parent Hamiltonian. For the case of on-site symmetries $U_g=u_g^{\otimes n}$ of a group $G$, the local terms can be taken as $h'= \sum_g (u_g \otimes u_g ) h (u_g^\dagger \otimes u_g^\dagger)/|G|$. 
The previous symmetrization can be generalized to MPO representations of ${C}^*$-weak Hopf algebras \cite{molnar22}. In particular, we use the existence of a positive element $\Omega \in  \mathcal{A}$, the so-called canonical left integral that for groups reduces to $\Omega = \sum_g g/|G|$, to symmetrize the local Hamiltonian terms such that the following is satisfied:
$$
    \begin{tikzpicture}
  		   \draw[red]  (1.35,0) --(2.9,0);
\node at (1.1,0.4) {$a$};
          \draw[red]  (1.35,0.3) --(2.9,0.3);
  		   \draw[red]  (1.35,-0.6) --(2.9,-0.6);
      \draw[hopf2] (1.5,-0.3)--(2.75,-0.3);
  		   \foreach \x in {1.75,2.5}{
 \node[tensor] (t\x) at (\x,0.3) {};
  		    \node[tensor] (t\x) at (\x,0) {};
  		    \draw (\x,-0.9) --++ (0,1.5);
  		    \node[tensor, fill=white] (t\x) at (\x,-0.6) {};
  		  }
        \draw[red] (2.9,0) to [out=0,in=0] (2.9,-0.6);
\draw[red] (1.35,0) to [out=180,in=180] (1.35,-0.6);
\node at (1,-0.5) {$B_\Omega$};
    \node[tensor, fill=blue] at (1.175,-0.3) {};
    \node[tensor, fill=green,scale=0.7] at (2.15,-0.6) {};
  		\end{tikzpicture}
=
    \begin{tikzpicture}
\node at (1.1,-1) {$a$};
          \draw[red]  (1.35,-0.9) --(2.9,-0.9);
  		   \draw[red]  (1.35,0) --(2.9,0);
  		   \draw[red]  (1.35,-0.6) --(2.9,-0.6);
      \draw[hopf2] (1.5,-0.3)--(2.75,-0.3);
  		   \foreach \x in {1.75,2.5}{
  		    \node[tensor] (t\x) at (\x,0) {};
  		    \draw (\x,-1.2) --++ (0,1.5);
\node[tensor] (t\x) at (\x,-0.9) {};
  		    \node[tensor, fill=white] (t\x) at (\x,-0.6) {};
  		  }
        \draw[red] (2.9,0) to [out=0,in=0] (2.9,-0.6);
\draw[red] (1.35,0) to [out=180,in=180] (1.35,-0.6);
\node at (1,-0.5) {$B_\Omega$};
    \node[tensor, fill=blue] at (1.175,-0.3) {};
    \node[tensor, fill=green,scale=0.7] at (2.15,-0.6) {};
  		\end{tikzpicture}
   \ \forall a,
$$
where the white tensors are the MPO tensors under the action of a linear map and the green dot comes from the coproduct of the white tensors, see \cite{molnar22} for details. This construction ensures that first, the new Hamiltonian commutes with the symmetry and second, the new Hamiltonian is still hermitian and finally that it is non-zero; see Appendix \ref{app:sumnonull} for a proof.

\subsection{Parent Hamiltonian restriction and further  simplifications} \label{sec:parentHrestric}

If two Hamiltonians are symmetric under the same representation of the symmetry (so that the local Hilbert space is the same), one could be interested in proposing a gapped path which remains symmetric under the same representation without embedding both systems. If such gapped and symmetric path exists, $\tilde{H}(\gamma)$, then they are also in the same phase under our definition of a symmetric phase. 
This can be seen by using the Hamiltonian interpolation ${H}(\gamma)= \tilde{H}(\gamma) + \Sigma(\gamma)$, where  
$\Sigma(\gamma)$ is a one-body qubit Hamiltonian with ground state $\sqrt{1-\gamma}|0\rangle + \sqrt{\gamma}|1\rangle$. We have enlarged the whole Hilbert space by one qubit and the symmetry is ${O}_a\otimes \id_2 \approx {O}_a \oplus {O}_a$ for all $a\in \mathcal{C}$ so that it satisfies our definition.

Using the same argument we can see that a symmetric gapped Hamiltonian $H'$ with exact MPS ground space and its symmetric parent Hamiltonian $H$ are in the same phase. This is because the interpolating path ${H}(\gamma)= (1-\gamma){H'}+ \gamma H$ is symmetric and gapped \cite{Schuch11}. 

Then, we can focus on the classification of parent Hamiltonians of MPSs. Furthermore, since we have assumed that through the interpolating path the ground spaces are MPSs, we can also focus on parent Hamiltonian paths. Due to the one-to-one correspondence between an MPS and its parent Hamiltonian we will use MPS paths without loss of generality.

As we commented in section \ref{sec:data}, we can assume that the MPS tensors are injective and orthogonal between them. Then, for every injective block $x$, there is a left inverse $\myinv{A}_x$ such that $\myinv{A}_x A_x = \id_{D_x}\otimes \id_{D_x}$, what including the orthogonality of the blocks is represented graphically as :

\begin{equation}\label{leftinv}
  \begin{tikzpicture}
   \node at (0,0.5) {$\myinv{A}_x$};
   \node at (0,-0.5) {$A_y$};
    \draw[] (-0.45,0.25)--(0.45,0.25);
    \draw (-0.45,-0.25)--(0.45,-0.25);
    \node[tensor] (t1) at (0,0.25) {};
    \node[tensor] (t2) at (0,-0.25) {};
    \draw (0,-0.2) -- (0,0.2);
  \end{tikzpicture} 
=
\delta_{x,y}
  \begin{tikzpicture}
\draw[rounded corners] (-0.4,0.25)--(-0.05,0.25)--(-0.05,-0.25)--(-0.4,-0.25);
\draw[rounded corners] (0.4,0.25)--(0.05,0.25)--(0.05,-0.25)--(0.4,-0.25);
  \end{tikzpicture} \ .
\end{equation}

\subsection{MPSs with inequivalent $L$-symbols belong to different phases}

In this section we show that Hamiltonians which remain invariant under the same fusion category and whose MPS ground spaces (satisfying Eq.\ \eqref{fusiontensors2}) are characterized by different classes of $L$-symbols belong to distinct phases. 

We will show that a continuous change of the MPS tensor, which is invariant under the same symmetry operators, will modify the action tensors in such a way that the new $L$-symbols belong to the same class as the previous one. This shows that a continuous transformation of the MPS cannot change the class of the $L$-symbols, so that MPSs with inequivalent $L$-symbols cannot be connected via a well-behaved path. Crucially, a continuous gapped path of local parent Hamiltonians $H(\gamma)$ induces a continuous transformation of the tensor in every block, see Ref.\ \cite{Schuch11}. Therefore, the impossibility of connecting MPSs implies the impossibility of connecting the corresponding parent Hamiltonians and then the inequivalence of their symmetric phases.

We remark that the path that is used in this section refers to the embedding path in the definition of phase \ref{defphase}. Then, by assumption, the symmetry operators are the same along such path.

We start by fixing a point $\gamma$ of the path, and study the continuity of the tensor $A(\gamma)$ in some environment $\gamma+d\gamma$. We consider a continuous change of the tensor in the form $A({\gamma +d\gamma}) = A({\gamma })+dA$ for every block, and we assume for now that the bond dimension does not change. The final goal is to compare the $L$-symbols of $A({\gamma +d\gamma})$ and $A({\gamma})$, since both corresponding MPSs are symmetric under the same symmetry operators. We first consider the following equation:
$$
  \begin{tikzpicture}
   \node at (0,0.8) {$A_y({\gamma +d\gamma})^{-1}$};
   \node at (0,-0.8) {$A_x({\gamma +d\gamma})$};
    \draw[red] (-0.5,0)--(0.5,0);
    \draw[] (-0.5,0.5)--(0.5,0.5);
     \node[irrep] at (-0.55,0) {$a$};
    \draw (-0.5,-0.5)--(0.5,-0.5);
    \node[tensor] (t1) at (0,0) {};
    \node[tensor] (t2) at (0,-0.5) {};
    \node[tensor] (t3) at (0,0.5) {};
    \draw (0,-0.5) -- (0,0.5);
  \end{tikzpicture} 
=
\sum_{i}
  \begin{tikzpicture}
    \pic[] (v) at (0.5,0) {V3=y/a/x};
    \draw (v-mid)--++(0,0.8)--++(0.6,0);
    \node[anchor=north,inner sep=2pt] at (v-bottom) {$i$};
    \pic[] (w) at (-0.5,0) {W3=y/a/x};
     \draw (w-mid)--++(0,0.8)--++(-0.6,0);
    \node[anchor=north,inner sep=2pt] at (w-bottom) {$i$};
  \end{tikzpicture} \ 
  =  \sum_i V_{a,x}^{y,i} \otimes \hat{V}_{a,x}^{y,i} \ , 
$$
where we have used equations \eqref{fusiontensors2} and \eqref{leftinv}. Since one can always find a left inverse tensor such that it is continuous we can find without loss of generality a basis change, where $V_{a,x}^{y,i}$ is also continuous (using the fact that the action tensors are linearly independent). Let us absorb this basis change and assume (see below for generalization) that the action tensors are differentiable:
$$   
\begin{tikzpicture}
    \pic[] (v) at (0.5,0) {V3=y/a/x};
    \node[anchor=north,inner sep=2pt] at (v-bottom) {$i$};
     \end{tikzpicture}
     =
     \begin{tikzpicture}
    \pic[] (v) at (0.5,0) {V2=y/a/x};
    \node[anchor=north,inner sep=2pt] at (v-bottom) {$i$};
     \end{tikzpicture}
     + \sum_{i'} (X_{ax}^y)_{i'}^i
          \begin{tikzpicture}
    \pic[] (v) at (0.5,0) {V2=y/a/x};
    \node[anchor=north,inner sep=2pt] at (v-bottom) {$i'$};
     \end{tikzpicture}
     d\gamma \ ,
    $$
where we have spanned the term in $d\gamma$ using the orthogonal set of action tensors $\{  V_{a,x}^{y,i}, i=1,{\cdots}, M_{a,x}^y\}$ of the subspace of the block $y$. Due to the associativity of the MPO action on the MPS blocks, a new ${L}'$-symbol can be defined: 
  \begin{equation*}
    \begin{tikzpicture}
    \pic (w1) at (0,0) {V3=y/a/z};
    \pic[] (w2) at (w1-down) {V3=/b/x};
    \node[anchor=north,inner sep=2pt] at (w1-bottom) {$i$};
    \node[anchor=north,inner sep=2pt] at (w2-bottom) {$j$};
        \end{tikzpicture} 
    =
    \sum_{c,\mu,k} \left({{L}'}_{abx}^y\right)_{c\mu k}^{zij}
    \begin{tikzpicture}[baseline=0.5mm]
        \pic[] (w1) at (0,0) {V3=y/c/x};
    \pic (w2) at (w1-up) {V1=/a/b};
    \node[anchor=north,inner sep=2pt] at (w1-bottom) {$k$};
    \node[anchor=south,inner sep=2pt] at (w2-top) {$\mu$};
    \end{tikzpicture} \ ,
    \end{equation*}
    where  
    \begin{equation}\label{eq:F_symbol3}
    \left({{L}'}_{abx}^y \right)_{c k \mu}^{z i j}  \id_y =
    \begin{tikzpicture}
     \pic (v1) at (0,0) {V3=y/a/z};
     \pic[] (v2) at (v1-down) {V3=/b/x};
    \pic[] (w1) at (2.2,-0.3) {W3=y/c/x};
    \pic (w2) at (w1-up) {W1=//};
            \node[anchor=north,inner sep=2pt] at (w1-bottom) {$k$};
    \node[anchor=south,inner sep=2pt] at (w2-top) {$\mu$};
    \node[anchor=north,inner sep=2pt] at (v1-bottom) {$i$};
    \node[anchor=north,inner sep=2pt] at (v2-bottom) {$j$};
    \draw[red] (v1-up)-- (w2-up);
    \draw[] (v2-down)-- (w1-down);
     \draw[red] (v2-up) to [out=0,in=180] (w2-down);
    \end{tikzpicture} \ .
    \end{equation}

Since the ${L}'$-symbols are a linear combination of differentiable action tensors, we conclude that these are differentiable too and we can write:

$$ \left({{L}'}_{abx}^y\right)_{c\mu k}^{zij}  =   \left({{L}}_{abx}^y\right)_{c\mu k}^{zij} +   \left({G}_{abx}^y\right)_{c\mu k}^{zij} d\gamma \ .
$$
In what follows we show that $\left({G}_{abx}^y\right)_{c\mu k}^{zij}$ is a gauge transformation. Let us consider the first-order terms in $d\gamma$ of Eq.\ \eqref{eq:F_symbol3}: the LHS of Eq.\ \eqref{eq:F_symbol3} is
$$
\sum_{i'} (X_{bx}^z)_{j'}^{j}
  \begin{tikzpicture}
    \pic (w1) at (0,0) {V2=y/a/z};
    \pic[] (w2) at (w1-down) {V2=/b/x};
    \node[anchor=north,inner sep=2pt] at (w1-bottom) {$i$};
    \node[anchor=north,inner sep=2pt] at (w2-bottom) {$j'$};
    \end{tikzpicture} 
    +
    \sum_{j'} (X_{az}^y)_{i'}^{i}
      \begin{tikzpicture}
    \pic (w1) at (0,0) {V2=y/a/z};
    \pic[] (w2) at (w1-down) {V2=/b/x};
    \node[anchor=north,inner sep=2pt] at (w1-bottom) {$i'$};
    \node[anchor=north,inner sep=2pt] at (w2-bottom) {$j$};
    \end{tikzpicture} \ ,
$$
and the RHS reads
\begin{align*}
    \sum_{c,\mu,k}  & \left({G}_{abx}^y\right)_{c\mu k}^{zij}  \  \begin{tikzpicture}[baseline=0.5mm]
        \pic[] (w1) at (0,0) {V2=y/c/x};
    \pic (w2) at (w1-up) {V1=/a/b};
    \node[anchor=north,inner sep=2pt] at (w1-bottom) {$k$};
    \node[anchor=south,inner sep=2pt] at (w2-top) {$\mu$};
    \end{tikzpicture}
    \ + \\
     & \sum_{c,\mu,k,k'} (X_{cx}^y)_{k'}^{k} \left({L}_{abx}^y\right)_{c\mu k}^{zij}
    \begin{tikzpicture}[baseline=0.5mm]
        \pic[] (w1) at (0,0) {V2=y/c/x};
    \pic (w2) at (w1-up) {V1=/a/b};
    \node[anchor=north,inner sep=2pt] at (w1-bottom) {$k'$};
    \node[anchor=south,inner sep=2pt] at (w2-top) {$\mu$};
    \end{tikzpicture} . 
\end{align*}
We apply the following operator
$$ \begin{tikzpicture}
    \pic (w1) at (0,0) {W2=y/a/\hat{z}};
    \pic[] (w2) at (w1-down) {W2=/b/x};
    \node[anchor=north,inner sep=2pt] at (w1-bottom) {$\hat{i}$};
    \node[anchor=north,inner sep=2pt] at (w2-bottom) {$\hat{j}$};
        \end{tikzpicture} 
    =
    \sum_{\hat{c},\hat{\mu},\hat{k}} 
    \left( {{{L}}_{abx}^y}^{-1} \right)^{\hat{c}\hat{\mu} \hat{k}}_{\hat{z}\hat{i}\hat{j}}
    \begin{tikzpicture}[baseline=0.5mm]
        \pic[] (w1) at (0,0) {W2=y/\hat{c}/x};
    \pic (w2) at (w1-up) {W1=/a/b};
    \node[anchor=north,inner sep=2pt] at (w1-bottom) {$\hat{k}$};
    \node[anchor=south,inner sep=2pt] at (w2-top) {$\hat{\mu}$};
    \end{tikzpicture},
$$
so that the LHS now is
$$
(X_{bx}^z)_{\hat{j}}^{j} {\cdot} \delta_{i,\hat{i}} \delta_{z,\hat{z}}  \id_y+(X_{az}^y)_{\hat{i}}^{i} {\cdot} \delta_{j,\hat{j}}\delta_{z,\hat{z}}\id_y 
$$
and the RHS can be written as:
\begin{align*}
 \sum_{c,\mu,k,\hat{k}}
\left({{{L}}_{abx}^y}^{-1}\right)_{\hat{z}\hat{i}\hat{j}}^{{c}{\mu} \hat{k}}
  & (X_{cx}^y)_{\hat{k}}^{k}  \left({L}_{abx}^y\right)_{c\mu k}^{zij} \id_y \\
+ &
 \sum_{c,\mu,k}
   \left({{{L}}_{abx}^y}^{-1}\right)^{{c}{\mu} {k}}_{\hat{z}\hat{i}\hat{j}} \left({{G}_{abx}^y}\right)_{{c}{\mu} {k}}^{z{i}{j}}\id_y = \\
      \sum_{c,\mu,k} \left({{{L}}_{abx}^y}^{-1}\right)_{\hat{z}\hat{i}\hat{j}}^{{c}{\mu} {k}} & \left [ 
    \sum_{\hat{k}}  (X_{cx}^y)^{\hat{k}}_{k}  \left({L}_{abx}^y\right)_{c\mu \hat{k}}^{zij}  
+
  \left({{G}_{abx}^y}\right)_{{c}{\mu} {k}}^{z{i}{j}} 
   \right ]\id_y.
   \end{align*}

By multiplying now with $\sum_{\{\hat{z},\hat{i},\hat{j}\}} \left({{{L}}_{abx}^y}\right)^{\hat{z}\hat{i}\hat{j}}_{{c}{\mu} {k}}$ we end up in 
\begin{align*}  
\sum_{\hat{z},\hat{i},\hat{j}} \left({{{L}}_{abx}^y}\right)^{\hat{z}\hat{i}\hat{j}}_{{c}{\mu} {k}}& \left [
(X_{bx}^z)_{\hat{j}}^{j} {\cdot} \delta_{i,\hat{i}} \delta_{z,\hat{z}}  +(X_{az}^y)_{\hat{i}}^{i} {\cdot} \delta_{j,\hat{j}}\delta_{z,\hat{z}}
\right]\id_y \\
=& \big[  \sum_{\hat{k}} (X_{cx}^y)^{\hat{k}}_{k}  \left({L}_{abx}^y\right)_{c\mu \hat{k}}^{zij}  
+
  \left({{G}_{abx}^y}\right)_{{c}{\mu} {k}}^{z{i}{j}} 
  \big ] \id_y
\end{align*}
Reordering the previous equation and restricting ourselves to the subspace of the $y$-block we can conclude that $\left({{G}_{abx}^y}\right)_{{c}{\mu} {k}}^{z{i}{j}}$ is equal to

\begin{align*}
\sum_{\hat{i},\hat{j},\hat{k}} & \left({{{L}}_{abx}^y}\right)^{{z}\hat{i}\hat{j}}_{{c}{\mu} \hat{k}} \big [
(X_{bx}^z)_{\hat{j}}^{j} {\cdot} \delta_{i,\hat{i}}\delta_{k,\hat{k}} 
\\  + \  
& (X_{az}^y)_{\hat{i}}^{i} {\cdot} \delta_{j,\hat{j}}\delta_{k,\hat{k}} 
-(X_{cx}^y)^{\hat{k}}_{k} {\cdot} \delta_{i,\hat{i}}\delta_{j,\hat{j}} 
    \big ] \ .
 \end{align*}
 
Therefore
 \begin{align*}
 \left({{{L}'}_{abx}^y}\right)_{{c}{\mu} {k}}^{z{i}{j}}  = & 
 \sum_{\hat{i},\hat{j},\hat{k}} \left({{{L}}_{abx}^y}\right)^{{z}\hat{i}\hat{j}}_{{c}{\mu} \hat{k}} \big [
 \delta_{i,\hat{i}}\delta_{j,\hat{j}}\delta_{k,\hat{k}}
+(X_{bx}^z)_{\hat{j}}^{j} {\cdot} \delta_{i,\hat{i}}\delta_{k,\hat{k}} 
\\
&
+(X_{az}^y)_{\hat{i}}^{i} {\cdot} \delta_{j,\hat{j}}\delta_{k,\hat{k}} 
-(X_{cx}^y)^{\hat{k}}_{k} {\cdot} \delta_{i,\hat{i}}\delta_{j,\hat{j}} 
    \big],
\end{align*}

which means that the ${L}'$-symbol is in the same class as the ${L}$-symbol since they are related by a (differentiable) gauge transformation. Similar results can be proven, using the arguments of Ref.\ \cite{Schuch11}, for the cases when the action tensors are just continuous (not differentiable) and when $A({\gamma +d\gamma})$ is supported in a bigger virtual space.

The previous argument is also valid the MPOs studied in section \ref{sec:PBC}. This is because, for a long enough chain, the following is satisfied:
$$
  \begin{tikzpicture}
   \node at (0,0.8) {$A_{y}^{-1}$};
   \node at (0,-0.8) {$A_x$};
     \node[irrep] at (-0.3,0) {$g$};
    \foreach \x in {0,0.5,1,1.5}{
    \foreach \y in {-0.5, 0.5}{
        \draw (\x-0.25,\y)--(\x +0.25,\y);
    \draw (\x,0.5)--(\x,-0.5);
    \node[tensor] at (\x,\y) {};
            \draw[red] (\x-0.25,0)--(\x +0.25,0);
    \node[tensor] at (\x,0) {};
    }}
  \end{tikzpicture} 
=
  \begin{tikzpicture}
    \pic[] (v) at (0.5,0) {V2=y/g/x};
    \draw (v-mid)--++(0,0.8)--++(0.6,0);
    \pic[] (w) at (-0.5,0) {W2=y/g/x};
     \draw (w-mid)--++(0,0.8)--++(-0.6,0);
  \end{tikzpicture} \ 
$$
and then, one proceeds similarly as above.

\subsection{Connecting two MPSs with same $L$-symbols}

In this section, we show that two MPSs, $|\psi_{A_0}\rangle$ and $|\psi_{A_1}\rangle$, that are symmetric under different representations, that generate possibly different MPO algebras, of the same fusion category $\mathcal{C}$ (and equivalent $F$-symbols) and whose symmetry action is characterized by the same class of $L$-symbols belong to the same phase. We do so by explicitly constructing a continuous and symmetric embedding path that connects both MPSs and therefore, their parent Hamiltonians. For the sake of simplicity we first deal with the case of unique ground state. Later we show that the general case with ground state degeneracy is a straightforward generalization.

Without loss of generality we restrict the local Hilbert space $\mathcal{H}_p$ ($p=0,1$) to the subspace spanned by the virtual MPS tensor $A_p$ so that $\mathcal{H} \simeq\mathbb{C}^{D_p} \otimes \mathbb{C}^{D_p}$. The embedding MPS tensor $A(\gamma)$ that we use to interpolate between $A_0$ and $A_1$ has physical dimension $(\mathbb{C}^{D_0} \oplus  \mathbb{C}^{D_1}) \otimes (\mathbb{C}^{D_0} \oplus  \mathbb{C}^{D_1})$
and bond dimension $D = D_0+D_1$. We define the matrices associated to the tensor $A(\gamma)$ by
\begin{equation}\label{defAgamma}
 A^{i,j}(\gamma)= A^{i,j}  W(\gamma) , 
\end{equation}
where  $W(\gamma) = (1-\gamma)\id_{D_0}\oplus \gamma\id_{D_1}$ and the matrices $\{ A^{i,j}\  1\leq i,j \leq D \}$ have the following form
\begin{equation*}
A^{i,j} = \left\{
    \begin{array}{cc}
        A_0^{i,j}, &  1 \leq i,j \leq D_0 \\
        |i\rangle \langle j |, &   i\leq D_0 \ \& \  j > D_0 \\
        |i\rangle \langle j |, &   j\leq D_0 \ \& \ i > D_0 \\
         A_1^{i-D_0,j-D_0}, &  D_0 < i,j \leq D
    \end{array} \right\} . 
\end{equation*}

The MPS constructed with the tensor $A(\gamma)$ has an MPO symmetry given by the tensor $T_a$ with bond dimension $\chi^0_a + \chi^1_a$, for every $a \in \mathcal{C}$, with the form
$$ T_a = \left(\begin{array}{c c} 
	T^0_a \ & \ T^{0,1}_a 
 \vspace{5pt}\\ 
	T^{1,0}_a \ & \ T^1_a  
\end{array}\right) \ ,  $$
with
$$
T^{0,1}_a =
  \begin{tikzpicture}
    \pic[] (v) at (0.3,0) {V2p=/a/};
    \draw[rounded corners] (v-mid)--++(-0.1,0)--++(0,0.6);
    \draw[rounded corners] (v-down)--++(0.1,0)--++(0,-0.35);
    \node[anchor=south,inner sep=2pt] at (v-top) {$0$};
    \pic[] (w) at (-0.3,0) {W2p=/a/};
    \draw[rounded corners] (w-mid)--++(0.1,0)--++(0,0.6);
    \draw[rounded corners] (w-down)--++(-0.1,0)--++(0,-0.35);
    \node[anchor=south,inner sep=2pt] at (w-top) {$1$};
  \end{tikzpicture}
\ \ \& \ \ 
T^{1,0}_a =
  \begin{tikzpicture}
    \pic[] (v) at (0.3,0) {V2p=/a/};
    \draw[rounded corners] (v-mid)--++(-0.1,0)--++(0,0.55);
    \draw[rounded corners] (v-down)--++(0.1,0)--++(0,-0.4);
    \node[anchor=south,inner sep=2pt] at (v-top) {$1$};
    \pic[] (w) at (-0.3,0) {W2p=/a/};
    \draw[rounded corners] (w-mid)--++(0.1,0)--++(0,0.55);
    \draw[rounded corners] (w-down)--++(-0.1,0)--++(0,-0.4);
    \node[anchor=south,inner sep=2pt] at (w-top) {$0$};
  \end{tikzpicture}  \ , $$
where the upperscript $p=0,1$ denotes the action tensor corresponding to the MPS tensor $A_p$.
The MPO symmetry can be seen by the local relation between the  tensors:
\begin{equation}\label{Agammasym}
T_a A(\gamma) = A(\gamma)
\left (
\begin{tikzpicture}
 \pic[scale=0.8] (w) at (0,0) {W2p=/a/};
    \draw[rounded corners] (w-mid)--++(0.2,0);
    \draw[rounded corners] (w-down)--++(-0.3,0);
    \node[anchor=south,inner sep=2pt] at (w-top) {$0$};
    \end{tikzpicture} 
    \oplus
    \begin{tikzpicture}
     \pic[scale=0.8] (w) at (0,0) {W2p=/a/};
    \draw[rounded corners] (w-mid)--++(0.2,0);
    \draw[rounded corners] (w-down)--++(-0.3,0);
    \node[anchor=south,inner sep=2pt] at (w-top) {$1$};
    \end{tikzpicture} 
\right ) \otimes \left (
  \begin{tikzpicture}
    \pic[scale=0.8] (v) at (0,0) {V2p=/a/};
    \draw[rounded corners] (v-mid)--++(-0.2,0);
    \draw[rounded corners] (v-down)--++(0.3,0);
    \node[anchor=south,inner sep=2pt] at (v-top) {$0$};
    \end{tikzpicture} 
    \oplus
     \begin{tikzpicture}
        \pic[scale=0.8] (v) at (0,0) {V2p=/a/};
    \draw[rounded corners] (v-mid)--++(-0.2,0);
    \draw[rounded corners] (v-down)--++(0.3,0);
    \node[anchor=south,inner sep=2pt] at (v-top) {$1$};
\end{tikzpicture} 
\right ) \ ,
\end{equation}
where the direct sum is over every virtual leg individually of the action tensors.

Some comments are in order. Importantly, the previous MPO tensor defines an MPO representation of $\mathcal{C}$ if the ${L}$-symbols are exactly the same, $L^0=L^1$  (which can be obtained by a gauge transformations if they are equivalent) and if $F^0=F^1$; the fusion tensors of $T_a^0$ and $T_a^1$ are chosen in the appropriate gauge such that the $F$-symbols that they define are the same. Importanlty the action tensors of Eq.\ \eqref{Agammasym} define a set of $L$-symbols equal to $L^0=L^1$ through the whole path. The previous requirements can be seen from the fact that the part of the MPO tensor $T_a$ acting on $(\mathbb{C}^{D_0} \otimes \mathbb{C}^{D_1}) \oplus (\mathbb{C}^{D_1} \otimes \mathbb{C}^{D_0})$ is $T_a^{0,1} \oplus T_a^{1,0}$ and has fusion tensors    
$$ \left (
\begin{tikzpicture} 
\pic (v) at (0,0) {W1=c/a/b};
\node[anchor=south,inner sep=2pt] at (v-top) {$0$};
\end{tikzpicture}\oplus \begin{tikzpicture} 
\pic (v) at (0,0) {W1=c/a/b};
\node[anchor=south,inner sep=2pt] at (v-top) {$1$};
\end{tikzpicture} \right )
\otimes \left (
\begin{tikzpicture} 
\pic (w) at (0,0) {V1=c/a/b};
\node[anchor=south,inner sep=2pt] at (w-top) {$0$};
\end{tikzpicture}\oplus \begin{tikzpicture} 
\pic (w) at (0,0) {V1=c/a/b};
\node[anchor=south,inner sep=2pt] at (w-top) {$1$};
\end{tikzpicture} \right ) 
$$
only if the $L$-symbols are the same (to use Eq.\ \eqref{eq:F_symbol2}). Then, these fusion tensors define a set of $F$-symbols, satisfying Eq.\ \eqref{Fsymbolsdef}, only if $F^0=F^1\equiv F$ since we are able to factor out $F^0 \oplus F^1$ as $F$.

In what follows we show that the symmetric path described by $A(\gamma)$ and $\{T_a\}$ satisfied all the required properties of our definition. Let us first show that $T_a$ is injective. Since we $T_a^0$ and $T_a^1$ are injective we only need to show that $T_a^{0,1} \oplus T_a^{1,0}$ are injective or equivalently, that there is a left inverse acting on the physical level. For that we map $T_a^{0,1} \oplus T_a^{1,0}$ to $A_0\otimes A_1( T_a^{0,1} \oplus T_a^{1,0}) \myinv{A}_0\otimes \myinv{A}_1$ under an appropriate rearrangement of the indices. This results in $T_a^{0} A_0\myinv{A}_0   \oplus T_a^{1} A_1\myinv{A}_1$ which is invertible since we have restricted to the subspace generated by the tensors so that $A_p\myinv{A}_p= \id_{\mathcal{H}_p}$.

It is clear that the tensor $A(\gamma)$ defined in Eq.\ \eqref{defAgamma} is injective for $\gamma \in (0,1)$, so that the parent Hamiltonian $H_{A(\gamma)} \equiv H(\gamma)$ is gapped and has as unique ground state $|\psi_{A(\gamma)} \rangle$ within that region. 
In what follows we show that the previous path converges for $\gamma =p$ ($p=0,1$) and the limit is a parent Hamiltonian of $|\psi_{A_p} \rangle$. We remind that the local Hamiltonian terms are defined by $h_\gamma = \id - \Pi_{S_\gamma}$. The subspace $\mathcal{S}_\gamma = \{ \sum_{i,j}\tr[A^i(\gamma) A^j(\gamma) X]|ij \rangle, \ X\in \mathcal{M}_D \}$ is equal to $\mathcal{S}'_\gamma = \{ \sum_{i,j}\tr[A^i W(\gamma) A^j X]|ij \rangle, \ X\in \mathcal{M}_D \}$ for $\gamma \in (0,1)$ since $W(\gamma) \mathcal{M}_D = \mathcal{M}_D$ because $W(\gamma)$ is full rank in that interval. The subspace $\mathcal{S}'_\gamma$ is $D^2$-dimensional for any $\gamma\in [0,1]$. Let us fix the matrices $\{X_i, i=0,\cdots, D^2\}$ with whom we construct an orthonormal basis (ONB) $\{ v_i(0) \}$ in $\mathcal{S}'_0$. Let us define the set of vectors $\{ v_i(\gamma) \} \in \mathcal{S}'_\gamma$ for $\gamma \in (0,1)$ with such matrices, which need not to be orthogonal but satisfy $\lim_{\gamma \to 0} v_i(\gamma) = v_i(0)$. Crucially, we can orthonormalize this set, by using the Gram-Schmidt process, to construct an ONB $\{ w_i(\gamma) \} \in \mathcal{S}'_\gamma$ that satisfy $\lim_{\gamma \to 0} w_i(\gamma) = v_i(0)$. Then, $\lim_{\gamma \to 0} \Pi_{S_\gamma} = \Pi_{S'_0}$ and similarly for $\gamma = 1$. It is easy to see that the Hamiltonian $H(\gamma = p) = \sum_i (\id - \Pi_{S'_p})$ with PBC is gapped and has as unique ground state $|\psi_{A(0)} \rangle$.

We now deal with the case of systems with ground state degeneracy. We want to connect two block-injective MPSs (with the same number of blocks) that share the $L$-symbols under the action of two MPO algebras with the same fusion category $\mathcal{C}$ and same $F$-symbols. For this we first have to establish a correspondence between the blocks of each MPS, $x:x_0 \to x_1$, and the degeneracies compatible with the action of $\mathcal{C}$ to associate $\left({L^0}_{abx_0}^{y_0} \right)_{c,k_0 \mu}^{z_0,i_0j_0} = \left({L^01}_{abx_1}^{y_1} \right)_{c,k_1 \mu}^{z_1,i_1j_1} $. The embedding MPS tensor has the block-diagonal form $A(\gamma)=\bigoplus_x A_x(\gamma)$. For every block $x$ the bond dimension is $D^0_x+D^1_x$ and it is defined by
$$A_x(\gamma)= A_x \cdot W_x(\gamma), \ A_x = 
  \begin{pmatrix}
    A^0_x &  B^{01}_x  \\
     B_x^{10} & A^1_x\\
  \end{pmatrix}, $$
where $B_x^{pq}= \id_{D^p_x}\otimes \id_{D_x^q}$ and $W_x(\gamma) = (1-\gamma)\id_{D_x^0}\oplus \gamma\id_{D_x^1}$.

The MPO tensor $T_a$ is defined as in the injective case, with bond dimension $\chi^0_a + \chi^1_a$ for every $a \in \mathcal{C}$, but now we have to define the action on all blocks (equivalently for $T^{1,0}_a$):
$$
T^{0,1}_a =
\bigoplus_{x,y,i}
  \begin{tikzpicture}
    \pic[] (v) at (0.35,0) {V2p=y_0/a/x_0};
    \draw[rounded corners] (v-mid)--++(-0.1,0)--++(0,0.6);
    \draw[rounded corners] (v-down)--++(0.1,0)--++(0,-0.35);
    \node[anchor=south,inner sep=2pt] at (v-top) {$0$};
        \node[anchor=north,inner sep=2pt] at (v-bottom) {$i_0$};
    \pic[] (w) at (-0.35,0) {W2p={y_1}/a/{x_1}};
    \draw[rounded corners] (w-mid)--++(0.1,0)--++(0,0.6);
    \draw[rounded corners] (w-down)--++(-0.1,0)--++(0,-0.35);
    \node[anchor=south,inner sep=2pt] at (w-top) {$1$};
            \node[anchor=north,inner sep=2pt] at (w-bottom) {$i_1$};
  \end{tikzpicture}
 \ ,  $$
where the direct sum is over the physical indices and we have used the correspondence between the blocks of $A_0$ and $A_1$ and also between the action of MPOs tensors $T_a^0$ and $T_a^1$. The previous MPO tensors define an MPO representation of $\mathcal{C}$ if $L^0=L^1$ and $F^0=F^1$, this can be seen as in the injective case. Also, it corresponds to an MPO symmetry of $|\psi_{A(\gamma)}\rangle$ since it satisfies 
\begin{equation*}
  \begin{tikzpicture}
    \draw[red] (-0.5,0)--(0.5,0);
    \draw (-0.5,-0.5)--(0.5,-0.5);
    \node[tensor] (t1) at (0,0) {};
    \node[tensor] (t) at (0,-0.5) {};
    \node[] at (0.3,0.2) {$T_a$};
\node[] at (0,-0.75) {$A_x(\gamma)$};
    \draw (0,-0.5) -- (0,0.3);
  \end{tikzpicture} =
  \sum_{i,y}
\left (
\begin{tikzpicture}
 \pic[scale=0.8] (w) at (0,0) {W2p=/a/x_0};
 \node[irrep] at (0.2,0) {$y_0$};
    \draw[rounded corners] (w-mid)--++(0.2,0);
    \draw[rounded corners] (w-down)--++(-0.3,0);
    \node[anchor=south,inner sep=2pt] at (w-top) {$0$};
            \node[anchor=north,inner sep=2pt] at (w-bottom) {$i_0$};
    \end{tikzpicture} 
    \oplus
    \begin{tikzpicture}
     \pic[scale=0.8] (w) at (0,0) {W2p=/a/x_1};
     \node[irrep] at (0.2,0) {$y_1$};
    \draw[rounded corners] (w-mid)--++(0.2,0);
    \draw[rounded corners] (w-down)--++(-0.3,0);
    \node[anchor=south,inner sep=2pt] at (w-top) {$1$};
            \node[anchor=north,inner sep=2pt] at (w-bottom) {$i_1$};
    \end{tikzpicture} 
\right )
  \begin{tikzpicture}
    \node[tensor] (t) at (0,0) {};
    \draw (-0.3,0) -- (0.3,0);
    \node[] at (0,-0.25) {$A_y(\gamma)$};
    \draw (0,0) -- (0,0.3);
  \end{tikzpicture} 
  \left (
  \begin{tikzpicture}
    \pic[scale=0.8] (v) at (0,0) {V2p=/a/x_0};
    \node[irrep] at (-0.2,0) {$y_0$};
    \draw[rounded corners] (v-mid)--++(-0.2,0);
    \draw[rounded corners] (v-down)--++(0.3,0);
    \node[anchor=south,inner sep=2pt] at (v-top) {$0$};
        \node[anchor=north,inner sep=2pt] at (v-bottom) {$i_0$};
    \end{tikzpicture} 
    \oplus
     \begin{tikzpicture}
        \pic[scale=0.8] (v) at (0,0) {V2p=/a/x_1};
        \node[irrep] at (-0.2,0) {$y_1$};
    \draw[rounded corners] (v-mid)--++(-0.2,0);
    \draw[rounded corners] (v-down)--++(0.3,0);
    \node[anchor=south,inner sep=2pt] at (v-top) {$1$};
            \node[anchor=north,inner sep=2pt] at (v-bottom) {$i_1$};
\end{tikzpicture} 
\right ) 
\end{equation*}
which is the analogous version of  Eq.\ \eqref{fusiontensors2} here, so that the action it is characterized by $L^0 = L^1$. The injectivity of every block of $A(\gamma)$ and the injectivity of $T_a$ is justified in the same way as in the previous case. Moreover, $H(\gamma)$ is well-behaved in the whole path its ground space is spanned by $\{ |\psi_{A_x(\gamma)}\rangle\}$.

\subsection{Connection with gapped boundaries of (2+1)d topological phases}
In this section we show how the connection of gapped boundaries of (2+1)d topological phases and our setting, MPO symmetries of (1+1)d systems, can be made explicit using tensor network states. Gapped boundaries of (2+1)d topological phases correspond to inserting boundary gapped Hamiltonian terms compatible with the bulk topological Hamiltonian in a contractible region, see Refs. \cite{Bravyi98,Beigi11}. It turns out, that the different gapped boundaries are in correspondence with module categories of the fusion category that describes the topologically ordered phase \cite{Kitaev12}.

\begin{figure}
 \includegraphics[scale=1.1]{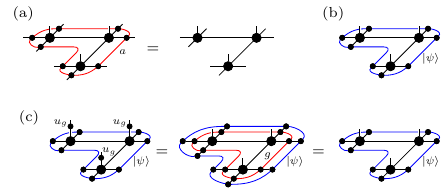}
\caption{(a) Virtual symmetry of the tensors of the PEPS, virtual level in black, in a small region under the MPO, virtual level in red, labeled by $a\in \mathcal{A}$. (b) The boundary state $|\psi\rangle$ is a MPS, virtual level in blue. (c) In the SPT case, the on-site symmetry of the PEPS is translated into an MPO acting at the virtual level. The boundary MPS $|\psi\rangle$ is invariant under the action of that MPO. }\label{fig:bound}
\label{sketch}
\end{figure}

As was pointed out before when the MPO algebra corresponds to a fusion category, our classification of MPSs which remain invariant under these MPOs also corresponds to module categories of fusion categories. In order to make explicit this mathematical connection, let us consider projected entangled pair states (PEPS) \cite{Verstraete04, reviewPEPS}, the two-dimensional version of MPSs. PEPS that realize topologically ordered phases are described by tensors which are invariant under MPO symmetries acting just at the virtual level \cite{Schuch10, Sahinoglu14}, see Fig.\ \ref{fig:bound}(a). For a PEPS on a disc, a natural candidate for a boundary is of the form of an MPS, see Fig.\ \ref{fig:bound}(b). The physical Hilbert space of the MPS is the virtual level of the PEPS, so that the MPO virtual symmetries of the PEPS tensor acts physically on the MPSs describing the boundary. An MPS boundary compatible with the bulk topological order of the PEPS is then, an MPS that is invariant under the virtual symmetries of the PEPS tensor. This situation also arises when considering boundaries of (2+1)d SPT phases described by PEPS, since the global on-site symmetry of the bulk is translated into an MPO acting on the virtual level of the boundary \cite{Chen11A,Molnar18B}, so that a compatible MPS boundary satisfy the relation in Fig.\ \ref{fig:bound}(c).

Here we show that in every subspace $\mathcal{S}_A$, see Eq.\ \eqref{MPSssubs}, which is  invariant under the MPO algebra of a fusion category $\mathcal{A}$, that is for every module category, there is a PBC MPS $|\psi \rangle$ that satisfies 
\begin{equation}\label{symboundary}
    O_a |\psi \rangle = r_a |\psi \rangle \ \forall a\in \mathcal{A},
\end{equation} 
where $r_a>0$. Therefore, $|\psi \rangle$ is a MPS boundary compatible with any PEPS which has as virtual MPO symmetry $\mathcal{A}$. This is also the case when $\mathcal{A}$ is a group and then, the MPS boundary is compatible with the bulk SPT order of the PEPS, see Fig.\ref{fig:bound}(c).

Let us recall that for PBC we have $O_a |\psi_{A_\alpha}\rangle =  \sum_\beta M_{a,\alpha}^\beta |\psi_{A_\beta}\rangle$. Let us define the matrices $M_a= \sum_{\alpha,\beta} M_{a,\alpha}^{\beta}|\beta\rangle \langle \alpha|$ for any block $a$ of the MPO tensor where $\alpha,\beta$ runs over the different MPSs blocks. Since $M_a$ has only non-negative entries, by virtue of Perron-Frobenius Theorem its spectral radius $r_a$ is an eigenvalue of $M_a$. Moreover, using the fact that the algebra represents a fusion category \cite{EGNObook}, there is a vector with positive entries such that it is the common eigenstate of all $M_a$ with eigenvalue $r_a$: $M_a {\cdot} v = r_a v$. The previous equation can be written as $ \sum_\alpha M_{a,\alpha}^\beta v_\alpha = r_a v_\beta$. Then, the MPS defined as $|\psi \rangle = \sum_\alpha v_\alpha |\psi_{A_\alpha} \rangle$ satisfies
$ O_a |\psi \rangle = \sum_\alpha  v_\alpha O_a|\psi_{A_\alpha} \rangle = 
\sum_\alpha  v_\alpha \sum_\beta M_{a,\alpha}^\beta |\psi_{A_\beta}\rangle =
\sum_\beta (\sum_\alpha  v_\alpha  M_{a,\alpha}^\beta) |\psi_{A_\beta}\rangle 
=r_a |\psi \rangle \ \forall a$. Two important examples are: $r_g =  1$ if the MPO algebra describes a group algebra and, $r_a=d_a$ (the quantum dimension of $a$) when the number of blocks of the MPO and the MPS coincides ($M_a=N_a$ and also ${F}={L}$).

We remark that considering symmetric MPSs as boundaries of 2d systems, SPT or topologically ordered, is more restrictive than MPSs which remain invariant under general MPO algebras. This is because in this section we have imposed that the MPO describe a fusion category and thus even the scenarios covered by Section \ref{sec:PBC} lie outside of this framework.

\section{Examples}\label{sec:examples}

\subsection{Solutions of the ${L}$-symbols}
In this section we study the possible phases of certain MPO algebras. We first start with MPO algebras of finite groups and then we move on and study MPO representations of the fusion categories $Rep(S_3)$, $su(2)_4$ and Fib. We emphasize that in this section we only focus on the fusion category part, the PBC elements, of the arbitrary BC MPO algebra since the different phases only depend on that \cite{etingof2003finite}.

\subsubsection{MPO algebras representing $\mathbb{Z}_2$}

Let us analyze the case when the MPO algebra is a representation of $ \mathbb{Z}_2=\{e,g; g^2=e\}$ which has two classes of 3-cocycles, $\mathcal{H}^3(\mathbb{Z}_2,U(1))=\mathbb{Z}_2$.

For the trivial class, $\omega = 1$, there are two phases. They are characterized by the choices $H=\mathbb{Z}_2$ and $H=\mathbb{Z}_1$, both with trivial projective representation so that ${L}$ restricted to $H$ is trivial. The former corresponds to a unique ground state and the latter corresponds to a two-fold degenerate ground space encoded by a MPS with two blocks $\{x_0,x_1\}$ with the permutation $g{\cdot} x_0= x_1$. The coupled pentagon equations, see \eqref{pentagongroups}, result in the following relations for the ${L}$-symbols:
\begin{align*}
 {L}_{e,e}^{x_0} = {L}_{e,g}^{x_0} = {L}_{g,e}^{x_1}, \\  
  {L}_{e,e}^{x_1} = {L}_{e,g}^{x_1} = {L}_{g,e}^{x_0},  \\ 
  \frac{{L}_{g,g}^{x_0}}{{L}_{g,g}^{x_1}} = (+1)\cdot \frac{{L}_{e,g}^{x_0}}{{L}_{g,e}^{x_0}} \ .
\end{align*}
Since we can normalize the $L$-symbols, {\it i.e.} we can always choose ${L}_{g,h}^x=1$ whenever $g=e$ or $h=e$, we can write the previous relations as 
$${L}_{g,g}^{x_0}= (+1) \cdot {L}_{g,g}^{x_1}\ .$$
A realization of these two phases is given by the disordered and ordered phases of the Ising transverse field Hamiltonian $H=-\sum_i (Z_i Z_{i+1}+\lambda X_i)$ with symmetry $U_g=X^{\otimes n}$.

Let us consider now MPO representations of $\mathbb{Z}_2$ that corresponds to the non-trivial $3$-cocycle which is given by $\omega(g,g,g) = -1$ and the other elements equal to one. Since there is no non-trivial subgroup $H$ trivializing $\omega$, there is a unique phase with two-blocks $\{x_0,x_1\}$ where $g{\cdot} x_0= x_1$. One can solve the coupled pentagon equations and obtain, with the previously fixed gauge in the $3$-cocycle, that
\begin{align*}
 {L}_{e,e}^{x_0} = {L}_{e,g}^{x_0} = {L}_{g,e}^{x_1}, \\  
  {L}_{e,e}^{x_1} = {L}_{e,g}^{x_1} = {L}_{g,e}^{x_0},  \\ 
  \frac{{L}_{g,g}^{x_0}}{{L}_{g,g}^{x_1}} = (-1)\cdot \frac{{L}_{e,g}^{x_0}}{{L}_{g,e}^{x_0}} \ .
\end{align*}
By choosing the always possible gauge that normalizes the ${L}$-symbols, {\it i.e.} ${L}_{g,h}^x=1$ whenever $g=e$ or $h=e$, the previous relations reduce to  
$${L}_{g,g}^{x_0}= (-1) \cdot {L}_{g,g}^{x_1}\ .$$
A particular solution for this case is ${L}_{a,b}^{c{\cdot} x} = \omega(a,b,c)$, for $x\in \{x_0,x_1\}$ and $a,b,c\in \mathbb{Z}_2$. 

A realization of this phase can be found in Ref.\ \cite{Roose19} with Hamiltonian $H=\sum_i CZ_{i,i+2}X_{i+1}-\mu Z_i Z_{i+1}$ for the parameter range $\mu>0$. The CZ gate acts on two qubits as $CZ|ab\rangle = (-1)^{a\cdot b}|ab\rangle$ for $a,b=\{0,1\}$ corresponding to the $Z$ basis. This Hamiltonian is invariant under $U=\prod CZ_{i,i+1}Z_i \prod X_i$ that is an MPO representation of $\mathbb{Z}_2$ with the non-trivial cocycle. The gapped phase in the range $\mu<0$ corresponds to a paramagnetic phase where the translation symmetry is broken in one site; we leave for future work the interplay between lattice symmetry and MPO symmetry.

\subsubsection{MPO algebras representing  $\mathbb{Z}_2 \times \mathbb{Z}_2$}

Let us consider $G = \mathbb{Z}_2 {\times} \mathbb{Z}_2 =\{e,a,b,ab\}$ with $\mathcal{H}^3(\mathbb{Z}_2 {\times} \mathbb{Z}_2,U(1))= \mathbb{Z}_2 {\times} \mathbb{Z}_2 {\times} \mathbb{Z}_2$. The different $3$-cocycles can be decomposed into three components \cite{ThesisMark}: $$\omega = (\omega_I^{(1)})^{p_1^I} \cdot (\omega_I^{(2)})^{p_2^I} \cdot (\omega_{II})^{p^{II}},$$
where $p_1^I$,$p_2^I$,$p^{II}$ can have values $0,1$ and the triplet $(p_1^I,p_2^I,p^{II})$ determines the element of $\mathbb{Z}_2 {\times} \mathbb{Z}_2 {\times} \mathbb{Z}_2$. A particular gauge for every component is $\omega_I^{(1)}(a b^i,a b^j, a b^k) = -1 $, $\omega_I^{(2)}(a^i b,a^j b, a^k b) = -1 $ and $\omega_{II}(ab^i,a^jb,a^k b) = -1$ for all $i,j,k=\{0,1\}$, and where the rest of the cocycle elements are equal to $+1$.

The different phases for each $3$-cocycle $\omega$ are given by the pairs $(H,\mathcal{H}^2(H,U(1)))$, where the subgroup $H{\subset} G$  trivializes $\omega$, $\omega|_H =+1$. Since all the proper subgroups of $G$ have trivial second cohomology group, the phases for non-trivial $3$-cocycles are just determined by the subgroup $H$ and have $|G|/|H|$ blocks in the MPS. 

Let $H_g$ denote the subgroup of $G$ generated by $g$ and let $\omega_g$ denote $\omega(g,g,g)$. It can be checked that $\{ \omega_g \}_{g\in G}$ can distinguish between the different classes of $3$-cocycles in this case. The following table shows all the phases of the different $3$-cocycles of $\mathbb{Z}_2 {\times} \mathbb{Z}_2$:

    \begin{center}
          \begin{tabular*}{\linewidth}{@{\extracolsep{\fill}}|c || c | c | }
            \hline\hline
        $(p_1^I,p_2^I,p^{II})$& $(\omega_a,\omega_b ,\omega_{ab})$ & $\{ H; \omega|_H=1 \}$   \\
        \hline \hline
        $(0,0,0)$ & $(+1,+1,+1)$ & $\{ H \subseteq G \}$    \\
        \hline
        $(1,0,0)$ & $(-1,+1,-1)$ & $\{ H_e,H_b \}$    \\
        $(0,1,0)$ & $(+1,-1,-1)$ & $\{ H_e,H_a \}$     \\
        $(1,1,0)$ & $(-1,-1,+1)$ & $\{ H_e,H_{ab} \}$     \\
        \hline
        $(0,0,1)$ & $(+1,+1,-1)$ & $\{ H_e, H_a,H_b \}$    \\
        $(0,1,1)$ & $(+1,-1,+1)$ & $\{ H_e,H_a,H_{ab} \}$    \\
        $(1,0,1)$ & $(-1,+1,+1)$ & $\{ H_e,H_b,H_{ab} \}$     \\
        \hline
        $(1,1,1)$ & $(-1,-1,-1)$ & $\{ H_e \}$     \\
        \hline
        \hline
        \end{tabular*}
    \end{center}

The first line of the table, the case of trivial $3$-cocycle, corresponds to the standard SPT phases. There are six of them, see Ref.\ \cite{Schuch11}, four phases are characterized simply by the unbroken groups $H_e,H_a,H_b,H_{ab}$ and there are two phases with unique ground state corresponding to $H=G$. These two phases come from the two distinct classes of $2$-cocycles of $G$: the non-trivial one characterizes the Haldane phase.    
    
There are $7$ non-trivial $3$-cocycles but the possible phases of those can be grouped in three types (by relabeling the elements of the $3$-cocycles)

First, there are MPO representations with $3$-cocycles that can only be trivialized by one proper subgroup, so there is only one phase with two blocks. This case can be understood as taking a tensor product of two $\mathbb{Z}_2$ MPO representations, one with non-trivial $3$-cocycle and the other with trivial $3$-cocycle that does not generate symmetry breaking.

Second, there are MPO representations with $3$-cocycles hosting two different phases with two-fold degeneracy each.

As a realization of one of these $3$-cocycles we can use the PBC $\mathbb{Z}_2\times \mathbb{Z}_2$ MPO representation $\{\id,U_a,U_b,U_{ab}\}$ defined by $U_a=\prod_i CZ_{i,i+1}Z_i$, $U_b= \prod_i X_i$ and $U_{ab}= U_a \cdot U_b$. In this realization the only non-trivial $3$-cocycle element is $\omega(ab,ab,ab)=-1$.

The explicit MPSs are given by noticing that the operators $U_a$ and $U_b$ leave invariant the subspaces generated by $\{ | 0\rangle^{\otimes n}, | 1\rangle^{\otimes n} \}$ and $\{ | + \rangle^{\otimes n} ,| \tilde{\it cs} \rangle \}$ respectively, where $| \tilde{\it cs} \rangle$ is the cluster-like state $U_a | + \rangle^{\otimes n}$. It is easy to check that for normalized ${L}$-symbols, in both phases  ${L}_{ab,ab}^{x} = - {L}_{ab,ab}^{y}$ is satisfied and for $H=H_a$, we have ${L}_{a,a}^{x} =  {L}_{a,a}^{y}$ and for $H=H_b$, ${L}_{b,b}^{x} =  {L}_{b,b}^{y}$.

Finally, the MPO representations with the $3$-cocycle corresponding to $(p_1^I,p_2^I,p^{II}) = (1,1,1)$ hosts only one phase with four MPS blocks. In this case, a solution for the ${L}$-symbols corresponds to ${L}_{a,b}^{c{\cdot} x} = \omega(a,b,c)$, for $x\in \{x_0,x_1,x_2,x_3\}$ and $a,b,c\in \mathbb{Z}_2 \times \mathbb{Z}_2 $. 

\subsubsection{MPOs representing $su(2)_4$}
In this example we consider MPOs whose blocks correspond to the algebra of $s u (2)_4$, with block labels $\{ 0, \frac{1}{2}, 1, \frac{3}{2}, 2\}$, and fusion rules:
    \begin{center}
          \begin{tabular*}{0.9\linewidth}{@{\extracolsep{\fill}}|c || c | c | c | c | c |}
            \hline\hline
        ${\times}$ & $0$ & $\frac{1}{2}$ & $1$ & $\frac{3}{2}$ & $2$  \\
        \hline
        $0$ & $0$ & $\frac{1}{2}$ & $1$ & $\frac{3}{2}$ & $2$   \\
        $\frac{1}{2}$ & $\frac{1}{2}$ & $0+1$  &  $\frac{1}{2}+\frac{3}{2}$ &  $1+2$ &  $\frac{3}{2}$  \\
        $1$ & $1$ &  $\frac{1}{2}+\frac{3}{2}$ &  $0+1+2$ &  $\frac{1}{2}+\frac{3}{2}$ &  $1$  \\
        $\frac{3}{2}$ & $\frac{3}{2}$& $1+2$  &  $\frac{1}{2}+\frac{3}{2}$ &  $0+1$ &  $\frac{1}{2}$  \\
        $2$ & $2$ & $\frac{3}{2}$  &  $1$ &  $\frac{1}{2}$ &  $0$  \\
        \hline
        \hline
        \end{tabular*}
    \end{center}
The ${F}$-symbols can be found in \cite{Ardonne2010}, where we consider the unique solution with positive Frobenius-Schur indicators. There are only two different phases associated to this MPO algebra (left module categories of $su(2)_4$ Ref.\ \cite{ostrik2001module}):

\begin{itemize}
\item  A phase is characterized by associating every MPS block (there are five of them) with an MPO block such that the ${F}$-symbols and the ${L}$-symbols coincide.

\item The second phase is associated to the so-called module category $\mathcal{M}_{TY}$ and it has been studied before in Refs.\ \cite{Runkel2000,vanhove2021topological}, where the ${L}$-symbols can also be found. The MPS has four blocks $\{ x, y, z, s\}$ and they transform under the action of the MPO blocks as follows: 
    \begin{center}
          \begin{tabular*}{1\linewidth}{@{\extracolsep{\fill}}|c || c | c | c | c |}
            \hline\hline
        ${\cdot}$ & $x$ & $y$ & $z$ & $s$  \\
        \hline
        $0$ & $x$&  $y$  &  $z$ &  $s$  \\
        $\frac{1}{2}$ & $s$ & $s$  &  $s$ &  $x+y+z$ \\
        $1$ & $y+z$ &  $x+z$ &  $x+y$ &  $2s$ \\
        $\frac{3}{2}$ & $s$& $s$  &  $s$ &  $x+y+z$ \\
        $2$ & $x$& $y$  &  $z$ &  $s$ \\
        \hline
        \hline
        \end{tabular*}
    \end{center}
 
We observe that the subalgebra generated by $\{ 0, 1, 2\}$ of $su(2)_4$ leaves invariant the set of MPS blocks $\{ x,y,z\}$ and also the block $\{ s \}$ independently.

The previous observation helps us to understand the phase diagram obtained in \cite{Gils13} for $S = 1$ in $su(2)_4$ anyonic spins chains. There, the authors study the phase diagrams of integer and half-integer sectors independently, which effectively implies that only the subalgebra $\{ 0, 1, 2\}$ of $su(2)_4$ is enforced as the 'topological symmetry' (since $\frac{1}{2},\frac{3}{2} \in su(2)_4 $ permute between the sectors). This explains the fact that in Ref.\ \cite{Gils13}, they found gapped phases with $3$-fold degeneracy in the integer sector (and unique ground state for the half-integer sector). This situation is forbidden when imposing the whole $su(2)_4$ symmetry.

\end{itemize}

\subsubsection{MPO representations of $Rep(S_3)$}
In this example we consider an MPO with three blocks $\{1,\pi, \psi \}$ that fuse as the irreps of $S_3$: $\pi\times \pi = 1+\pi+\psi$, $\psi \times \psi = 1$ and $\psi \times \pi = \pi \times \psi = \pi$. This corresponds to the fusion category $Rep(S_3)$. We notice that $Rep(S_3)$ is a subcategory of $su(2)_4$ by associating $\{ 0, 1, 2\}$ with $\{ 1, \pi, \psi \}$, we will use this fact to provide the ${L}$-symbols of this case based on the previous one. Since we are working with the irreps of a group, the ${F}$-symbols are associated with the Wigner $6j$-symbols satisfying the Biedenharn-Elliot identity as the pentagon equation \cite{Lootens21A}. 

In general, the possible phases that are MPO symmetric of $Rep(G)$ are characterized by the subgroups of $G$ and their second cohomology group \cite{etingof2003finite}. All the subgroups of $S_3$ have trivial second cohomology group, so in this case every phase is characterized by a subgroup $H\subset S_3$ only. Moreover, the MPS blocks can be associated to the elements of the category $Rep(H)$, the irreps of $H$: see \cite{etingof2003finite}. Let us consider the different subgroups $H$ of $S_3$ and study their phases:

\begin{itemize}
    \item $H= \mathbb{Z}_1$. In this case there is only one block, labelled $s$, in the MPS. The block $s$ is invariant under
    the action of the MPO blocks $1$ and $\psi$ and it transforms as $\pi{\cdot} s = 2 s$ under the action of the block $\pi$. We notice that this does not contradict the fact that a non-trivial MPO cannot have unique ground state, as derived in section \ref{nonuniqueGS}, since we assume that every action has multiplicity one. The $L$-symbols describing this phase can be chosen to be the same as the $L$-symbols in the $\mathcal{M}_{TY}$ phase restricted to the $s$ block.

    \item $H= \mathbb{Z}_2$. This phase is characterized by two MPS blocks, $x$ and $y$, and the action of the MPO blocks is given by:
    \begin{center}
       \begin{tabular*}{0.5\linewidth}{@{\extracolsep{\fill}}|c || c | c |}
            \hline\hline
        ${\cdot}$ & $x$ & $y$  \\
        \hline
        1 & $x$&  $y$  \\
        $\pi$ & $x+y$& $x+y$   \\
        $\psi$ & $y$ &  $x$ \\
        \hline
        \hline
        \end{tabular*}
    \end{center}
    
     The $L$-symbols of this phase can be chosen the same as the $L$-symbols of the first phase analyzed in the $su(2)_4$ case. To do so, we have to associate the blocks $\{x,y\}$ with blocks labeled by $\{\frac{1}{2}, \frac{3}{2}\}$ in that case.
     
    \item $H= \mathbb{Z}_3$. This phase is characterized by three MPS blocks, $x$, $y$ and $z$, where the action of the MPO blocks is given by:
    \begin{center}
          \begin{tabular*}{0.7\linewidth}{@{\extracolsep{\fill}}|c || c | c | c |}
            \hline\hline
        ${\cdot}$ & $x$ & $y$ & $z$  \\
        \hline
        1 & $x$&  $y$  &  $z$ \\
        $\pi$ & $y+z$& $x+z$  &  $x+y$  \\
        $\psi$ & $x$ &  $z$ &  $y$ \\
        \hline
        \hline
        \end{tabular*}
    \end{center}
    Again, the $L$-symbols can be obtained by restricting the $L$-symbols of the phase $\mathcal{M}_{TY}$ invariant under $su(2)_4$ when restricting to the blocks $\{x,y,z\}$.
    
    \item $H= {S_3}$. This phase is characterized by three MPS blocks, $x$, $y$ and $z$, and the action of the MPO blocks is given by:
    \begin{center}
          \begin{tabular*}{0.6\linewidth}{@{\extracolsep{\fill}}|c || c | c | c |}
            \hline\hline
        ${\cdot}$ & $x$ & $y$ & $z$  \\
        \hline
        1 & $x$&  $y$  &  $z$ \\
        $\pi$ & $y$& $x+y+z$  &  $y$  \\
        $\psi$ & $z$ &  $y$ &  $x$ \\
        \hline
        \hline
        \end{tabular*}
    \end{center}
Notice that this action reproduces the fusion rules of $Rep(S_3)$, so the MPS blocks can be labeled by the irreps of $S_3$ too and then the ${L}$-symbols coincide with the ${F}$-symbols ( {\it i.e.} the Wigner $6j$-symbols).
\end{itemize}

\subsubsection{MPO representing the Fib fusion category}

We consider an MPO representing the Fibonacci fusion category $\{ 1, \tau ; \  \tau {\times} \tau = 1+ \tau\}$, see an explicit construction and its $F$-symbols in Ref.\ \cite{Bultinck17A}. The only solution for an invariant MPS subspace is characterized by two blocks  $\{x_1, x_\tau\}$ that transform under $\tau$ as $\tau {\cdot} x_1 = x_\tau$ and $\tau {\cdot} x_\tau = x_1 + x_\tau$. Then, the ${L}$-symbols coincide with the ${F}$-symbols of the Fibonacci fusion category.

\subsection{Examples of explicit MPSs and MPO representations}\label{sec:expexam}
In this section we consider MPOs whose blocks correspond to elements of a finite group $G$ and the $F$-symbols are given by a $3$-cocycle of $G$.

We first consider examples satisfying Eq.\ \eqref{fusiontensorG}, {\it i.e.} MPOs that form an algebra even with arbitrary BC. These examples are based on Ref.\ \cite{Bultinck17A}. Then, we construct examples for MPO representations of $G$, where the algebra is defined only for arbitrary BC as in Sec.\ref{sec:PBC}.

\subsubsection{Arbitrary boundary condition case}

We construct examples where the local Hilbert space where the MPOs act on is $\mathbb{C}[G]\otimes \mathbb{C}[G]$. Also the virtual space is $\mathbb{C}[G]\otimes \mathbb{C}[G]$, bond dimension $|G|^2$, where $\mathbb{C}[G]$ is the group algebra of $G$ and the basis is labeled by the group elements $g \equiv |g\rangle$. The non-zero elements of the MPO tensors are
$$
\begin{tikzpicture}
\draw[red] (-0.75, 0.25) -- (0.75, 0.25);
\node at (-0.95, 0.25) {$g$};
\node at (0.95, 0.25) {$g$};
\draw[red] (-0.75, -0.25) -- (0.75, -0.25);
\node at (-0.95, -0.25) {$k$};
\node at (0.95, -0.25) {$l$};
\draw (-0.25, 0.75) -- (-0.25, -0.75);
\node at (-0.35, -0.9) {$k$};
\node at (0.35, -0.9) {$l$};
\draw (0.25, 0.75) -- (0.25, -0.75);
\node at (-0.25, 0.95) {$gk$};
\node at (0.25, 0.95) {$gl$};
\draw[rounded corners, fill= black] (-0.5,-0.5) rectangle (0.5,0.5);
\end{tikzpicture}
= \omega(g,l,l^{-1}k) \ , 
$$
where the blocks are labeled by the elements $g$ above and every block has bond dimension $|G|$. The non-zero elements of the fusion tensors are given by 
\begin{equation}\label{ftexam}
\begin{tikzpicture}
\draw[red] (-0.5, 0.25) -- (0, 0.25);
\node at (-0.75, 0.25) {$gh$};
\draw[red] (-0.5, -0.25) -- (0, -0.25);
\node at (-0.7, -0.25) {$k$};
\draw[red] (0.5, 0.55) -- (0, 0.55);
\node at (0.7, 0.55) {$g$};
\draw[red] (0.5, 0.25) -- (0, 0.25);
\node at (0.75, 0.25) {$hk$};
\draw[red] (0.5, -0.55) -- (0, -0.55);
\node at (0.7, -0.55) {$k$};
\draw[red] (0.5, -0.25) -- (0, -0.25);
\node at (0.7, -0.25) {$h$};
\draw[rounded corners, fill= black] (-0.25,-0.75) rectangle (0.25,0.75);
\end{tikzpicture}
= \omega(g,h,k)^{-1} \ .
\end{equation}
It is easy to verify that these fusion tensors satisfy Eq.\ \eqref{3cocygroup} with the $3$-cocycle $\omega(g,h,k)^{-1}$.

It is instructive to rewrite the tensors corresponding to each block $g\in G$ (with bond dimension $|G|$) as
$$T_g = 
\begin{tikzpicture}
\draw (-0.3,-0.5)--(-0.3,0.75);
\draw (0.3,-0.5)--(0.3,0.75);
\draw (-0.3,0.25)--(0.3,0.25);
 \node[tensor,label=below:$\omega_g$]  at (0,0.25) {};
  \node[tensor,label=right:$L_g$]  at (0.3,0.5) {};
   \node[tensor,label=left:$L_g$]  at (-0.3,0.5) {};
\draw[red] (-0.3,-0.2)--(-0.6,-0.2);
\draw[red] (0.3,-0.2)--(0.6,-0.2);
\end{tikzpicture} \ , 
$$
where $\omega_g$ is the $|G|\times|G|$ matrix giving by $(\omega_g)_{k,l}= \omega(g,l,l^{-1}k)$ and every three line intersection corresponds to a $|G|$-dimensional delta tensor. Let us define the matrix $W_g$ by
$$W_g = 
\begin{tikzpicture}
\draw (-0.3,-0.25)--(-0.3,0.25);
\draw (-0.3,0) -- (0.3,0);
\draw (0.3,-0.25)--(0.3,0.25);
 \node[tensor,label=below:$\omega_g$]  at (0,0) {};
\end{tikzpicture} \ .
$$
This matrix is diagonal in the  $\mathbb{C}[G]\otimes \mathbb{C}[G]$ basis: $W_g|k,l\rangle = \omega(g,l,l^{-1} k)|k,l\rangle$. In particular $W_e=\id$ for normalized 3-cocycles. 

Let us show this construction for a particular example. We choose $G=\mathbb{Z}_2$ with the only non-trivial 3-cocycle given by $\omega(g,g,g)=-1$, $W_g= CZ(\id\otimes Z)$ and $L_g= X$.

Let $O_g$ denotes the PBC MPO described by the MPO tensor $T_g$. In particular, the PBC MPO corresponding to the trivial element $e$ is a projector
$$ 
O_e = \bigotimes_i \mathcal{P}^{[i]_r,[i+1]_l} = \ 
\cdots \
\begin{tikzpicture}
\foreach \x in {-1,0,1}{
\draw[densely dotted,rounded corners] (\x+0.2,-0.35) rectangle (\x+0.8,0.35);
}
\node at (0.5, 0.45) {$i$};
\node at (0.25, -0.5) {$[i]_l$};
\node at (0.85, -0.5) {$[i]_r$};
\foreach \x in {-1,0,1,2}{
\begin{scope}[shift={(\x,0)}]
\draw (-0.3,-0.25)--(-0.3,0.25);
\draw[red] (-0.3,0) -- (0.3,0);
\draw (0.3,-0.25)--(0.3,0.25);
\end{scope}}
\end{tikzpicture}
\ \cdots,
$$
where $\mathcal{P}= \sum_g |g,g\rangle\langle g,g|$. Note that the projector $\mathcal{P}$ acts on two consecutive sites, $i$ and $i+1$. These sites have an internal tensor product structure and $\mathcal{P}$ only acts on the right tensor component of site $i$ (labeled by $i_r$), and only on the left tensor component of the site $i+1$ (labeled by $i+1_r$). For $g\neq e$ we can write
$$ O_g = \bigotimes_i (L_g^{[i]_l}\otimes L_g^{[i]_r}){\cdot} W_g^{[i]_l,[i]_r} {\cdot} \mathcal{P}^{[i]_r,[i+1]_l}. $$

It is easy to verify that these MPOs satisfy $O_g O_h= O_{gh}$ and $O_g^\dagger =O_{g^{-1}}$, so they form a representation of $G$ on the subspace $O_e (\mathbb{C}[G]\otimes \mathbb{C}[G])^{\otimes n } $ of the full Hilbert space $(\mathbb{C}[G]\otimes \mathbb{C}[G]) ^{\otimes n }$.

Let us now construct MPSs that are invariant under the action of these MPOs.  These MPSs satisfy Eq.\ \eqref{MPOsymG}, for a given solution of the ${L}$-symbols of Eq.\ \eqref{pentagongroups}. Let $H$ be a subgroup of $G$ that trivializes the $3$-cocycle defined by the MPO. We will construct a MPS with unbroken symmetry group $H$, so that the MPS has $|G/H|$ blocks.

Let us denote as $\{k_i\}$ a complete set of coset representatives such that $G= \bigcup_i k_i H $. We associate every MPS block $x_i$ with a coset representative $k_i$, such that the action of the element $g$  on the MPS block $x_i$ is $g\cdot x_i = x_{i'}$, where ${i'}$ is determined by $gk_i = k_{i'} h'$.  In our construction every block $\alpha$ has bond dimension $|G/H|$, so that the total bond dimension is $D=|G/H|^2$, and the non-zero elements of the MPS tensor are
$$
\begin{tikzpicture}
\draw (-0.75, 0.1) -- (0.75, 0.1);
\node at (-1, 0.3) {${k}{\cdot} x$};
\node at (1, 0.3) {${l}{\cdot} x$};
\draw (-0.75, -0.25) -- (0.75, -0.25);
\node at (-0.9, -0.25) {$x$};
\node at (0.9, -0.25) {$x$};
\draw (-0.25, 0.75) -- (-0.25, 0);
\draw (0.25, 0.75) -- (0.25, 0);
\node at (-0.25, 0.9) {$\myinv{k}$};
\node at (0.25, 0.9) {$\myinv{l}$};
\draw[rounded corners, fill= black] (-0.5,-0.5) rectangle (0.5,0.5);
\end{tikzpicture}
= \left({L}^{{k}{\cdot} x}_{\myinv{l},l\myinv{k}} \right)^{-1} \ . 
$$
The non-zero elements of action tensors are given by 
$$
\begin{tikzpicture}
\draw (-0.5, 0.25) -- (0, 0.25);
\node at (-0.75, 0.45) {$h{\cdot} x$};
\draw (-0.5, -0.25) -- (0, -0.25);
\node at (-0.7, -0.45) {$g{\cdot} x$};
\draw[red] (0.5, 0.55) -- (0, 0.55);
\node at (0.65, 0.55) {$g$};
\draw[red] (0.5, 0.25) -- (0, 0.25);
\node at (0.75, 0.25) {$\myinv{h}$};
\draw (0.5, -0.55) -- (0, -0.55);
\node at (0.65, -0.55) {$x$};
\draw (0.5, -0.25) -- (0, -0.25);
\node at (0.8, -0.25) {$h{\cdot} x$};
\draw[rounded corners, fill=gray ] (-0.25,-0.75) rectangle (0.25,0.75);
\end{tikzpicture}
= \left({L}^{h{\cdot} x}_{g,\myinv{h}} \right)^{-1} \ .
$$

It is easy to verify that the MPO tensor, the action tensor and the MPS tensor satisfy Eq.\ \eqref{MPOsymG}, {\it i.e.}
$$
\begin{tikzpicture}
\begin{scope}[yshift = 1.5cm]
\draw[red] (-0.75, 0.25) -- (0.75, 0.25);
\node at (-0.95, 0.25) {$g$};
\node at (0.95, 0.25) {$g$};
\draw[red] (-0.75, -0.25) -- (0.75, -0.25);
\node at (-0.95, -0.25) {$k$};
\node at (0.95, -0.25) {$l$};
\draw (-0.25, 0.75) -- (-0.25, -0.75);
\draw (0.25, 0.75) -- (0.25, -0.75);
\node at (-0.25, 0.95) {$gk$};
\node at (0.25, 0.95) {$gl$};
\draw[rounded corners, fill= black] (-0.5,-0.5) rectangle (0.5,0.5);
\end{scope}
\draw (-0.75, 0.1) -- (0.75, 0.1);
\node at (-0.85, 0.3) {$\myinv{k}{\cdot} x$};
\node at (0.85, 0.3) {$\myinv{l}{\cdot} x$};
\draw (-0.75, -0.25) -- (0.75, -0.25);
\node at (-0.9, -0.25) {$x$};
\node at (0.9, -0.25) {$x$};
\draw (-0.25, 0.75) -- (-0.25, 0);
\draw (0.25, 0.75) -- (0.25, 0);
\node at (-0.35, 0.75) {${k}$};
\node at (0.35, 0.75) {${l}$};
\draw[rounded corners, fill= black] (-0.5,-0.5) rectangle (0.5,0.5);
\end{tikzpicture}
=
\begin{tikzpicture}
\draw (-1.25,0.2)-- (1.25,0.2);
\draw (-1.25,-0.2)-- (1.25,-0.2);
\draw[red] (-1.75,0.55)-- (-2.1,0.55);
\draw[red] (-1.75,0.25)-- (-2.1,0.25);
\draw[] (-1.75,-0.55)-- (-2.1,-0.55);
\draw[] (-1.75,-0.25)-- (-2.1,-0.25);
\draw[red] (1.75,0.55)-- (2.1,0.55);
\draw[red] (1.75,0.25)-- (2.1,0.25);
\draw[] (1.75,-0.55)-- (2.1,-0.55);
\draw[] (1.75,-0.25)-- (2.1,-0.25);
\node at (1.9, 0.65) {$g$};
\node at (1.9, 0.35) {$l$};
\node at (2.1, -0.1) {$\myinv{l}{\cdot} x$};
\node at (1.9, -0.45) {$x$};
\node at (-1.9, 0.65) {$g$};
\node at (-1.9, 0.35) {$k$};
\node at (-2.1, -0.1) {$\myinv{k}{\cdot} x$};
\node at (-1.9, -0.45) {$x$};

\draw[rounded corners, fill= black] (-0.5,-0.5) rectangle (0.5,0.5);
\draw[rounded corners, fill=gray ] (-1.75,-0.75) rectangle (-1.25,0.75);
\draw[rounded corners, fill=gray ] (1.75,-0.75) rectangle (1.25,0.75);
\node at (-0.9, 0.35) {$\myinv{k}{\cdot} x$};
\node at (0.9, 0.35) {$\myinv{l}{\cdot} x$};
\node at (-0.9, -0.35) {$g{\cdot} x$};
\node at (0.9, -0.35) {$g{\cdot} x$};
\draw (-0.25, 0.75) -- (-0.25, 0);
\draw (0.25, 0.75) -- (0.25, 0);
\node at (-0.25, 0.9) {$gk$};
\node at (0.25, 0.9) {$g{l}$};
\end{tikzpicture}
$$

\subsubsection{Periodic boundary condition case}

Let us construct now a PBC MPO representation of $G$ by modifying the previous examples. To do so we can map $O_e (\mathbb{C}^{|G|} \otimes \mathbb{C}^{|G|})^{\otimes n}$ to $\mathbb{C}^{|G|}$ by using the isometry transformation $\Gamma: |g,g\rangle_{[i]_r,[i+1]_l} \to |g\rangle_i$ so that ${U}_g = \bigotimes_i \Gamma_{[i]_r,[i+1]_l}^\dagger (U_g) \bigotimes_i \Gamma_{[i]_r,[i+1]_l}$ and then, ${U}_e= \id_{|G|}^{\otimes n}$. We notice that this transformation acts between sites such that $\bigotimes_i \Gamma_{[i]_r,[i+1]_l}$ is a $2$-body isometry circuit of depth $n$. The final MPOs are a representation of the group $G$ of the form
$$ {U}_g = \bigotimes_i (L_g^{i}\otimes L_g^{i+1}){\cdot} W_g^{i,i+1} \ . $$
The MPO tensor describing this operator $U_g$ is 
$$
\hat{T}_g = 
\begin{tikzpicture}
\draw[red] (-0.2,0.25)--(0.3,0.25);
\draw[red] (0.3,0)--(0.6,0);
\draw (0.3,-0.25)--(0.3,0.75);
 \node[tensor,label=below:$\omega_g$]  at (0.05,0.25) {};
  \node[tensor,label=right:$L_g$]  at (0.3,0.5) {};
\end{tikzpicture}
\equiv
\begin{tikzpicture}
      \draw (0,-0.3) -- (0,0.3);
      \draw[red] (-0.3,0) -- (0.3,0);
    \node[tensor] (t) at (0,0) {};
\end{tikzpicture} \ .
$$

For example, for the non-trivial 3-cocycle of $\mathbb{Z}_2$ ${U}_g=\prod CZ_{i,i+1}Z_i \prod X_i$.  These MPO tensors satisfy Eq.\ \eqref{fusiontensorG2} with fusion tensors defined in Eq.\ \eqref{ftexam}. We notice that they also satisfy the left zipper equation:

\begin{equation*}
  \begin{tikzpicture}
\draw[red] (0, 0) -- (0.3, 0);
\draw[red] (0, -0.6) -- (0.3, -0.6);
     \pic (v) at (-0.5,-0.3) {V1=gh/g/h};
    \node[tensor] (t) at (0,0) {};
    \node[tensor] (t) at (0,-0.6) {};
    \draw (0,-0.9) -- (0,0.3);
  \end{tikzpicture} =
  \begin{tikzpicture}[baseline=-1mm]
 \pic (w) at (0.5,0) {V1=/g/h};
      \draw (0,-0.3) -- (0,0.3);
      \draw[red] (-0.3,0) -- (0.3,0);
    \node[tensor] (t) at (0,0) {};
   \node[irrep] at (0.25,0.05) {$gh$};
     \end{tikzpicture}\ ,
\end{equation*}
but they satisfy neither the right zipper equation nor Eq.\ \eqref{fusiontensorG}. Instead, they satisfy the following equation

\begin{equation*}
  \begin{tikzpicture}
    \draw[red]  (-0.95,0)--(0.5,0);
    \draw[red]  (-0.95,-0.5)--(0.5,-0.5);
    \draw[thick, fill=white] (-0.7,-0.7) rectangle (-0.5,0.2);
    \node[irrep] at (-0.25,0) {$g$};
    \node[irrep] at (-0.25,-0.5) {$h$};
        \node[irrep] at (-0.95,0) {$g$};
    \node[irrep] at (-0.95,-0.5) {$h$};
    \node[] at (-0.6,0.4) {$P_{g,h}$};
    \node[tensor] (t) at (0,0) {};
    \node[tensor] (t) at (0,-0.5) {};
    \draw (0,-0.8) -- (0,0.3);
  \end{tikzpicture} =
  \begin{tikzpicture}[baseline=0mm]
    \node[tensor] (t) at (0,0) {};
    \draw (0,-0.3) -- (0,0.3);
    \pic (v) at (0.5,0) {V1=gh/g/h};
    \pic (w) at (-0.5,0) {W1=gh/g/h};
  \end{tikzpicture}\ ,
\end{equation*}
where $P_{g,h}$ is a projector with components
\begin{equation*}
\begin{tikzpicture}

\draw[red] (0.5, 0.55) -- (-0.5, 0.55);
\node at (0.7, 0.55) {$g$};
\node at (-0.7, 0.55) {$g$};
\draw[red] (0.5, 0.25) -- (-0.5, 0.25);
\node at (0.75, 0.25) {$k'$};
\node at (-0.75, 0.25) {$k'$};
\draw[red] (0.5, -0.55) -- (-0.5, -0.55);
\node at (0.7, -0.55) {$k$};
\node at (-0.7, -0.55) {$k$};
\draw[red] (0.5, -0.25) -- (-0.5, -0.25);
\node at (0.7, -0.25) {$h$};
\node at (-0.7, -0.25) {$h$};
\draw[thick, fill= white] (-0.25,-0.75) rectangle (0.25,0.75);
\end{tikzpicture}
= \delta_{k',hk} \ .
\end{equation*}
We note that this projector is supported on the PBC virtual space so it acts as the identity when closing the boundaries.

\section{Conclusions and outlook}\label{sec:outlook}

In this work we have classified the quantum phases of ground spaces generated by MPSs that are invariant under the action of MPO algebras. The different phases are in one-to-one correspondence with the inequivalent classes of ${L}$-symbols, classified by the coupled pentagon equation \eqref{coupledpent}. In particular, when the MPO algebra describes on-site representations of groups, we recover the SPT classification: there the ${L}$-symbol corresponds to a $2$-cocycle phase factor. Our phase classification matches with previous categorical approaches \cite{Kitaev12,Thorngren19} obtaining module categories of fusion categories, but benefits of lattice realizations, facilitates their numerical study and extends it outside of renormalization fixed points. Even though our results are derived using tensor network approaches, we expect that our classification is applicable in more general cases as well, as it was the case for SPT phases \cite{ogata2021classification}.

We have made an explicit connection between the phases we study and the boundaries of two-dimensional systems. In the light of this connection, our classification could be used to detect the bulk topological order (either intrinsic or symmetry protected). Concretely, one can use the class of the ${L}$-symbols of the gapped boundary theory to obtain the unique associated intrinsic topological bulk order \cite{Kong17} by detecting the $F$-symbols. For SPT phases, such protocols have been used to detect the bulk phase in Refs. \cite{Chen11A,Else14,Zaletel14,Kawagoe21}, using the boundary. It would be interesting to use our work to generalize the results of Ref. \cite{Kawagoe21} for the cases where the boundary does not fully break the symmetry.

It is interesting to note that, as it can be seen from the analysis in Sec.\ref{sec:TRS}, time reversal symmetry could restrict the form of the ${F}$-symbol of the MPO representation, and more surprisingly, the form of the ${L}$-symbols. This result shows that some topologically ordered states in the presence of time reversal symmetry cannot host some of the gapped boundaries previously allowed without TRS. As gapped boundaries are in one-to-one correspondence with anyon condensation patterns \cite{Kitaev12}, it could be an interesting research project to understand how time reversal symmetry affects these anyon condensation patterns.

Non-trivial SPT phases are resources for measurement based quantum computation \cite{Else12}. The same protocol can be used in the phases we study, since the same projective action protecting a virtual subspace appears. Interestingly, Ref.\ \cite{Wang15A} studied the relation between MPO symmetries of abelian groups and domain wall (DW) excitations. We wonder how one can use interactions between DW excitations, arising from general MPO algebras, to design protocols to obtain non-trivial gates.

\subsection*{Acknowledgements}
We thank Norbert Schuch and Frank Verstraete for inspiring  discussions. JGR thanks Alex Turzillo for helpful comments on the manuscript. This work has been partially supported by the European Research Council (ERC) under the European Union’s Horizon 2020 research and innovation programme through the ERC-CoG SEQUAM (Grant Agreement No. 863476). L.L. is supported by a PhD fellowship from the Research Foundation Flanders (FWO)

\appendix
\section{Existence and properties of the fusion and action tensors}\label{ap:proofs}

In this section we prove the existence of the fusion and action tensors and the properties that they satisfy. We start by assuming the closedness condition of Eq.\ \eqref{algcond}. That is, there is a linear map $B: \mathcal{M}_\chi \otimes \mathcal{M}_\chi \to \mathcal{M}_\chi$ (indendent of $n$), where $\mathcal{M}_\chi$ is the space of $ \chi \times \chi$ matrices, that for every $X,Y \in \mathcal{M}_\chi$ there is a $Z \in \mathcal{M}_\chi$ such that $B(X\otimes Y) = Z$. 

For a linear map we can always express it as its minimal rank decomposition: $B=\sum_\mu W^\mu \otimes V^\mu$ so that $B(X\otimes Y)= \sum_\mu W^\mu (X\otimes Y) V^\mu$ where the matrices $W$ and $V$ are not unique. These matrices correspond to the fusion tensors.

The matrices of the map $B$ are supported on the virtual level of the MPO. Since the virtual level has a block decomposition, the matrices in $\mathcal{M}_\chi$ can also be decomposed in that way. Let us first define the projectors $P_a$ onto the subspace supported by the block $a$ (with dimension $\chi_a$), that is $P_a=\id_a\oplus 0_{\rm rest}$. With these projectors we can define $W_{ab}^{c,\mu} = P_cW^\mu (P_a\otimes P_b)$. In general, given $a,b,c$ not all $W_{ab}^{c,\mu}$ are linear independent, some of them are $0$, instead only $N_{ab}^c$ of them are. For any matrix $Z\in \mathcal{M}_\chi$, we denote by $Z_c= P_c Z P_c$ so we can write $Z_c =  \sum_{\mu=1}^{N_{ab}^c} W_{ab}^{c,\mu}(X_a\otimes Y_b) V_{ab}^{c,\mu}$. From the block structure of the MPO tensors and Eq.\ \eqref{algcond} (the assumption) we have that $O^n_{T_a,X_a}  {\cdot} O^n_{T_b,Y_b} = \sum_c O^n_{T_c,Z_c}$, for every $n$, and from the previous relation this is equal to $ \sum_c  \sum_{\mu=1}^{N_{ab}^c}  O^n_{T_c,W_{ab}^{c,\mu}(X_a\otimes Y_b) V_{ab}^{c,\mu}} $. Since the previous equation is true for all $X_a,Y_b$ the following is satisfied

\begin{equation}\label{decompopen}
			\begin{tikzpicture}
		  \draw[red] (0,0) -- (2,0);
		  \draw[red] (0,0.4) -- (2,0.4);
		  \foreach \x in {0.5, 1, 1.5}{
		    \node[tensor] (t\x) at (\x,0) {};
		    \node[tensor] (t\x) at (\x,0.4) {};
		    \draw (\x,-0.3) --++ (0,1);
		    \node[] at (-0.1,0.4) {$a$};
		    \node[] at (-0.1,0) {$b$};
      }
		\end{tikzpicture}  \   
		= \sum_c \sum_{\mu=1}^{N_{ab}^c}
		\begin{tikzpicture}
				  \draw[red] (0,0) -- (2,0);
		  \foreach \x in {0.5, 1, 1.5}{
		    \node[tensor] (t\x) at (\x,0) {};
		    \draw (\x,-0.3) --++ (0,0.6);
      }
          \pic (w2) at (2,0) {V1=c/a/b};
    \node[anchor=north,inner sep=2pt] at (w2-bottom) {$\mu$};
              \pic (w1) at (0,0) {W1=c/a/b};
    \node[anchor=north,inner sep=2pt] at (w1-bottom) {$\mu$};
		\end{tikzpicture} \ ,
	\end{equation}
for every length $n$ of the MPO. In particular this equation is true for $n=1$: 
\begin{equation*}
  \begin{tikzpicture}
    \draw[red]  (-0.5,0)--(0.5,0);
    \draw[red]  (-0.5,-0.5)--(0.5,-0.5);
    \node[irrep] at (-0.35,0) {$a$};
    \node[irrep] at (-0.35,-0.5) {$b$};
    \node[tensor] (t) at (0,0) {};
    \node[tensor] (t) at (0,-0.5) {};
    \draw (0,-0.8) -- (0,0.3);
  \end{tikzpicture} =
  \sum_{c,\mu}
  \begin{tikzpicture}[baseline=-1mm]
    \node[tensor] (t) at (0,0) {};
    \draw (0,-0.3) -- (0,0.3);
    \pic (v) at (0.5,0) {V1=c/a/b};
    \pic (w) at (-0.5,0) {W1=c/a/b};
     \node[anchor=south,inner sep=2pt] at (v-top) {$\mu$};
    \node[anchor=south,inner sep=2pt] at (w-top) {$\mu$};
  \end{tikzpicture} \ , 
\end{equation*}
which is Eq.\ \eqref{fusiontensors}. 
Eq.\ \eqref{decompopen} for $n=2$ reads 
\begin{align*}
			\begin{tikzpicture}
		  \draw[red] (0,0) -- (1.5,0);
		  \draw[red] (0,0.4) -- (1.5,0.4);
		  \foreach \x in {0.5, 1}{
		    \node[tensor] (t\x) at (\x,0) {};
		    \node[tensor] (t\x) at (\x,0.4) {};
		    \draw (\x,-0.3) --++ (0,1);
		    \node[] at (-0.1,0.4) {$a$};
		    \node[] at (-0.1,0) {$b$};
      }
		\end{tikzpicture}  \   
		 &= \sum_c \sum_{\mu=1}^{N_{ab}^c}
		\begin{tikzpicture}
				  \draw[red] (0,0) -- (1.5,0);
		  \foreach \x in {0.5, 1}{
		    \node[tensor] (t\x) at (\x,0) {};
		    \draw (\x,-0.3) --++ (0,0.6);
      }
          \pic (w2) at (1.5,0) {V1=c/a/b};
    \node[anchor=north,inner sep=2pt] at (w2-bottom) {$\mu$};
              \pic (w1) at (0,0) {W1=c/a/b};
    \node[anchor=north,inner sep=2pt] at (w1-bottom) {$\mu$};
		\end{tikzpicture} \ .
\end{align*}
Also, using twice Eq.\ \eqref{decompopen} for $n=1$, we can write 
\begin{align*}
			\begin{tikzpicture}
		  \draw[red] (0,0) -- (1.5,0);
		  \draw[red] (0,0.4) -- (1.5,0.4);
		  \foreach \x in {0.5, 1}{
		    \node[tensor] (t\x) at (\x,0) {};
		    \node[tensor] (t\x) at (\x,0.4) {};
		    \draw (\x,-0.3) --++ (0,1);
		    \node[] at (-0.1,0.4) {$a$};
		    \node[] at (-0.1,0) {$b$};
      }
		\end{tikzpicture}  \   
		&= \sum_{c,d} \sum_{\mu,\nu =1}^{N_{ab}^c}
      		\begin{tikzpicture}
						  \foreach \x in {-0.5, 1.1}{
		 \draw[red] (\x-0.4,0) --++ (0.8,0);
		    \node[tensor] (t\x) at (\x,0) {};
		    \draw (\x,-0.3) --++ (0,0.6);
      }
		 \pic (w1) at (-1,0) {W1=c/a/b};
        \node[anchor=north,inner sep=2pt] at (w1-bottom) {$\mu$};
                \pic (w2) at (0.0,0) {V1=c/a/b};
    \node[anchor=north,inner sep=2pt] at (w2-bottom) {$\mu$};
    		 \pic (w3) at (0.6,0) {W1=d//};
        \node[anchor=north,inner sep=2pt] at (w3-bottom) {$\nu$};
        		 \pic (w4) at (1.6,0) {V1=d/a/b};
        \node[anchor=north,inner sep=2pt] at (w4-bottom) {$\nu$};
		\end{tikzpicture}.
	\end{align*}
We can now use the injectivity of every block of the MPO tensor, and apply on the previous equation $T_c^{-1}$ and $T_d^{-1}$ on the first and second tensor respectively and obtain:
$$ \delta_{c,d}  \sum_{\mu=1}^{N_{ab}^c}
		\begin{tikzpicture}
		 \draw[red] (0.6,0) -- (0.9,0);
		\node[anchor=south,irrep] at (0.75,0) {$c$};
          \pic (w2) at (1.5,0) {V1=c/a/b};
    \node[anchor=north,inner sep=2pt] at (w2-bottom) {$\mu$};
              \pic (w1) at (0,0) {W1=c/a/b};
    \node[anchor=north,inner sep=2pt] at (w1-bottom) {$\mu$};
		\end{tikzpicture} 
		= \sum_{\mu,\nu =1}^{N_{ab}^c}
      		\begin{tikzpicture}
		 \pic (w1) at (-1,0) {W1=c/a/b};
        \node[anchor=north,inner sep=2pt] at (w1-bottom) {$\mu$};
                \pic (w2) at (0.0,0) {V1=c/a/b};
    \node[anchor=north,inner sep=2pt] at (w2-bottom) {$\mu$};
    		 \pic (w3) at (0.6,0) {W1=d//};
        \node[anchor=north,inner sep=2pt] at (w3-bottom) {$\nu$};
        		 \pic (w4) at (1.6,0) {V1=d/a/b};
        \node[anchor=north,inner sep=2pt] at (w4-bottom) {$\nu$};
		\end{tikzpicture}.
	$$
The linear independence of the fusion tensors implies that 
\begin{equation*}
   \begin{tikzpicture}\label{eq:ortho}
    \pic (v) at (0,0) {V1=d/a/b};
    \pic (w) at (0.6,0) {W1=c//};
     \node[anchor=north, inner sep=2pt] at (v-bottom) {$\mu$};
    \node[anchor=north, inner sep=2pt] at (w-bottom) {$\nu$};
   \end{tikzpicture} 
  = 
  		\begin{tikzpicture}
		 \draw[red] (0.3,0) -- (0.9,0);
		\node[anchor=south,irrep] at (0.6,0) {$c$};
		\node[scale=1.2] at (1.6,0) {$ {\cdot} \delta_{c,d}\delta_{\mu,\nu}$};
		\end{tikzpicture} ,
\end{equation*}
which corresponds to the orthogonality relations of the fusion tensors. 

With the same procedure we can prove that the condition of Eq.\ \eqref{eq:compatible} implies the existence of the action tensors and that they satisfy Eq.\ \eqref{fusiontensors2} and their orthogonality relations.\\

\section{The symmetrized parent Hamiltonian is non-zero}\label{app:sumnonull}

Given an MPS tensor $A$, the parent Hamiltonian $H=\sum_i h_i$ is constructed by defining the translationally invariant local terms $h = \id - \Pi_S$, where $\Pi_\mathcal{S}$ is the orthogonal proyector onto $\mathcal{S}$:
$$\mathcal{S} = \left\{ \sum_{\{i_1, i_2\}} \tr[XA^{i_1} A^{i_2}]|{i_1}, {i_2}\rangle, X\in \mathcal{M}_{D} \right\}.$$

If the MPS is invariant under an on-site symmetry $U_g=u_g^{\otimes n}$ of a group $G$ it is clear that $u^{\otimes 2}_g \mathcal{S} = \mathcal{S}$. 
In the following we show that the symmetrized local term proyector  $h' = \frac{1}{|G|}\sum_g u^{\otimes 2}_g  h u^{\otimes 2}_{\myinv{g}}$ is a non-zero proyector onto $\mathcal{S}^{\perp}$. To do that we first notice that $h' = \id- \frac{1}{|G|}\sum_g u^{\otimes 2}_g  \Pi_\mathcal{S} u^{\otimes 2}_{\myinv{g}} \equiv \id - \Pi'_{\mathcal{S}'}$. It is clear that $\tr[\Pi'_{\mathcal{S}'}] = \tr[\Pi_S]$ so it is different from zero by assumption and $\Pi'_{\mathcal{S}'}$ and $\Pi_{\mathcal{S}}$ have the same rank. We can also see that $\Pi'_{\mathcal{S}'}$ is actually a proyector onto $\mathcal{S}$. For that we take any vector $v$ in $\mathcal{S}$. Since $u^{\otimes 2}_{\myinv{g}} v $ belongs to $\mathcal{S}$ then $\Pi_\mathcal{S} u^{\otimes 2}_{\myinv{g}}v = u^{\otimes 2}_{\myinv{g}} v $, so that $\Pi'_{\mathcal{S}'} v = \frac{1}{|G|}\sum_g u^{\otimes 2}_g  \Pi_\mathcal{S} u^{\otimes 2}_{\myinv{g}} v = \frac{1}{|G|}\sum_g u^{\otimes 2}_g  u^{\otimes 2}_{\myinv{g}} v = v $.

If the MPS is invariant under an MPO representation of a ${C}^*$-WHA we use the canonical left integral $\Omega$ and the linear map $T$, defined in \cite{molnar22}, to symmetrize the Hamiltonian local terms. Concretely we map $\Pi_{\mathcal{S}}$ to $\Pi'_{\mathcal{S}'} \equiv \sum \Delta(\Omega_{(1)}) \Pi_{\mathcal{S}} \Delta(T(\Omega_{(2)}))$, where $\Delta$ corresponds to the coproduct of the ${C}^*$-WHA and we have used Sweedler notation; $\Delta(x)=\sum x_{(1)}\otimes x_{(2)}$.  With the properties of the aforementioned objects we can check that $\Pi'_{\mathcal{S}'}$ is hermitian, $\tr[\Pi'_{\mathcal{S}'}] =\tr[ \Delta(1) \Pi_{\mathcal{S}}]$ so it is non-zero and that $\Pi'_{\mathcal{S}'}$ is also a proyector on $\mathcal{S}$. The main property to use is $\sum \Delta(\Omega_{(1)}) \Delta(T(\Omega_{(2)})) = \sum \Delta(\Omega_{(2)}) \Delta(T(\Omega_{(1)})) = \Delta(1)$.

\bibliographystyle{unsrtnat}
\bibliography{bibliography.bib}

\end{document}